\documentclass[a4paper,11pt]{article}    

\usepackage{amsmath,amsthm,wasysym,amssymb,amsfonts,cite,enumerate,rotating,graphicx}
\usepackage{bm} 
\def\d{\mathrm{d}}
\newcommand{\bgmetric}[2]{\bar{g}_{#1 #2}}
\def \boldu {\mbox{\boldmath$u$}}                   
\def \boldE {\mbox{\boldmath$E$}}                   
\def \boldZ {\mbox{\boldmath$Z$}}        

\usepackage{bm} 

\def \hbl {\mbox{\boldmath{$\hat \ell$}}}
\def \bn {\mbox{\boldmath{$n$}}}
\def \hbn {\mbox{\boldmath{$\hat n$}}}

\def \bu {\mbox{\boldmath{$u$}}}
\def \bg {\mbox{\boldmath{$g$}}}

\def \hbm #1 {\mbox{\boldmath{$\hat m^{(#1)}$}}}
\def \bmd #1 {\mbox{\boldmath{$m_{(#1)}$}}}
\def \bT {\mbox{\boldmath{$T$}}}
\def \bR {\mbox{\boldmath{$R$}}}
\def \bF {\mbox{\boldmath{$F$}}}
\def \bC {\mbox{\boldmath{$C$}}}

\def \bk {\mbox{\boldmath{$k$}}}
\def \bl {\mbox{\boldmath{$\ell$}}}
\def \bM {\mbox{\boldmath{$m$}}}

\newcommand{\be}{\begin{equation}}
\newcommand{\ee}{\end{equation}}
\newcommand{\bea}{\begin{eqnarray}}
\newcommand{\eea}{\end{eqnarray}}

\newcommand{\pul}{{\textstyle{\frac{1}{2}}}}

\newtheorem{defn}{Definition}[section]
\newtheorem{prop}{Proposition}[section]

\newcommand{\Phis}{\Phi^\mathrm{S}} 
\newcommand{\Phia}{\Phi^\mathrm{A}} 
\newcommand{\M}[1]{{\stackrel{#1}{M}}}  
\newcommand{\tho}{{\textrm{\thorn}}}
\newcommand{\rhob}{\bm{\rho}}

\newcommand{\beqn}{\begin{eqnarray}}
\newcommand{\eeqn}{\end{eqnarray}}
\newcommand{\pa}{\partial}
\newcommand{\ba}{\begin{array}}
\newcommand{\ea}{\end{array}}
\newcommand{\pp}{{\it pp\,}-}
\newcommand{\beqnn}{\begin{eqnarray*}}
\newcommand{\eeqnn}{\end{eqnarray*}}

\newcommand{\ind}{$\mathcal{I}$-non-degenerate}

\newcommand{\del}{{\delta}}
\newcommand{\Del}{{\Delta}}

\newcommand{\kap}{{\kappa}}
\newcommand{\la}{{\lambda}}

\newcommand{\Om}{\Omega} 
\newcommand{\Ps}{\Psi}   

\newcommand{\taub}{{\boldsymbol \tau}}
\newcommand{\kapb}{{\boldsymbol \kap}}

\def \lb {\mbox{\boldmath{$\ell$}}}
\def \nb {\mbox{\boldmath{$n$}}}
\def \mb #1 {\mbox{\boldmath{$m^{(#1)}$}}}
\newcommand{\eb}{\mathbf{e}}
\newcommand{\ebh}{\mathbf{\hat e}}

\numberwithin{equation}{section}

\begin{document}

\title{Algebraic classification of higher dimensional spacetimes based on null alignment}

\author{Marcello Ortaggio\thanks{ortaggio@math.cas.cz}, Vojt\v ech Pravda\thanks{pravda@math.cas.cz}, Alena Pravdov\'a\thanks{pravdova@math.cas.cz} \\
Institute of Mathematics, Academy of Sciences of the Czech Republic \\ \v Zitn\' a 25, 115 67 Prague 1, Czech Republic}

\maketitle

\begin{abstract}
{We review recent developments and applications of the classification of the Weyl tensor in higher dimensional Lorentzian geometries. First, we discuss the general setup, i.e. main definitions and methods for the classification, some refinements and the generalized Newman-Penrose and Geroch-Held-Penrose formalisms. Next, we summarize general results, such as a partial extension of the Goldberg-Sachs theorem, characterization of spacetimes with vanishing (or constant) curvature invariants and the peeling behaviour in asymptotically flat spacetimes. Finally, we discuss certain invariantly defined {families} of metrics and their relation with the Weyl tensor classification, including: Kundt and Robinson-Trautman spacetimes; the Kerr-Schild ansatz in a constant-curvature background; purely electric and purely magnetic spacetimes; direct and (some) warped products; and geometries with certain symmetries. To conclude, some applications to quadratic gravity are also overviewed.}
\end{abstract}

\tableofcontents

\section{Introduction}

Almost a decade has passed since a classification scheme for the Weyl tensor of higher dimensional spaces with Lorentzian signature was put forward  \cite{Milsonetal05,Coleyetal04}. This is based on the concept of null alignment (as explained below) and extends to any dimensions $n>4$ the well-known Petrov classification  \cite{Petrov,Stephanibook}, to which it reduces for $n=4$. Over the past few years, a deeper geometric understanding of the null alignment method has been achieved, along with several developments, and a number of applications have been presented. The aim of the present paper is thus to review those new results which are, in our view, most important. Already published proofs and extended discussions are not  repeated here, and  readers will be referred to related references for more details.  

Already in 2008, there appeared a review on the classification of the Weyl tensor in higher dimensions \cite{Coley08}, where some useful information complementary to the one given here can be found {(see also \cite{Reall_algrev} for a recent introductory review).}
However, several new results have been published since then, in particular on the Geroch-Held-Penrose formalism \cite{Durkeeetal10}, on spacetimes ``characterized by their invariants'' \cite{Hervik11}, on the Goldberg-Sachs theorem \cite{DurRea09,Ortaggioetal12,OrtPraPra12}, on alternative approaches to the classification \cite{Ortaggio09,Senovilla10,ColHer09,Coleyetal12},  {on perturbations of near-horizon geometries  \cite{DurRea11a,DurRea11b,Godazgar:2011sn}} and on other aspects. We thus believe that summarizing {some of} these and other recent developments will be useful.

The plan of the paper is already illustrated in detail by the table of contents, and here it will suffice to just comment on the general structure. The first part (sections \ref{sec_classific} and \ref{sec_NP_GHP}) is devoted to presenting the formalism. First, a null alignment classification is set up that can be applied to any tensor. Then, this is  specialized to the Weyl tensor, for which refinements and alternative approaches are also mentioned. The Newman-Penrose (NP) and Geroch-Held-Penrose (GHP) formalisms are also described since these are extremely useful computational tools, especially for algebraically special spacetimes, {and have been already used to study various distinct problems mentioned in this review}.
In the ``central'' part (sections~\ref{sec_GS}--\ref{sec_asympt}), we discuss general applications of the classification: an extension of the Goldberg-Sachs theorem; a characterization of spacetimes for which all curvature invariants vanish or are constant; and the asymptotic and peeling behaviour of the Weyl tensor, with special interest for asymptotically flat spacetimes. The final part of the paper (sections~\ref{sec_shearfree}--\ref{sec_QG}) summarizes more specific applications, describing certain important classes of spacetimes which can be defined invariantly assuming certain geometric conditions (e.g., their ``geometric optics'', some discrete symmetry properties of the Weyl tensor, or some continuous symmetries of the spacetime). No field equations are generally assumed, so that most of the results are purely geometric. However, in some cases,  emphasis is put on Einstein spacetimes (defined by $R_{ab}=Rg_{ab}/n$, {with cosmological constant normalized as $(n-2)R=2n\Lambda$}), since these represent vacuum solutions in general relativity. A partial exception to this is section~\ref{sec_QG}, where some algebraically special solutions to quadratic gravity are considered.

\section{Classification of tensors via null alignment}

\label{sec_classific}

\subsection{General tensors}

The focus of this paper is the algebraic classification of the Weyl tensor {based on null alignment, as put forward in \cite{Milsonetal05,Coleyetal04},} and its applications {in higher dimensional gravity}. However, the {scheme} of \cite{Milsonetal05,Coleyetal04} applies, {more generally,} to arbitrary tensors on Lorentzian manifolds. Thus, in this section, we develop the classification for a general tensor $\bT$. The classification of the Weyl tensor then follows immediately {as a special case}.  

\subsubsection{Null frames} 

In the tangent space, we introduce  a (null) frame
\begin{equation}
  \{\lb \equiv \eb_{(0)}=\eb^{(1)},
    \nb \equiv \eb_{(1)} = \eb^{(0)},
   \ \bmd{i} = \mb{i} \equiv \eb_{(i)} = \eb^{(i)} \}, \ 
    \label{frame}
\end{equation}
with two null vectors $\bl$ and $\bn$ and $n-2$ spacelike vectors $\mbox{\boldmath{$m^{(i)}$}}=\mbox{\boldmath{$m_{(i)}$}} $ ({hereafter indices $i,j,\ldots$ take values from 2 to $n-1$}) obeying
\be
\ell^a \ell_a= n^a n_a = 0, \qquad   \ell^a n_a = 1, \qquad \ m^{(i)a}m^{(j)}_a=\delta_{ij}.  \label{ortbasis}
\ee
Frame indices  denoted by $\hat{a},\hat{b}, \dots $ and coordinate indices denoted by $ a, b, \dots $  take values from 0 to  $n-1$. The corresponding form of the  metric, 
\be
g_{a b} = 2\ell_{(a}n_{b)} + \delta_{ij} m^{(i)}_a m^{(j)}_b , \label{metric} 
\ee
is preserved under Lorentz transformations. 
The group of (real) proper  orthochronous Lorentz transformations  is generated by {\em null rotations} of one of the null frame vectors  about the other, {i.e.,}
\bea
&&\hbl =  \mbox{\boldmath{$\ell$}} +z_i {\mbox{\boldmath{$m^{(i)}$}}} -\frac{1}{2} z^i z_i\, \bn ,    
\qquad \hbn =  \bn,  \qquad 
    \mbox{\boldmath{$\hat m^{(i)}$}} =  \mbox{\boldmath{$m^{(i)}$}} - z_i \bn ,\label{nullrot}\\
 &&    \hbn =  \bn +z'_i {\mbox{\boldmath{$m^{(i)}$}}} -\pul z'^{i} z'_i\, \mbox{\boldmath{$\ell$}},  
\qquad 
\hbl =  \mbox{\boldmath{$\ell$}} ,    
\qquad
    \mbox{\boldmath{$\hat m^{(i)}$}} =  \mbox{\boldmath{$m^{(i)}$}} - z'_i \bl ,\label{nullrot_l}
\eea
with $2(n-2)$ real parameters $z_i$ and $z'_i$, {\em spins} described by an $SO(n-2)$  matrix $X^i_{\ j}$ 
\be
\hbl =  \bl, \qquad   \hbn = \bn, \qquad   \hbm{i} =  X^{i}_{\ j} \mbox{\boldmath{$m^{(j)}$}}  , \label{spin}
\ee
and {\em boosts} with a parameter {$\lambda> 0$}
\be
\hbl = \lambda \bl, \qquad    \hbn = \lambda^{-1} \bn, \qquad   \hbm{i} = \mbox{\boldmath{$m^{(i)}$}}  \label{boost}.
\ee

\subsubsection{Definitions and general theory}

{Before defining the algebraic types of $\bT$ based on alignment, we  need to} give several  useful definitions introduced in \cite{Milsonetal05,Coleyetal04}.

\begin{defn}[Boost weight (b.w.)]
A quantity  $q$  has  a {boost weight} {$($b.w.$)$}  ${\rm b}$  if it transforms under a boost~\eqref{boost} according to
\be
\hat q = \lambda^{\rm b} q . 
\ee 
\end{defn}
For a frame component $T_{\hat{a}\hat{b}\dots\hat{c}} = e^a_{(\hat{a})} e^b_{(\hat{b})} \dots e^c_{(\hat{c})} T_{ab\dots c}
$ of a tensor $\bT$, it follows that
\be
{\hat T}_{\hat{a}\dots \hat{b}}\equiv T(\ebh_{(\hat{a})},\dots \ebh_{(\hat{b})}) =\lambda^{{\rm b}(T_{\hat{a}\dots \hat{b}})}T_{\hat{a}\dots \hat{b}},    \ \  
\ee
and thus b$(T_{\hat{a}\dots \hat{b}})$ can be conveniently expressed as number of  0's minus number of 1's
in the frame component indices. 
Sometimes we will simply use the symbol ${\rm b}$. 

\begin{defn}[Boost order]
{The} boost order of a tensor $\bT$ with respect to  the null frame $\bl$, $\bn$, ${\mbox{\boldmath{$m^{(i)}$}}}$ is the maximum
b.w. of its frame components
\be
{\rm bo}(\bT)={\rm max}\left\{ {\rm b}(T_{\hat{a}\dots \hat{b}})\ |\ T_{\hat{a}\dots \hat{b}}\not= 0\right\}.
\ee
\end{defn}

Looking at transformations \eqref{nullrot_l}--\eqref{boost},  it is straightforward to show \cite{Milsonetal05} that

\begin{prop}[Invariant character of boost order  \cite{Milsonetal05}]
\label{prop_bo}
Let $\bl$, $\bn$, $\mbox{\boldmath{$m^{(i)}$}} $ and $\hbl$, $\hbn$, $\hbm{i} $ be two null frames with $\bl$ and $\hbl$ being
scalar multiples of each other. Then, the boost order of a given tensor is the same relative to both frames.
\end{prop}

Thus, the boost order of a tensor depends only on the choice of null direction $\left\langle\bl\right\rangle$ and we will denote it ${\rm bo}_{\bl }(\bT)$. For components of the {\em highest b.w.},  
it also follows {that, under null rotations about $\bl$ \eqref{nullrot_l},}
\be
\hat T_{\hat{a}\dots \hat{b}}\equiv T(\hat \bM_{(\hat{a})}\dots \hat \bM_{(\hat{b})})=T( \bM_{(\hat{a})}\dots  \bM_{(\hat{b})})\equiv T_{\hat{a}\dots \hat{b}} .
\ee

\begin{defn}[ANDs and multiplicity]
  Let  ${\bT}$  be a tensor and let  ${\rm bo}_{\rm{max}}(\bT)$ denote the maximum value 
of  ${\rm bo}_{\bl }(\bT)$   taken over all null vectors~$\bl$, i.e.,
\be
{\rm bo}_{\rm{max}}(\bT)={\rm max} \{ {\rm bo}_{\bl } (\bT)\ |\ {\  \left\langle \bl\right\rangle   \rm{\ is \ null} } \}. 
\ee
We say  that a vector $\bl$ is an aligned null direction $(${\rm AND}$)$ of a tensor $\bT$ whenever   ${\rm bo}_{\bl}(\bT) < {\rm bo}_{\rm{max}}(\bT)$,  and  we call the integer  ${\rm bo}_{\rm{max}}(\bT)-{\rm bo}_{\bl }(\bT)$ its multiplicity.
\label{def_AND}
\end{defn}

\begin{defn}[PAT and SAT]

We define the principal alignment type $(${\rm PAT}$)$ of a tensor $\bT$ as the integer  
\be
	\mbox{\rm PAT}={\rm bo}_{\rm{max}}(\bT)-{\rm bo}_{\rm{min}}(\bT),
\ee
where
\be
{\rm bo}_{\rm{min}}(\bT)
={\rm min} \{ {\rm bo}_{\bl }(\bT)\ |\ \left\langle \bl\right\rangle  \rm{\ is \ null} \}. 
\ee

Choosing $\bl$ with the maximal multiplicity $($which is equal to  {\rm PAT}$)$, we define the secondary alignment type, {\rm SAT}, to be {the integer}  
\be
	\mbox{\rm SAT}={\rm bo}_{\rm{max}}(\bT)-{\tilde {\rm bo}}_{\rm{min}}(\bT),
\ee
with
\be
{\tilde {\rm bo}}_{\rm{min}}(\bT)={\rm{min}}  \{ {\rm bo}_{\bn }(\bT)\  |\ \left\langle \bn\right\rangle\ 
\rm{\ is \ null}, \ \langle\bn\rangle\not= \langle\bl\rangle \}.
\ee
\end{defn}             

\begin{defn}[Alignment type]\label{def_align_type} 
	The alignment type of an arbitrary tensor consists of {the pair of}  integers  $({\rm PAT},{\rm SAT})$.
\end{defn}

To determine the alignment type of a tensor one has to project the tensor $\bT$ on the null frame and sort its components by their b.w.
\be
\bT=\sum_{\rm b} (\bT)_{({\rm b})},\label{decompT}
\ee
where
\be
(\bT)_{({\rm b})}=\sum T_{\hat{a}\dots \hat{b}} \mbox{\boldmath{$m$}}^{(\hat{a})}\dots \mbox{\boldmath{$m$}}^{(\hat{b})}, \qquad
{\rm b}(T_{\hat{a}\dots \hat{b}})={\rm b}.
\ee
Then using null rotations \eqref{nullrot} {and \eqref{nullrot_l} about $\bn$ and $\bl$}, one has to set as many leading and trailing terms in \eqref{decompT} as possible to zero.

For a tensor $\bT$, we define the following algebraic types {in terms of its PAT (and SAT)} \cite{Milsonetal05,Coleyetal04,Hervik11,HerOrtWyl12}:
\begin{defn}[Algebraic types] 
\label{def_algT}
A {non-vanishing} tensor $\bT$ is of 
\begin{itemize}
\item
type G if {\rm PAT}$= 0$, i.e. for all frames   $(\bT)_{ ({\rm bo}_{\rm{max}} ( {{\bT}}  ))}  \not= 0$,  
\item
type I if  {\rm PAT}$ \geq 1$, i.e. there exists a frame such that $(\bT)_{({\rm bo}_{\rm{max}}(\bT))}= 0$,\footnote{Sometimes a different definition of type I is used, e.g. in \cite{ColHerPel09a}. However, in the case of the Weyl tensor, this definition becomes in fact equivalent to ours.}
\item
 type II if  {\rm PAT}$\geq {\rm bo}_{\rm{max}}(\bT)$, i.e. there exists a frame such that \mbox{$\bT=\sum_{{\rm b}\leq 0}(\bT)_{({\rm b})}$}, 
\item
type D if  {\rm PAT}$={\rm bo}_{\rm{max}}(\bT)=${\rm SAT}, i.e. there exists a frame such that  \mbox{$\bT=(\bT)_{(0)}$},
\item
type III if  {\rm PAT}$\geq {\rm bo}_{\rm{max}}(\bT)+1$, i.e. there exists a frame such that \mbox{$\bT=\sum_{{\rm b}< 0}(\bT)_{({\rm b})}$}, 
\item
type N if  {\rm PAT}$=2{\rm bo}_{\rm{max}}(\bT)$, i.e. there exists a frame such that \mbox{$\bT=(\bT)_{(-{\rm bo}_{\rm{max}}(  {\tiny{\bT}}  ))}$}.
\end{itemize}
\end{defn}

{Note that this definition can be {equivalently given} in terms of (the maximal possible) multiplicity of ANDs of  $\bT$.}
Let us remark that, although (for convenience) expressed in terms of a given frame, the above definitions of algebraic types are in fact frame independent and thus invariant (this follows from proposition~\ref{prop_bo}, see also sections \ref{Sec_BelDeb} and \ref{sec_superenergy} for alternative viewpoints in the case of the Weyl tensor).
{According to definition~\ref{def_algT},} type N is a  subcase of type III which is in turn a subcase of type II, etc. {Sometimes, we use the term {\it genuine type}\footnote{Note, however, that in the literature the algebraic {types} of the Weyl tensor have {a  ``mixed''} meaning, e.g., sometimes {``type II'' means ``type II or more special'' and sometimes ``genuine type II''} (note the use of ``genuine'' in table \ref{tab:types}).} 
{I, II, III, meaning that $\bT$ is of type I, II or III and {\em not} more special}, i.e. not II/D/III/N, D/III/N and N, respectively  {(nor zero).} At points where $\bT$ vanishes the corresponding type is dubbed O.}

Except for the type D, all the above {types} are {\em principal} types {(see a summary in table~\ref{tab_principal})} since they refer to the alignment properties of just one AND  {(i.e. SAT is not relevant)}. If there exist at least two distinct ANDs, i.e. SAT$ \geq 1$,  one may further use another  AND for refining the algebraic types using again definition \ref{def_algT}, leading e.g. to {{the} genuine} types I$_i$,  II$_i$  or III$_i$ for the {{\em secondary}} alignment types (1,1), (${\rm bo}_{\rm{max}}(\bT)$,1) and 
(${\rm bo}_{\rm{max}}(\bT)+1$,1)
, respectively.

\begin{table}[htb]
  \begin{center}
  \begin{tabular}{|c|c|l|} 
    \hline Genuine type & PAT & $\bT$   \\ \hline
    G   &  $0$  &    $\sum_{\rm b}(\bT)_{({\rm b})}$ \\ 
    I  &  {$\geq 1$, $ < {\rm bo}_{\rm max}$}  &  $\sum_{{\rm b}<  {\rm bo}_{max}}(\bT)_{({\rm b})}$   \\ 
    II  &  ${\rm bo}_{\rm{max}}(\bT)$  &  $\sum_{{\rm b}\leq 0}(\bT)_{({\rm b})}$  \\ 
    III &  {$\ge{\rm bo}_{\rm{max}}(\bT)+1$, $<2{\rm bo}_{\rm{max}}$}  &  $\sum_{{\rm b}< 0}(\bT)_{({\rm b})}$   \\ 
     N &  $2{\rm bo}_{\rm{max}}(\bT)$  &$\sum_{{\rm b}=- {\rm bo}_{\rm{max}}}(\bT)_{({\rm b})}$   \\
\hline
  \end{tabular}
  \caption{(Genuine) principal alignment types for a tensor $\bT$. Recall that the (secondary) type D is a subtype of type II such that {PAT}$={\rm bo}_{\rm{max}}(\bT)=${SAT}, i.e. $\bT=\sum_{{\rm b}=0}(\bT)_{({\rm b})}$.}
  \label{tab_principal}
   \end{center}
\end{table}

Before moving to the Weyl tensor, let us illustrate this classification on {simpler} examples.

\subsubsection{Application to vectors, symmetric rank-2 tensors and bivectors}

\label{subsubsec_vectors_etc}

A vector 
\be
\mbox{\boldmath{$v$}}=v_0\bn +v_i \mbox{\boldmath{$m^{(i)}$}}+v_1\bl\ \ 
\ee
has ${\rm bo}_{\rm{max}}(\mbox{\boldmath{$v$}})=1$. {The three algebraic classes of vectors are the following.}
(i) A timelike  vector ($v^a v_a<0$) is of  type G (alignment type $(0,0)$), i.e. there are no ANDs.   
(ii) A spacelike vector ($v^a v_a>0$) is of type  D (alignment type $(1,1)$). (iii) A null vector ($v^a v_a=0$) is of  type N (alignment type $(2,0)$).

The decomposition of a symmetric rank-2 tensor $\bR$ (such as the Ricci tensor), {with ${\rm bo}_{\rm{max}}(\bR)=2$,}
is 
\bea
  R_{ab} = \overbrace{R_{00}\, n_{a}n_b}^{+2} +\overbrace{2R_{0i}\, n_{(a} m^{(i)}{}_{b)}}^{+1} +
  \overbrace{2R_{01}\, n_{(a}\ell_{b)}+ R_{ij}\, m^{(i)}{}_{(a} m^{(j)}{}_{b)}}^0\nonumber\\
   +\overbrace{2R_{1i}\, \ell_{(a} m^{(i)}{}_{b)}}^{-1}+\overbrace{R_{11}\, \ell_{a}\ell_b}^{-2} .  
\eea
Possible types are G - $(0,0)$, I$_i$ - $(1,1)$ (note that the {alignment type $(1,0)$} is forbidden by proposition~4.10 of \cite{HerOrtWyl12}), II - $(2,0)$, II$_i$ -  $(2,1)$, D - $(2,2)$, III -  $(3,0)$, III$_i$ - $(3,1)$, N - $(4,0)$. In the case of four dimensions, the relation with Segre types is discussed in \cite{Milsonetal05}, while a discussion in arbitrary dimensions can be found in appendix~B of \cite{HerOrtWyl12}. Note, for example, that for proper Einstein spacetimes the Ricci tensor is of type D.

A bivector (an antisymmetric  tensor of rank two)  $\bF$  can be decomposed as
\bea
  F_{ab} = \overbrace{2F_{0i}\, n_{[a} m^{(i)}{}_{b]}}^{+1} +
  \overbrace{2F_{01}\, n_{[a}\ell_{b]}+ F_{ij}\, m^{(i)}{}_{[a}
    m^{(j)}{}_{b]}}^0+ 
   \overbrace{2F_{1i}\, \ell_{[a} m^{(i)}{}_{b]}}^{-1}.  
\eea
Thus ${\rm bo}_{\rm{max}}(\bF)=1$.
In general dimension, the following cases occur:
type G  - $(0,0)$,  
II - $(1,0)$, 
D -  $(1,1)$ 
and N - $(2,0)$. 
In even dimensions, there always exists an AND, and thus, a bivector is of type II or more special (see the paragraph before lemma~4.6 of \cite{BerSen01} and proposition~4.4 in \cite{Milson04}).
In four dimensions, only  two cases exist \cite{Stephanibook,Hallbook}:
type D - $(1,1)$  
with canonical form $F_{ab}=\lambda m^{(2)}{}_{[a} m^{(3)}{}_{b]}+\mu n_{[a}\ell_{b]}$
and type N - $(2,0)$  
with canonical form $F_{ab}=\lambda \ell_{[a} m^{(2)}{}_{b]}$ (in~\cite{Hallbook}, the relation with the Segre types is also discussed). 
See appendix~B of \cite{HerOrtWyl12} for related discussions (including canonical forms and Segre types of $F_{ab}$) in arbitrary dimension.

\subsection{{Classification of} the Weyl tensor}

\label{subsec_Weyl}

We can now finally apply the scheme described above to the Weyl tensor {(${\rm bo}_{\rm{max}}(\bC)=2$)},  {which} satisfies the identities
\be
C_{abcd}=C_{\{ a bc d\} } \equiv \pul (C_{[a b] [c d]}+ C_{[c d] [ab]}), \qquad
{C^{c}}_{acb}=0, \qquad C_{a[bcd]}=0 .
\label{Weyl_symm}
\ee
{Thanks to these, in $n$ spacetime dimensions, $C_{abcd}$ has $(n+2)(n+1)n(n-3)/12$ independent components. In the frame~\eqref{frame} it admits} the decomposition
\bea
   C_{abcd}\ = &&\phantom{+}
  \overbrace{
    4 C_{0i0j}\, n^{}_{\{a} m^{(i)}_{\, b}  n^{}_{c}  m^{(j)}_{\, d\: \}}}^{\mathrm{\scriptscriptstyle{ boost\ weight\ +2 
    }}}   
  +\overbrace{
    8C_{010i}\, n^{}_{\{a} l^{}_b n^{}_c m^{(i)}_{\, d\: \}} +
    4C_{0ijk}\, n^{}_{\{a} m^{(i)}_{\, b} m^{(j)}_{\, c} m^{(k)}_{\, d\: \}}}^{\scriptscriptstyle{+1
    }}  
  \nonumber \\
    && \left. \begin{array}{l}
      + 4 C_{0101}\, \, n^{}_{\{a} l^{}_{ b} n^{}_{ c} l^{}_{\, d\: \}} 
\;  + \;  4 C_{01ij}\, \, n^{}_{\{a} l^{}_{ b} m^{(i)}_{\, c} m^{(j)}_{\, d\: \}}  \\
      +8 C_{0i1j}\, \, n^{}_{\{a} m^{(i)}_{\, b} l^{}_{c} m^{(j)}_{\, d\: \}}
   +  C_{ijkl}\, \, m^{(i)}_{\{a} m^{(j)}_{\, b} m^{(k)}_{\, c} m^{(l)}_{\, d\: \}}
    \end{array} \right\} \scriptscriptstyle{ 0 
    }
  \label{eq:rscalars}\\[1mm] 
    &&+  \overbrace{
    8 C_{101i}\, l^{}_{\{a} n^{}_b l^{}_c m^{(i)}_{\, d\: \}} +
    4 C_{1ijk}\, l^{}_{\{a} m^{(i)}_{\, b} m^{(j)}_{\, c} m^{(k)}_{\, d\: \}}}^{\scriptscriptstyle{ {-1   
     }}} 
  +\overbrace{
      4 C_{1i1j}\, l^{}_{\{a} m^{(i)}_{\, b}  l^{}_{c}  m^{(j)}_{\, d\: \}}}^{\scriptscriptstyle{ { -2       
      }  }} ,\nonumber
\eea
{where the various components have been ordered by b.w.} {Throughout the paper the Weyl components will be expressed in the GHP notation of \cite{Durkeeetal10}, summarized in table~\ref{tab:weyl} along with} identities following from \eqref{Weyl_symm}. {For transformations {of the components} under null rotations, boosts and spins see \cite{ColHer09,Durkeeetal10}}. 

\begin{table}[htb]
  \begin{center}
  \begin{tabular}{|c|c|c|c|l|c|}
    \hline ${\rm b}$ & Compt. & Notation & $s$ & Identities & Independent compts. \\\hline
    2 & $C_{0i0j}$& $\Om_{ij}$   & 2 & $\Om_{ij} = \Om_{ji}$, $\Om_{ii}=0$ & $\pul n(n-3)$\\\hline
    1 & $C_{0ijk}$& $\Ps_{ijk}$  & 3 & $\Ps_{ijk} = -\Ps_{ikj}$, $\Ps_{[ijk]}=0$ 
                                                            & $\frac{1}{3} (n-1)(n-2)(n-3)$\\
      & $C_{010i}$& $\Ps_{i}$    & 1 & $\Ps_i = \Ps_{kik}$. & \\\hline
    0 & $C_{ijkl}$& $\Phi_{ijkl}$& 4 & $\Phi_{ijkl} = \Phi_{[ij][kl]} = \Phi_{klij}$, $\Phi_{i[jkl]}=0$
                                                          & $\frac{1}{12}(n-1)(n-2)^2(n-3)$\\
      & $C_{0i1j}$& $\Phi_{ij}$  & 2 & $\Phi_{(ij)} \equiv \Phis_{ij} = -\pul\Phi_{ikjk}$ & \\
      & $C_{01ij}$& $2\Phia_{ij}$& 2 & $\Phia_{ij} \equiv \Phi_{[ij]}$ & $\pul (n-2)(n-3)$ \\
      & $C_{0101}$& $\Phi$       & 0 & $\Phi=\Phi_{ii}$ & \\\hline
    -1& $C_{1ijk}$& $\Ps'_{ijk}$ & 3 & $\Ps'_{ijk} = -\Ps'_{ikj}$, $\Ps'_{[ijk]}=0$
                                                             & $\frac{1}{3} (n-1)(n-2)(n-3)$\\
      & $C_{101i}$& $\Ps'_{i}$   & 1 & $\Ps'_i = \Ps'_{kik}$. & \\\hline
    -2& $C_{1i1j}$& $\Om'_{ij}$  & 2 & $\Om'_{ij} = \Om'_{ji}$, $\Om'_{ii}=0$ & $\pul n(n-3)$\\
\hline
  \end{tabular}
    \caption{GHP notation \cite{Durkeeetal10} for the Weyl tensor components of boost weight ${\rm b}$ and spin $s$ for an $n\geq4$ dimensional spacetime. 
    \label{tab:weyl}}
  \end{center}
\end{table}

{In the case of the Weyl tensor, it is useful to specialize definition~\ref{def_AND} to}
\begin{defn}[WAND]
A null vector field $\bl$ is a  Weyl aligned null direction $($WAND$)$ if it is an aligned null direction of the Weyl tensor. A WAND $\bl$ is a multiple WAND {$($mWAND$)$} if its multiplicity is greater than $1$.  
\end{defn}

{It will also be useful to introduce a special nomenclature for Weyl tensors of type II or more special: 

\begin{defn}[{Algebraically special Weyl tensor}]
\label{def_algspec}
A {Weyl tensor} is said to be algebraically special if it admits a multiple WAND.\footnote{Note that this is different from the definition originally used in \cite{Milsonetal05,Coleyetal04} and later in some other papers. There type~I was also considered ``special'' when $n>4$.   { However, in analogy with the $n=4$ case (where algebraically special means admitting a {\em multiple} PND), {and} for other reasons \cite{GodRea09,Hervik11}, we use the terminology of definition~\ref{def_algspec}.}} 
\end{defn}

If the Weyl tensor is of the same algebraic type at all points of the spacetime, then the spacetime is said to be of the corresponding algebraic (Weyl) type (and similarly for an open region of the spacetime).

The {{\em genuine}} algebraic types of the Weyl tensor (``Weyl types'', {see the text following definition \ref{def_algT})} following from definition \ref{def_algT} and the comparison with  Petrov types in four dimensions  are given in table \ref{tab:types}.
Note that in four dimensions this algebraic classification is equivalent to the Petrov classification,  and the notion of WAND coincides with that of a principal null direction (PND). However, the case $n=4$ is somewhat special in that there exist always exactly four PNDs (possibly repeated), so that the type G does not exist, and there are even fewer possible types since I=I$_i$, II=II$_i$, III=III$_i$ \cite{Milsonetal05}. By contrast, an $n>4$ spacetime may admit no WANDs (type G, which is the generic situation \cite{Milsonetal05}), a finite number of WANDs, or infinitely many.  
In any dimension, there is a {\em unique} multiple WAND for the {genuine} types II (double), III (triple) and N (quadruple), whereas there exist at least two (both double) for type D. {However, in the latter case, there may exist also an infinity of mWANDs.}\footnote{As shown in \cite{Wylleman12}, for a general $n$, the set of mWANDs of a type D Weyl tensor is homeomorphic to a sphere S$^k$, the dimension $k$ being at most $n-4$. When $k=0$ the sphere S$^0$ reduces to two points, representing the unique pair of mWANDs. This is in fact the generic situation in any dimension and the only possibility for $n=4$.\label{foot_Wyll}} {For a given type, further subtypes can be defined when some (but not all) of the components of the Weyl tensor having maximal b.w. vanish \cite{Coleyetal04,Ortaggio09,ColHer09}, see section~\ref{subsubsec_spin} {and table~\ref{tab:equiv class}}.}

\vspace{1mm}
\begin{table}[!ht]
\begin{center}
\begin{tabular}{|cc||c|}
\hline
$n>$4 dimensions & & 4 dimensions \\
\hline
Weyl type & alignment type& Petrov type   \\
\hline
G     & (0,0) &      \\
I     & (1,0)   &  \\
${\rm I}_{i}$ & (1,1) & I \\
II    & (2,0)  &  \\
${\rm II}_{i}$ & (2,1) & II \\
D & (2,2) & D \\
III & (3,0) &  \\
${\rm III}_{i}$ & (3,1) & III \\
N & (4,0) & N \\
\hline
\end{tabular}
\caption{Possible {Weyl/alignment} {(genuine)} types in higher dimensions {(definitions~\ref{def_algT} and \ref{def_align_type})} compared to the four-dimensional case  \cite{Coleyetal04}.\label{tab:types}}
\end{center}
\end{table}

\subsection{Spin types: a refinement}

\label{subsubsec_spin}

From the viewpoint of the null alignment classification, for any given Weyl type more special than G, it is natural to employ a null frame \eqref{frame}, where $\bl$ has maximal multiplicity.\footnote{For type G, there is no aligned null direction, and for type I$_i$,  there are more than one with the same multiplicity. In order to be able to define the ``spin type'' at a point, for those types, one thus additionally needs to define a ``total ordering'' between the spin types associated to different null directions -- see \cite{Coleyetal12} for more details and for an explicit choice of total ordering in 5D.} This frame is defined up to null rotations about $\bl$ {\eqref{nullrot_l}}, spins \eqref{spin} and boosts \eqref{boost}. 
Boosts act ``trivially'' on the frame Weyl tensor components (in the sense that they produce a simple rescaling, at most), while null rotations leave Weyl components of maximal b.w. unchanged. On the other hand, components of fixed b.w. can be characterized in terms of basic constituents which transform under irreducible representations of the spins (for instance, for b.w. $-2$ components, these reduce to the tracefree symmetric matrix $\Omega'_{ij}$; for b.w. 0 components the constituents are the tracefree part of $\Phi_{ijkl}$, $\Phi$, $\Phi_{[ij]}$, and the tracefree part of $\Phi_{(ij)}$; and similarly for other b.w.s  -- cf. table~\ref{tab:weyl}). It follows that spin-invariant quantities
defined by the highest b.w. constituents are properties associated with $\bl$. As proposed in \cite{ColHer09} {(but see also \cite{Coleyetal04})}, it is thus sound to build {a} refinement of the alignment classification based on geometric relations between the {\em highest b.w. constituents}. 
The types arising are referred to as \emph{spin types}. In particular, these include the coarser subtypes (such as I(a), II(a), II(b), etc.) defined in \cite{Coleyetal04} (cf. also \cite{Ortaggio09,ColHer09}), {see below}.

In four dimensions, spins reduce to a U(1) transformation so that this refinement is trivial (except for type II(D), where it takes into account the possible vanishing of the real/purely imaginary part of $\Psi_2$). However, in any higher dimensions, the spin-type refinement indeed enables one to discriminate between Weyl tensors having the same alignment type. As a peculiar feature of five dimensions, the highest b.w. constituents are represented only by square matrices, vectors and a single scalar \cite{ColHer09,Coleyetal12} {(see also below)}. This makes it possible to carry out the 5D spin-type refinement, and its intersections with the null alignment and the bivector operator {(section~\ref{subsec_bivector})} schemes, in a fully explicit manner \cite{Coleyetal12}. As it turns out, spin types can be used, for instance, to discriminate between different type N (or type III) Weyl operators having the same Segre type. The same method works in principle in any higher dimensions but working out all details explicitly may become cumbersome. In particular, the tracefree part of $\Phi_{ijkl}$ is itself a ``Weyl tensor'' (in an $(n-2)$-dimensional Riemannian space), so that clearly the classification cannot be given in the closed form for a generic $n$. Nevertheless this scheme may be very useful for particular purposes. {Here, we simply summarize the coarser subclassification consisting of the ``subtypes'' mentioned in \cite{Coleyetal04,Ortaggio09,ColHer09}: although this does not cover all spin types, it has the advantage that it applies to arbitrary dimensions.

\paragraph{Type I} 
Type I is characterized by b.w. +1 components. These can be irreducibly} decomposed as \cite{ColHer09} 

\be
	\Ps_{ijk}=-{{\textstyle{\frac{1}{n-3}}}}(\delta_{ij}\Ps_k-\delta_{ik}\Ps_j)+\tilde\Ps_{ijk}, \qquad \tilde\Ps_{[ijk]}=0, \quad \tilde\Ps_{iji}=0, \quad \tilde\Ps_{i(jk)}=0.
\ee 
{Thus one can define two invariant subtypes:}
\begin{enumerate}[(a)]
\item
{I(a)}: $\Ps_i=0$ $ \Leftrightarrow$ $\Ps_{iji}=0$
\item
{I(b)}: $\tilde\Ps_{ijk}=0$ $ \Leftrightarrow$ $\Ps_{ijk}\Ps_{ijk}=\frac{2}{n-3}\Ps_{i}\Ps_{i}$.
\end{enumerate}

{In four dimensions, $\tilde\Ps_{ijk}$ vanishes identically, so that I$\equiv$I(b) and I(a)$\equiv$II. In five dimensions, $\tilde\Ps_{ijk}$ can be reduced by duality to a symmetric traceless matrix \cite{ColHer09,Coleyetal12}.}

\paragraph{Type II(D)} Zero b.w. components {are specified by $\Phia_{ij}$ and $\Phi_{ijkl}$. When $n>4$ the latter} can be decomposed like a Riemann tensor in an auxiliary $(n-2)$-dimensional Riemannian space (see also table~\ref{tab:weyl}): 
\be
	\Phi_{ijkl}=\tilde\Phi_{ijkl}-\frac{4}{n-4}\left( \delta_{i[k}\Phis_{l]j}-\delta_{j[k}\Phis_{l]i}\right)
+\frac{4}{(n-3)(n-4)}\Phi\delta_{i[k}{\delta}_{l]j} , \qquad \tilde\Phi_{ijkj}=0 \qquad (n>4) .
\ee
In turn one can decompose
\be
	\Phis_{ij}=\frac{\Phi}{n-2}\delta_{ij}+\tilde\Phi^S_{ij} , \qquad \tilde\Phi^S_{ii}=0 .
\ee

Therefore, the following {subtypes} appear: 
\begin{enumerate}[(a)]
	\item {II(a)}: $\Phi=0$,
	\item {II(b)}: $\tilde\Phi^S_{ij}=0$,
	\item {II(c)}: $\tilde\Phi_{ijkl}=0$,
	\item {II(d)}: $\Phia_{ij}=0$.
\end{enumerate}

{These subtypes can also combine if two or three of the above conditions hold simultaneously -- some of such} possible combinations are given in table~\ref{tab:equiv class}. 
The same subtypes apply to type D with a similar notation. In four dimensions, $\Phia_{ij}$ is fully specified by $\Phia_{23}$, and $\Phi_{ijkl}$ reduces to the only non-trivial component $\Phi_{2323}=-\Phi$, so that II$\equiv$II(bc) and II(ad)$\equiv$III. In five dimensions, $\tilde\Phi_{ijkl}$ vanishes identically, so that II$\equiv$II(c), and II(abd)$\equiv$III; moreover, by duality, $\Phia_{ij}$ can be reduced to a vector \cite{Coleyetal12}. In six dimensions,  $\tilde\Phi_{ijkl}$ and $\Phia_{ij}$ can be further decomposed into their (anti-)self-dual parts  \cite{ColHer09}.

\paragraph{Type III} {The subclassification of} type III is fully analogous to that of type I, with $\Ps'_{ijk}$ replacing $\Ps_{ijk}$, i.e. 
\begin{enumerate}[(a)]
\item
{III(a)}: $\Ps'_i=0$ $ \Leftrightarrow$ $\Ps'_{iji}=0$
\item
{III(b)}: $\tilde\Ps'_{ijk}=0$ $ \Leftrightarrow$ $\Ps'_{ijk}\Ps'_{ijk}=\frac{2}{n-3}\Ps'_{i}\Ps'_{i}$.
\end{enumerate}
Obviously, in four dimensions, III$\equiv$III(b) and III(a)$\equiv$N. In five dimensions, $\tilde\Ps'_{ijk}$ can be reduced to a symmetric traceless matrix \cite{ColHer09,Coleyetal12}.

\paragraph{Type G and N} {Type G and N spacetimes are characterized by b.w. $\pm2$ components, respectively, which} are represented by a symmetric traceless matrix $\Om_{ij}$/$\Om'_{ij}$. These types can thus be further classified in any dimension according to, e.g., {the Segre characteristic (or/and spin types, see above) of such matrices. See \cite{ColHer09,Coleyetal12} for discussions in four and five dimensions.}

\subsection{Bel-Debever criteria for the Weyl tensor} 

\label{Sec_BelDeb}

{The Weyl types are defined above in terms of the multiplicity of one (or two) WANDs, i.e. by the vanishing of certain Weyl components in a null frame. However, {by proposition~\ref{prop_bo}}, the multiplicity of a WAND does not depend on the chosen frame. Indeed, there exist manifestly frame-independent criteria that uniquely determine the multiplicity of a given null direction. These are either based on the Weyl tensor itself (the Bel-Debever criteria, this section), or on a superenergy tensor (built as a ``square'' of the Weyl tensor, see the next section).}
The Bel-Debever criteria represent a set of polynomial {conditions} for an unknown null vector $\bl$ to be a WAND of a given multiplicity. At present,
solving these polynomial equations (or prove the non-existence of a solution) seems to be  the simplest way how to determine the algebraic 
type of a given metric.

For a null vector $\bl$, the following equivalence holds in arbitrary dimension \cite{Milsonetal05,Ortaggio09}: 
\bea
 \bl \ {\rm \ is\ a\ WAND\ (type\ I)}   &\Leftrightarrow & \! \ell_{[e}C_{a]bc[d}\ell_{f]}\ell^b\ell^c   = 0,\label{BdB_I}\\
 \bl \ {\rm \ is\ a\ WAND\ of\ multiplicity\ } \geq 2 \ {\rm (type\ II)}  &\Leftrightarrow &\!    \ell_{[e}C_{a]b[cd}\ell_{f]}\ell^b  = 0,\label{BdB_II} \\
  \bl \ {\rm \ is\ a\ WAND\ of\ multiplicity\ } \geq 3 \ {\rm (type\ III)}    &\Leftrightarrow &  \!  \ell_{[e}C_{ab][cd}\ell_{f]}=0=C_{abc[d}\ell_{e]}\ell^c, \ \ \label{BdB_III}\\
\bl \ {\rm \ is\ a\ WAND\ of\ multiplicity\ } 4   \ {\rm (type\ N)}   &\Leftrightarrow & \!  C_{ab[cd}\ell_{e]}= 0,\   \ \label{BdB_N}
\eea
{where it is understood that each type is a special subcase of the preceding types (i.e. each of the conditions on the r.h.s. implies the preceding conditions).}

{Similar conditions for several of the subtypes of section~\ref{subsubsec_spin} also take a simple form and are summarized in table~\ref{tab:equiv class}.}
Note that  \eqref{BdB_I} \cite{Milsonetal05} coincides with the standard Bel-Debever condition for type I in four dimensions, while relations \eqref{BdB_II}--\eqref{BdB_N} \cite{Ortaggio09} are distinct from their four-dimensional counterparts \cite{Bel59,Bel62,Debever59,Debever59b} (see also, e.g., \cite{Stephanibook,Hallbook}). However, they become in fact equivalent when $n=4$. This is due to the fact that various distinct Weyl subtypes are equivalent in the special case $n=4$ {(see section~\ref{subsubsec_spin})} and thus are the corresponding Bel-Debever criteria \cite{Ortaggio09}.

\subsection{WANDs as principal null directions of the superenergy tensor}

\label{sec_superenergy}

In higher dimensions, a superenergy tensor can be defined as \cite{Senovilla00,Senovilla10} 
\be
 {\cal T}_{abcd}=C_{aecf}{{{C_{b}}^{ e}}_{ d}}^{ f}+C_{aedf}{{{C_{b}}^{ e}}_{ c}}^{ f}
 -\pul g_{ab}C_{efcg}{{C^{ef}}_d}^g -\pul g_{cd} C_{aefg}{C_b}^{efg}
 +{{\textstyle{\frac{1}{8}}}}
g_{ab}g_{cd}C_{efgh}C^{efgh},\label{superen}
\ee
with symmetries
\be
{\cal T}_{abcd}={\cal T}_{(ab)(cd)}={\cal T}_{(cd)(ab)}.
\ee
In four and five dimensions, the superenergy tensor is completely symmetric \cite{Senovilla00}. In four dimensions,  {it is moreover traceless, and} it reduces \cite{EdgarWin2003,Senovilla00} to the well-known Bel-Robinson tensor \cite{Bel62,Robinson97} 

\be
 {\cal T}_{abcd}=C_{aecf}{{{C_{b}}^{e}}_{ d}}^{ f}-{{\textstyle{\frac{1}{8}}}}g_{ab}g_{cd}C_{efgh}C^{efgh} \qquad (n=4) .\label{BelRob}
\ee

Necessary and sufficient polynomial conditions for WANDs of a given  multiplicity can be  also elegantly expressed in terms of the superenergy tensor \cite{Senovilla10}
\bea
 \bl \ {\rm \ is\ a\ WAND }  \ \ &\Leftrightarrow& \ \  {\cal T}_{abcd}\ell^a\ell^b\ell^c\ell^d =0 , \label{superI} \\
 \bl \ {\rm \ is\ a\ WAND\ of\ multiplicity\ } \geq 2  \ \ &\Leftrightarrow & \ \  {\cal T}_{abcd}\ell^b\ell^c\ell^d =0, \label{superII} \\
  \bl \ {\rm \ is\ a\ WAND\ of\ multiplicity\ } \geq 3  \ \ &\Leftrightarrow & \ \  {\cal T}_{abcd}\ell^c\ell^d =0, \ \ \ \ \label{superIII}\\
\bl \ {\rm \ is\ a\ WAND\ of\ multiplicity\ } 4  \ \ &\Leftrightarrow & \ \  {\cal T}_{abcd}\ell^d =0. \   \ \ \ \ \ \label{superN}
\eea
These conditions are equivalent to the Bel-Debever conditions {of} section \ref{Sec_BelDeb} \cite{Senovilla10}. Applications {of similar ideas} to arbitrary tensors are also discussed in \cite{Senovilla10}.

\begin{table}[th!]
  \begin{center}
    \begin{tabular}{|l|l|l|c|c|c|c|}
    \hline Type & Bel-Debever  & superenergy & +2 & +1 & 0 & -1 \\\hline
    I & $\ell_{[e}C_{a]bc[d}\ell_{f]}\ell^b\ell^c   = 0$& ${\cal T}_{abcd}\ell^a\ell^b\ell^c\ell^d=0$    &$\Om_{ij}$  &  &  &     \\
    I(a)  & $\ell_{[e}C_{a]bcd}\ell^b\ell^c   = 0$   &   & $\Om_{ij}$ & $\Ps_i $ & &  \\
    I(b) &  &   & $\Om_{ij}$ &  $\tilde\Psi_{ijk}$ & &  \\ [1mm] \hline
   II & $\ell_{[e}C_{a]b[cd}\ell_{f]}\ell^b  = 0$ & ${\cal T}_{abcd}\ell^a\ell^b\ell^c=0$   & $\Om_{ij}$ & $\Psi_{ijk}$ ($\Psi_i$) & &  \\
  II(a)& $\ell_{[e}C_{a]b[cd}\ell_{f]}\ell^b  = 0$ & & $\Om_{ij}$ & $\Psi_{ijk}$ ($\Psi_i$) & $\Phi$ &  \\
      & $C_{abcd}\ell^b\ell^c  = 0$   &   &  &  & &  \\
   II(b)& &   & $\Om_{ij}$ & $\Psi_{ijk}$ ($\Psi_{i}$) & ${\tilde \Phi}^S_{ij}$&  \\
   II(c) & &  & $\Om_{ij}$ & $\Psi_{ijk}$ ($\Psi_{i}$) & ${\tilde \Phi}_{ijkl}$ &   \\
   II(d) & $C_{ab[cd}\ell_{e]}\ell^b   = 0$  &  & $\Om_{ij}$ & $\Psi_{ijk}$  ($\Psi_{i}$) & $\Phia_{ij}$ &  \\
   II(abc) & $\ell_{[e}C_{ab][cd}\ell_{f]}= 0$& & $\Om_{ij}$ & $\Psi_{ijk}$ ($\Psi_{i}$) & $\Phi_{ijkl}$ 
                          &   \\
                           &  & & & & ($\Phi$, $\Phis_{ij}$)  &  \\
   II(abd) & $C_{abc[d}\ell_{e]}\ell^c = 0$& ${\cal T}_{abcd}\ell^a\ell^b=0$ & $\Om_{ij}$ & $\Psi_{ijk}$ ($\Psi_{i}$) & $\Phi_{ij}$  &  \\
    &  & & & & ($\Phi$, $\Phia_{ij}$) &  \\
   II'(abd) & $C_{abcd}\ell^d=0$ & & $\Om_{ij}$ & $\Psi_{ijk}$ ($\Psi_{i}$) & $\Phi_{ij}$  & $\Ps'_i$ 
                    \\
                     &  & & & & ($\Phi$, $\Phia_{ij}$) &  \\ [1mm] \hline
  III & $\ell_{[e}C_{ab][cd}\ell_{f]}= 0$&  ${\cal T}_{abcd}\ell^a\ell^c=0$  & $\Om_{ij}$ & $\Psi_{ijk}$ ($\Psi_{i}$) & $\Phi_{ijkl}$,         $\Phia_{ij}$  &   \\
   & $C_{abc[d}\ell_{e]}\ell^c= 0$& & & & ($\Phi$, $\Phis_{ij}$)  & \\
   III(a) & $\ell_{[e}C_{ab][cd}\ell_{f]}= 0$&   & $\Om_{ij}$ & $\Psi_{ijk}$ ($\Psi_{i}$) & $\Phi_{ijkl}$, $\Phia_{ij}$ & $\Ps'_i$  \\
     & $C_{abcd}\ell^d= 0$& & & & ($\Phi$, $\Phis_{ij}$)  &  \\
    III(b) & &   & $\Om_{ij}$ & $\Psi_{ijk}$ ($\Psi_{i}$) & $\Phi_{ijkl}$, $\Phia_{ij}$  &  $\tilde\Psi'_{ijk}$  \\
      &  & & & & ($\Phi$, $\Phis_{ij}$)  &  \\ [1mm] \hline
 N &  $C_{ab[cd}\ell_{e]}= 0$ &  ${\cal T}_{abcd}\ell^a=0$  & $\Om_{ij}$ & $\Psi_{ijk}$ ($\Psi_{i}$) & $\Phi_{ijkl}$, $\Phia_{ij}$ & $\Ps'_{ijk}$ \\       
 & & & & & ($\Phi$, $\Phis_{ij}$)  & ($\Ps'_i$)  \\   
\hline
  \end{tabular}
    \caption{Equivalent criteria for various algebraic types of the Weyl tensor. Note that Bel-Debever criteria or conditions involving the superenergy tensor  have not been worked out for all subcases.  The last four columns list vanishing Weyl components for a given (sub)type ordered by boost weight (components vanishing   due to the identities given in table~\ref{tab:weyl} are in  parentheses). {A useful ``mixed'' subtype of type II, namely II'(abd) (for which also some b.w. $-1$ components vanish), has also been defined (see also section~\ref{subsubsec_othersymm}).} 
    The same conditions can be also used for  the secondary classification (e.g., the condition for type II applied to a {second mWAND} $\bn$ in type D spacetimes). {Recall (section~\ref{subsubsec_spin})} that  in four dimensions the following equivalences hold: I(a)$=$II$=$II(b)$=$II(c), II(abc)$=$II(a), II(abd)$=$III and II'(abd)$=$III(a)$=$N \cite{Ortaggio09}.  
\label{tab:equiv class}}
  \end{center}
\end{table}

\subsection{Bivector operator approach and further refinements of the classification}

\label{subsec_bivector}

In four dimensions, the Petrov classification of the Weyl tensor admits various formulations that, while describing properties of different geometric objects (bivectors, null directions, spinors), are in fact equivalent (see, e.g., \cite{Stephanibook,penrosebook2,Hallbook} for reviews and references). It is thus natural to investigate whether such alternative methods also extend to higher dimensions, and whether they are still equivalent to the null alignment scheme. Two such approaches have been already summarized above in sections~\ref{Sec_BelDeb} and \ref{sec_superenergy}. Here, we discuss the bivector operator approach, originally proposed by Petrov himself \cite{Petrov} and recently studied in higher dimensions in \cite{ColHer09,Coleyetal12}.\footnote{Petrov already observed that the bivector method (in particular, a classification based on the Segre characteristic of a curvature operator acting on bivector space) was well applicable in any dimensions and with an arbitrary signature. But, as he noticed ``the number of possible types \ldots increases rapidly with $n$ and depends on the possible signatures'' \cite{Petrov},  he then naturally developed his scheme in full detail only in the $n=4$ case with Lorentzian signature.} As it turns out, and in sharp contrast with the $n=4$ case, this method is inequivalent to the null alignment classification when $n>4$. Since the latter is rather coarse, developing the algebraic classification of the Weyl bivector operator will thus also lead to a more refined scheme. A further refinement can be obtained by {intersecting this with the spin types described in \ref{subsubsec_spin}}.

Given a spacetime point $p\in M$, let $\wedge^2T_pM$ be the $N\equiv n(n-1)/2$-dimensional real vector space of
contravariant bivectors (antisymmetric two-tensor $F^{ab}=F^{[ab]}$) at $p$. By the first equation of~(\ref{Weyl_symm}), the map
\begin{equation}
\label{Cop}
{\sf C}:\quad F^{ab}\mapsto  C^{ab}{}_{cd}F^{cd}=F^{cd} C_{cd}{}^{ab} 
\end{equation}
determines a linear operator on $\wedge^2T_pM$, which we shall refer to as the {\em Weyl operator}. In a given basis of $\wedge^2T_pM$, ${\sf C}$ can be represented by an $N\times N$ matrix.  
One can thus study invariant properties of ${\sf C}$ in order to classify the possible algebraic structures of the corresponding Weyl tensor. As originally suggested by Petrov \cite{Petrov}, one can in particular characterize the Weyl operator in terms of its elementary divisors and the corresponding {\em Segre type}. It is well known that in four dimensions {the Petrov (i.e. Segre type) classification of the Weyl operator is fully equivalent to the null alignment type classification}: type I corresponds to $[111,\overline{111}]$, type D to $[(11)1,\overline{(11)1}]$, type II to $[21,\overline{21}]$, type N to $[(21,{21})]$ and type III to $[(3,{3})]$ \cite{Petrov,Stephanibook,Hallbook}. However, this appears to be a ``miracle'' of the $n=4$ case only, and in higher dimensions, a particular alignment type can allow
for different Segre types, and vice versa. For example, in five dimensions, the type N includes two different Segre types, namely [(2221111)] and [(22111111)] \cite{Coleyetal12} {(the former, in turn, consists of two different spin types)}. Furthermore, in higher dimensions, a type II Weyl operator can be nilpotent \cite{Coleyetal12} (in contrast to the 4D case): as a consequence, for instance, a nilpotent type D operator and a (certain) type III operator may have the same Segre type \cite{Coleyetal12}. It is clear that the two classification schemes are thus in general independent. 

This new feature of higher dimensions can in fact be exploited to arrive at a {\em refinement} of the null alignment classification: by intersecting the Weyl types with the Segre types, one can define a number of subtypes and, thus, discriminate between spacetimes which would be undistinguishable by considering only one of the two classification schemes. The examples mentioned above already demonstrate this. Additionally, let us mention that Myers-Perry black holes and Kerr black strings are both of type D, but they possess a different eigenvalue spectrum (see appendix~C of \cite{Coleyetal12} for details in the 5D case).

In addition to the Segre type, one may study other properties of the Weyl operator, such as rank, kernel, image, etc. 
This may be useful also for practical purposes, and for algebraically special Weyl types in 5D, it has been discussed in some detail in \cite{Coleyetal12} (but some of these results hold for any $n$). Let us just mention here, for example, that the two possible Segre types of type N Weyl operators can be also distinguished by just looking at rank$({\sf C})$. Furthermore, the difference between the indices of nilpotence serves as an easily testable criterion for distinguishing the alignment types~{III} and {N} (this was also previously noted in \cite{Coleyetal04vsi}), with no need to find the mWAND and counting its multiplicity.

Finally, let us observe  that the bivector operator approach has also proven fruitful in the study of spacetimes that can (or cannot) be characterized by their curvature invariants (see section~\ref{Sec_invars} for some results and references).

\subsection{Inequivalent  methods for the classification of the Weyl tensor}

\subsubsection{Spinor approach in five dimensions}

\label{subsubsec_spinors}

In four dimensions, Penrose presented a spinor classification of the Weyl tensor based on the multiplicity of the factors of the Weyl polynomial (which is constructed from the totally symmetric, four-index Weyl spinor) \cite{penrosespin}. Owing to the fact that an $SL(2,\mathbb{C})$ spinor naturally defines a null direction in spacetime, it is straightforward to see that Penrose's scheme is equivalent to the 4D null alignment (as well as to the Petrov-Segre type) classification (see also \cite{penrosebook2,Stephanibook}). 

As an extension of the Petrov-Penrose classification, De~Smet put forward a spinor classification of the Weyl tensor in 5D \cite{DeSmet02} (recently further investigated in \cite{Godazgar10}). In contrast to the 4D case, in 5D, the spinor classification is {\em not} equivalent to the null alignment classification considered in the present review (section~\ref{subsec_Weyl}). We thus only briefly summarize here a few basic facts about the spinor approach and mention some examples that demonstrate essential differences with respect to (wrt) the null alignment classification, referring the reader to the relevant references for more details.

Compared to 4D, a new feature of the 5D Weyl spinor is that, in general, the corresponding Weyl polynomial does not factorize (corresponding to the algebraically general ``type 4''). When factorization takes place one has an ``algebraically special'' spinor (or De~Smet) type. According to the possible multiplicity of the factors, there exist in principle twelve spinor types \cite{DeSmet02}. However, a careful analysis of the reality condition on the Weyl spinor reveals that only {\em eight} of those are actually permitted \cite{Godazgar10}. Explicit examples of most De~Smet types are known (see \cite{DeSmet02} for a definition of the notation): vacuum black ring (``type 4'' \cite{Godazgar10}); static black string (``type 22'' \cite{DeSmet02}); Schwarzschild and Myers-Perry black hole (``type $\underline{22}$'' \cite{DeSmet02,DeSmet04}); direct product of a 4D type N spacetime with a spacelike dimension (``type 1111'' \cite{Godazgar10}); and direct product of a timelike dimension with an Euclidean (self-dual and conformal to a K\"ahler metric) 4D Einstein space (``type $\underline{11}$ $\underline{11}$'' \cite{Godazgar10}).  In fact, the possible De~Smet types of general 5D direct/warped product spacetimes have been determined \cite{Godazgar10}.  See \cite{DeSmet02,DeSmet03,Godazgar10} for more examples.

The null alignment type of (some of) the above-mentioned metrics is given in table~\ref{tab_summary} so that one can compare how the two schemes work for such examples. More generally, the connection between the 5D null alignment and the De~Smet scheme has been analyzed systematically in \cite{Godazgar10}, and the possible De~Smet type of type D, III and N Weyl tensors (in the sense of section~\ref{subsec_Weyl}) has been determined. It turns out, for example, that type III and D Weyl tensors may be algebraically general in the spinor classification (such an explicit  example, pointed out in \cite{HervikPhD}, is the direct product of a Robinson-Trautman type III vacuum spacetime with a spacelike dimension), while type N is algebraically special also in the spinor classification. Conversely, any special spinor type {(except possibly for ``type 31'')} may correspond to null alignment type G: for instance, the ``homogeneous wrapped object'' is of De~Smet special type 22 \cite{DeSmet02} but of null alignment type G \cite{GodRea09}. It is thus clear that the two classification schemes are really distinct, and neither of those is a refinement of the other one. 

To our knowledge, in dimension higher that five, a spinor classification of the Weyl tensor has not yet been developed.

\subsubsection{Optical structure approach}

\label{subsubsec_TC}

The notion of a 4D null geodesic congruence with vanishing shear admits various higher dimensional generalizations, see, e.g., \cite{RobTra83,HugMas88} (and \cite{OrtPraPra12} for a recent overview). Consequently, it is conceivable that an extension of the Goldberg-Sachs theorem to higher dimensions might admit various, inequivalent formulations. One possible such extension was studied in \cite{Taghavi-Chabert11,Taghavi-Chabert11b} (a different one is discussed in section~\ref{sec_GS}). A definition of ``algebraically special'' spacetimes was proposed which, subject to a genericity assumption on the Weyl tensor (along with assumptions on the Ricci tensor allowing for Einstein spacetimes), implies the existence of an ``optical structure'' (see \cite{Taghavi-Chabert11,Taghavi-Chabert11b} for a definition). In four dimensions, this reduces to the standard type~II condition (and an optical structure indeed defines a geodesic shearfree null congruence in 4D). However, in five and higher dimensions, the ``algebraically special'' condition of \cite{Taghavi-Chabert11,Taghavi-Chabert11b} is {\em more restrictive} than {the condition} defined in section~\ref{subsec_Weyl} (definition~\ref{def_algspec}) since, in addition to the vanishing of positive b.w. components, it also constraints certain b.w. 0 and $-1$ components (see \cite{Ortaggioetal12} for details in five dimensions).
Counterexamples are known \cite{Taghavi-Chabert11,Taghavi-Chabert11b} showing that such ``algebraically special'' condition is {\em not necessary} for the existence of an optical structure. As shown in \cite{Ortaggioetal12,OrtPraPra12} (see proposition~\ref{prop_integrab} of the present paper), in 5D, this ``algebraically special'' condition can in fact be relaxed to the type II (or more special) condition defined in section~\ref{subsec_Weyl} (which is also not a necessary condition for the existence of an optical structure \cite{Ortaggioetal12}). By contrast, in more than five dimensions, this is likely not to be the case.

\subsection{Interpretation of different boost-weight components of the Weyl tensor using the equation of geodetic deviation}

Recently, the equation of geodetic deviation
\be
\label{EqGeoDev}
\frac{\mbox{D}^2 Z^a}{\mbox{d}\,\tau^2}=R^a_{\ bcd}\,u^b u^c Z^d ,
\ee
{which describes the relative acceleration of close free test particles (with no charge and no spin, 
and ``small'' relative velocities),}  
has been used for interpreting different b.w. components of the Weyl tensor {in higher dimensions \cite{PodolskySvarc2012} (thus extending the classical works \cite{piraniriem,piraniinvariant,szekerescompass} for $n=4$)}. Here, ${\boldu=u^a\partial_a }$, with ${u^a=\frac{\mbox{d} x^a}{\mbox{d}\tau}}$, is the velocity of the reference particle moving along a timelike geodesic $x^a(\tau)$, 
with $\tau$ being its proper time. The separation vector ${\boldZ=Z^a\partial_a}$ connects the reference particle with another  test particle moving along a nearby timelike geodesic.

In a frame with one timelike vector $\boldE_{[0]}=\boldu$ and $(n-1)$ spacelike vectors $\boldE_{[I]}$, $\frac{\mbox{d}^2 Z^{[0]}}{\mbox{d}\tau^2}=0$, and by  choosing initial conditions, one can set $Z^{[0]}=0$. Introducing a frame \eqref{frame} by
 \be
 \label{NullFrame}
 \bl=\frac{1}{\sqrt{2}}(\boldu+\boldE_{[1]})\,, \qquad \bn=\frac{1}{\sqrt{2}}(-\boldu+\boldE_{[1]})\,, \qquad \ \bmd{i} =\boldE_{[i]} ,
\ee
{and decomposing the Riemann tensor in terms of the Ricci and Weyl tensors (and using Einstein's equations to express the Ricci tensor in terms of the energy-momentum tensor $T_{ab}$)}, eq.~\eqref{EqGeoDev}   becomes 
\begin{eqnarray}
\ddot{Z}^{[1]}&=&  \frac{2\Lambda }{(n-1)(n-2)}\,Z^{[1]} -\Phi\,Z^{[1]}
       - \frac{1}{\sqrt{2}}\,(\,\Psi_{j}+\Psi'_{j})\,Z^{[j]} \nonumber\\
      &&  +\frac{8\pi}{n-2}\left[\,T_{[1][1]} \,Z^{[1]}+T_{[1][j]} \,Z^{[j]}-\Big(T_{[0][0]}+\frac{2}{n-1}\,T\Big)\, Z^{[1]}\,\right],\label{InvGeoDevFinal1}\\
\ddot{Z}^{[i]}&=&  \frac{2\Lambda }{(n-1)(n-2)}\,Z^{[i]} + \Phis_{ij}\,Z^{[j]}
       - \frac{1}{\sqrt{2}}\,(\,\Psi_{i}+\Psi'_{i})\,Z^{[1]} -\frac{1}{2}\,(\,\Omega_{{ij}}+\Omega'_{{ij}})\,Z^{[j]} \nonumber\\
      &&  +\frac{8\pi}{n-2}\left[\,T_{[i][1]} \,Z^{[1]}+T_{[i][j]} \,Z^{[j]}-\Big(T_{[0][0]}+\frac{2}{n-1}\,T\Big)\, Z^{[i]}\,\right].\label{InvGeoDevFinali}
\end{eqnarray}

{One can observe that, similarly as in 4D,} the influence of cosmological  constant $\Lambda$  is isotropic, the effect of $\Omega'_{ij}$ [$\Omega_{ij}$] is transverse
and corresponds to a transverse gravitational wave propagating in the $+\boldE_{[1]}$ [$-\boldE_{[1]}$]
direction, the components $\Psi'_{i}$ [$\Psi_{i}$]  
cause longitudinal deformations, and $\Phi$ and $\Phis_{ij}$  lead to ``Newton-Coulomb''-like tidal deformations. {Effects of $T_{ab}$ obviously depend on the specific matter content.} {Note that, in the approximations under which \eqref{EqGeoDev} applies, only (some of) the {\em electric} components \cite{Senovilla00,Senovilla01,HerOrtWyl12} of the Riemann tensor appear, and magnetic effects are thus not detectable at this level. See \cite{PodolskySvarc2012} for references going beyond this approximation.}

\section{Newman-Penrose and Geroch-Held-Penrose formalisms}

\label{sec_NP_GHP}

In four dimensions, Newman-Penrose (NP) and Geroch-Held-Penrose (GHP) formalisms are very useful computational tools, especially when studying spacetimes with a given Petrov type or admitting a null congruence with special geometric properties. These have been extended to higher dimensions 
in \cite{Pravdaetal04,OrtPraPra07,Coleyetal04vsi}\footnote{{{Note a}
minor correction to the paper \cite{OrtPraPra07}: {the} missing term $\frac{R}{(n-1)(n-2)}\delta_{i[k}\delta_{l]j}$ should be added on the r.h.s. of eq.~(11p). }}
 and \cite{Durkeeetal10}, respectively, and have been already proven {fruitful} in several studies. In addition to the applications reviewed in the present paper, let us mention that the GHP formalism has also been useful in the study of perturbations of near-horizon geometries {\cite{DurRea11a,DurRea11b,Godazgar:2011sn}.}  Let us now summarize these methods.

\subsection{Ricci rotation coefficients, optical matrix and optical constraint}

\label{subsec_ricci}

\subsubsection{Ricci rotation coefficients and optical matrix} 

We denote the covariant derivatives of the frame vectors as
\begin{equation}
 L_{ab} = \nabla_b \ell_a, \qquad N_{ab} = \nabla_b n_a, \qquad \M{i}_{ab} = \nabla_b m_{(i)a}.
 \label{LNM}
\end{equation}
The projections onto the basis are the Ricci rotation coefficients $L_{\hat{a}\hat{b}}$, $N_{\hat{a}\hat{b}}$, $\M{i}_{\hat{a}\hat{b}}$. 
Orthogonality properties of the basis~\eqref{ortbasis} imply 
\beqn
\label{eqn:ident1}
  & &  N_{0\hat{a}} + L_{1\hat{a}} = 0, \quad \M{i}_{0\hat{a}} + L_{i\hat{a}} = 0,
  \quad \M{i}_{1\hat{a}} + N_{i\hat{a}} = 0, \quad \M{i}_{j\hat{a}} + \M{j}_{i\hat{a}} = 0, \\
  & & L_{0\hat{a}} = N_{1\hat{a}} = \M{i}_{i\hat{a}} = 0. \label{eqn:ident2}
\eeqn
{The transformation properties of the Ricci rotation coefficients under  Lorentz transformations of the basis vectors are given in \cite{OrtPraPra07}.}

The vector field $\bl$ is tangent to a null geodesic congruence if, and only if,
\begin{equation}
  \kap_i \equiv L_{i0} \equiv L_{ab}m^{(i)a}\ell^b= 0. 
\end{equation}
In such  case, one can always choose an affine parameterization with $L_{10}=0$. {Moreover, one can pick up a frame {\em parallelly transported} along $\bl$, i.e. such that $\M{i}_{j0}=0=N_{i0}$ \cite{OrtPraPra07}. If $\det\rhob\neq 0$ {(see definition \eqref{rho})}, one can further set $\tau_i\equiv L_{i1}=0$ {by an appropriate choice of the frame} \cite{Durkeeetal10}.}

{In the space of the spacelike frame vectors, the {\em optical matrix} $\rhob$ is defined as}
\be
 \rho_{ij}\equiv L_{ij}\equiv L_{ab}m^{(i)a}m^{(j)b} . 
 \label{rho}
\ee
It is convenient to decompose $\rhob$ into its trace $\theta$ {(``\em expansion'')}, trace-free symmetric part $\sigma_{ij}$ and antisymmetric part $A_{ij}$
\bea
 \rho_{ij}=\sigma_{ij}+\theta\delta_{ij}+A_{ij} , \label{L_decomp} \qquad 
 \sigma_{ij}\equiv \rho_{(ij)}-\textstyle{\frac{1}{n-2}} \rho_{kk}\delta_{ij} , 
\qquad \theta\equiv\textstyle{\frac{1}{n-2}}{\rho_{kk}} , \qquad A_{ij}\equiv \rho_{[ij]} . \label{opt_matrices}
\eea
{\em Shear} and {\em twist} of $\bl$ are then given by the traces $\sigma^2\equiv\sigma^2_{ii}=\sigma_{ij}\sigma_{ji}$ and $\omega^2\equiv-A^2_{ii}=-A_{ij}A_{ji}$.
For an {\em affinely parametrized} $\bl$, the optical scalars can be expressed in terms of  $\bl$ as
\be
 \hspace{-1.3cm} \sigma^2=\ell_{(a;b)}\ell^{(a;b)}-\textstyle{\frac{1}{n-2}}\left(\ell^a_{\;;a}\right)^2 , \qquad \theta=\textstyle{\frac{1}{n-2}}\ell^a_{\;;a} , \qquad 
 \omega^2=\ell_{[a;b]}\ell^{a;b} .
\ee

Let us also introduce covariant derivatives along the frame vectors by
\be
D \equiv \ell^a \nabla_a, \qquad \bigtriangleup  \equiv n^a \nabla_a, \qquad \delta_i \equiv m^a_{(i)} \nabla_a .
\ee
{Their commutators can be found  in eqs.~(21)--(24) of \cite{Coleyetal04vsi}.}

\subsubsection{Sachs equation}

Contractions of the Ricci identity $v_{a;bc}-v_{a;cb}={R}_{sabc}v^s$ with various combinations of the frame vectors \eqref{frame} lead to the full set of Ricci identities (or ``NP equations'') given in \cite{OrtPraPra07}. Among these, the {\em Sachs equation} turns out to be particularly useful as it determines how $\rhob$ ``propagates'' along $\bl$. For a {\em geodesic}, affinely parametrized $\bl$ and using a frame parallelly transported along $\bl$, it reads simply \cite{Pravdaetal04,OrtPraPra07}
\be
	D\rho_{ij}=-\rho_{ik}\rho_{kj}-\Omega_{ij}-\frac{1}{n-2}R_{00}\delta_{ij} .
	\label{sachs}
\ee
This can be decomposed into its irreducible parts as \cite{OrtPraPra07} (see also \cite{LewPaw05} for previous related results)
\beqn
 & &  D\sigma_{ij}=-\left({\sigma^2}_{ij}-\textstyle{\frac{1}{n-2}}\sigma^2\delta_{ij}\right)-\left({A}^2_{ij}+{\frac{1}{n-2}}\omega^2\delta_{ij}\right)-2\theta\sigma_{ij}-\Omega_{ij} , \label{sachs_she} \\ 
 & &  D\theta=-{\frac{1}{n-2}}\sigma^2-\theta^2+\textstyle{\frac{1}{n-2}}\omega^2-{\frac{1}{n-2}}R_{00} , \label{sachs_exp} \\
 & &  DA_{ij}=-2\theta A_{ij}-2\sigma_{k[j}A_{i]k} . \label{sachs_twi}
\eeqn

\subsubsection{Optical constraint}
\label{sec_OC}

When the Riemann type is I (wrt $\bl$), the curvature tensor does not enter \eqref{sachs}. Under this assumption, this equation has been integrated in various special cases in order to determine the dependence of $\rhob$ on an affine parameter $r$ along $\bl$ in \cite{OrtPraPra10}; see also \cite{PodOrt06,OrtPraPra07,PraPra08,OrtPraPra09,OrtPraPra09b}. Certain important consequences will be discussed in an appropriate context in the following. It is however worth mentioning already here the case in which $\rhob$ obeys the 
so-called {\em optical constraint} (OC) \cite{OrtPraPra09} {(see \cite{Ortaggioetal12,OrtPraPra12} for recent discussions)}, i.e.
\be
	\rho_{ik} \rho_{jk} \propto \rho_{(ij)} . \label{OC}
\ee
{This} implies $[\rhob, \rhob^T]=0$ {so that} $\rhob$ is a {\em normal} matrix; {by} using spins,
it can be put into a block-diagonal form 
\bea
\rhob = \alpha{\rm diag}\left(1, \dots 1, 
\frac{1}{1+ \alpha^2 b_1^2}\left[\begin {array}{cc} 1 & -\alpha b_1 \\ 
   \alpha b_1 & 1  \label{canformL} \\
  \end {array}
 \right]
, \dots, 
\frac{1}{1+ \alpha^2 b_\nu^2} \left[\begin {array}{cc} 1 & -\alpha b_\nu \\ 
   \alpha b_\nu & 1 \\ \end {array}
 \right] 
 , 0, \dots ,0
\right).
\eea

This condition appears to be of interest in the study of algebraically special Einstein spacetimes and will occur several times in the following sections. In particular, it is satisfied by {the Kerr-Schild vector} of all (generalized) Kerr-Schild spacetimes \cite{OrtPraPra09,MalPra11} (section~\ref{subsec_KS}), by non-degenerate geodesic double WANDs in asymptotically flat type II vacuum spacetimes \cite{OrtPraPra09b} (section~\ref{subsec_asym_special}), by the  mWAND of type III/N  \cite{Pravdaetal04} (section~\ref{subsubsec_GS_III_N}) and (in five dimensions) genuine type II \cite{Ortaggioetal12,Wylleman_priv} Einstein spacetimes (see sections~\ref{subsubsec_GS_III_N} and \ref{subsubsec_GS_5d}, which also includes a discussion of type D). In 4D, this is a necessary condition for mWANDs \cite{Ortaggioetal12}, but it is not so in higher dimensions, in general \cite{Ortaggioetal12,OrtPraPra12}. See \cite{OrtPraPra09,OrtPraPra10,Ortaggioetal12} for  further discussions. Let us emphasize that we will {\em not} assume \eqref{OC} in what follows.

\subsection{GHP scalars}

By studying the transformation properties of $L_{\hat{a}\hat{b}}$ and $N_{\hat{a}\hat{b}}$  under  spins \eqref{spin} and boosts \eqref{boost} \cite{OrtPraPra07} we observe that, {with the exception of $L_{1\hat{a}}$ and $\M{i}_{j\hat{a}}$, the Ricci rotation coefficients} are GHP {scalars},\footnote{Note that GHP scalars are scalars under coordinate transformation, {but not} under frame transformations.} as defined in \cite{Durkeeetal10} by
\begin{defn}
  A quantity $q$ is a \emph{GHP scalar}  
  of spin $s$ and b.w. ${\rm b}$ if, and only if, it transforms as
  \begin{equation}
    T_{i_1...i_s} \mapsto X_{i_1 j_1}...X_{i_s j_s} T_{j_1...j_s} ,
    \label{Tspin}
  \end{equation}
  under spins $\mathbf{X}\in SO(d-2)$, and as
  \begin{equation}
    T_{i_1...i_s} \mapsto \la^{\rm b} T_{i_1...i_s} ,
    \label{Tboost}
  \end{equation}
  under boosts.
\end{defn}
GHP scalars appearing in the NP and GHP formalisms, their b.w. ${\rm b}$ and spin $s$ and their geometric interpretation 
are summarized in table \ref{tab:weights} \cite{Durkeeetal10}.

\begin{table}[htb]  
 \begin{center}
   \begin{tabular}{|c|c|c|c|l|}
    \hline NP & GHP & ${\rm b}$ & $s$ & Interpretation\\ [1mm]\hline
    $L_{ij}$  & $\rho_{ij}$  & 1  & 2 & expansion, shear and twist of $\lb$\\[1mm]
     $L_{ii}$  & $\rho=\rho_{ii}$  & 1  & 0 & expansion of $\lb$\\[1mm]
    $L_{i0}$  & $\kap_{i}$   & 2  & 1 & non-geodesity of $\lb$\\[1mm]
    $L_{i1}$  & $\tau_{i}$   & 0  & 1 & transport of $\lb$ along $\bn$\\[1mm]
    $N_{ij}$  & $\rho'_{ij}$ & $-1$ & 2 & expansion, shear and twist of $\bn$\\[1mm]
     $N_{ii}$  & $\rho'=\rho'_{ii}$ & $-1$ & 0 & expansion of $\bn$\\[1mm]
     $N_{i1}$  & $\kap'_{i}$  & $-2$ & 1 & non-geodesity of $\bn$\\[1mm]
    $N_{i0}$  & $\tau'_{i}$  & 0  & 1 & transport of $\bn$ along $\bl$\\[1mm]\hline
  \end{tabular}
  \caption{\label{tab:weights}List of those Ricci rotation coefficients $L_{\hat{a}\hat{b}}$ and $N_{\hat{a}\hat{b}}$ {(defined in \eqref{LNM})} which are also GHP scalars. {The first column gives the coefficients in the NP notation, the second column the corresponding GHP quantities. Columns ${\rm b}$ and $s$, give, respectively, the boost weight and spin of each quantity.}}
 \end{center}
\end{table}

If $q$ is a GHP scalar, then $D q$, $\Del q$ and $\del_i q$ are not. However, one can define new GHP differential operators, which are covariant, 
{as follows} \cite{Durkeeetal10}:
\begin{defn} \label{GHPderiv}
For a GHP scalar $T_{i_1 i_2...i_s}$ of b.w. ${\rm b}$ and spin $s$,
  the \emph{GHP derivative operators} $\tho$, $\tho'$, $\eth_i$ are defined  as
  \begin{eqnarray}
    \tho T_{i_1 i_2...i_s} &\equiv & D T_{i_1 i_2...i_s} - {\rm b} L_{10} T_{i_1 i_2...i_s} 
                                     + \sum_{r=1}^s \M{k}_{i_r 0} T_{i_1...i_{r-1} k i_{r+1}...i_s},\\
    \tho' T_{i_1 i_2...i_s} &\equiv & \Del T_{i_1 i_2...i_s} - {\rm b} L_{11} T_{i_1 i_2...i_s} 
                                     + \sum_{r=1}^s \M{k}_{i_r 1} T_{i_1...i_{r-1} k i_{r+1}...i_s},\\
    \eth_i T_{j_1 j_2...j_s} &\equiv & \del_i T_{j_1 j_2...j_s} - {\rm b} L_{1i} T_{j_1 j_2...j_s} 
                                     + \sum_{r=1}^s \M{k}_{j_r i} T_{j_1...j_{r-1} k j_{r+1}...j_s}.
  \end{eqnarray}
\end{defn}

As pointed out in \cite{Durkeeetal10}, $\tho$,  $\tho'$   and $\eth_i$ obey the Leibniz rule and annihilate $\del_{ij}$. Note that for a GHP scalar $T_{i_1 i_2...i_s}$ of b.w. ${\rm b}$ and spin $s$,  $\tho T_{i_1 i_2...i_s}$, $\tho' T_{i_1 i_2...i_s}$ and $\eth_j T_{i_1 i_2...i_s}$ are  GHP scalars, with {boost weights} 
(${\rm b}+1$, ${\rm b}-1$, ${\rm b}$) and spins ($s$,$s$,$s+1$), respectively. {Note also that the derivative operators $\tho$ and $D$ coincide if $\bl$ is geodesic and affinely parametrized and a parallelly transported frame is employed.}

For the Weyl components, we will follow the notation of \cite{Durkeeetal10}, summarized in table~\ref{tab:weyl} {(see \cite{Durkeeetal10}  for the  Ricci components)}.

{Contractions of the Bianchi identity $R_{ab[cd;e]}=0$ with various combinations of the frame vectors \eqref{frame} lead to the full set of Bianchi identities. They were first given in \cite{Pravdaetal04} and rewritten and simplified in GHP notation in \cite{Durkeeetal10} (where some redundancy in the equations of \cite{Pravdaetal04} was also removed\footnote{{L. Wylleman has observed that some further redundancy can still be removed (private communication).  }}).  The Ricci identities and the commutators have also been rewritten in GHP notation in \cite{Durkeeetal10}.

\section{Extensions of the Goldberg-Sachs theorem}

\label{sec_GS}

In four-dimensional general relativity, the Goldberg-Sachs (GS) theorem \cite{GolSac62} (see also \cite{NP}) has played an important role in the study of solutions of the Einstein equation with an algebraically special Weyl tensor. It is, in particular, the first step in exploiting the algebraically special property to solve the Einstein equation -- this is how the Kerr metric was discovered \cite{Kerr63}, for example. In this section we shall restrict to the case of Einstein (including Ricci-flat) spacetimes. If matter fields are not considered (but see, e.g., \cite{Stephanibook,penrosebook2,GovHilNur11} and references therein for generalizations), the 4D GS theorem can be stated as follows: in a non-conformally flat Einstein spacetime, a null vector field is a multiple PND if, and only if, it is geodesic and shearfree (i.e. $\kappa=0=\sigma$). It thus provides one with a fundamental connection between geometric optics of a null congruence and the algebraic structure of the Weyl tensor.

It is therefore natural to investigate whether a similar connection holds also in the case of higher dimensional Einstein spacetimes, with the notion of PND replaced by that of WAND (see section~\ref{subsec_Weyl}). However, it became clear already from the pioneering work of Myers and Perry \cite{MyePer86} (see also \cite{FroSto03}) that higher dimensional spinning black holes possess mWANDs with non-zero shear (although still geodesic), as opposed to the PNDs of the Kerr spacetime. More recent work has shown that in fact this behaviour is generic when $n>4$, {algebraically special} shearfree solutions being a very special subset of all algebraically special spacetimes. For instance, the mWAND is shearing in the case of direct products of an {expanding} algebraically special 4D solution with some flat direction (such as Schwarzschild or Kerr black {strings}/branes) \cite{PraPraOrt07}\footnote{{Let us here correct some typos and minor mistakes in \cite{PraPraOrt07}:  there should be 
$m=n-2$ instead of $n=m-2$  in the last sentence of section 2; in the third line of section 6, $m$ should be replaced by $n$; as pointed out in \cite{Durkee09} 
 there is a missing term $+\frac{1}{3}\Phia_{ij}L$ on the l.h.s. of eq. (44); in point~(2) after proposition~10 {the conditions on $\Phi^S_{ij}$/$\Phi^A_{ij}$ should be replaced by} $\Phi_{24}=0=\Phi_{34}=\Phi_{44}$;  there should be ``$A_{34}$ is arbitrary'' instead of $A_{23}$ in case (c) after proposition 11. Finally, the expression for $C_{ \hat{I} \hat{J} \hat{K} \hat{L}  }$ in (14) of [66] is valid also for $n_1=2$.}}
 and for {expanding} type III/N Einstein spacetimes \cite{Pravdaetal04}, and all geodesic twisting WANDs are shearing in odd dimensions \cite{OrtPraPra07}. Furthermore, it was observed in \cite{PraPraOrt07,Durkee09} that the higher dimensional Bianchi identities applied to {type II} spacetimes do not require the corresponding mWAND to be geodesic (as opposed to the 4D case), unless a ``genericity'' assumption on the Weyl tensor is made. While the first example of non-geodesic mWANDs was given in \cite{PraPraOrt07} in any $n\ge 7$ (see also \cite{Ortaggio07}), a number of solutions with the same property have been subsequently constructed using product geometries \cite{GodRea09,Durkee09}, also for $n=5,6$.

The above examples clearly demonstrate that the GS theorem does not extend in an obvious way to higher dimensions. A question thus arises as to whether an $n>4$ extension of the GS theorem exists at all, and if so, whether such an extension is unique, and what formulation(s) it admits. The most desirable form of a higher dimensional generalization of the GS theorem would be a statement of necessary and sufficient {\it algebraic} conditions on certain Ricci rotation coefficients (such as $\kapb$, $\rhob$)  for a null vector field $\lb$ to be a multiple WAND. Moreover, this statement should reduce (at least in some sense) to the standard GS theorem when $n=4$. 

However, recent work seems to indicate that there are no conditions which are both ``necessary and sufficient'' when $n>4$. Moreover, it turns out that in higher dimensions it is much more convenient to treat the ``geodesic part'' and the ``shearfree part'' of the ``would-be-GS theorem'' separately. The results obtained so far are summarized below.

\subsection{Geodesic part}

Preliminary results were obtained in \cite{Pravdaetal04,PraPraOrt07,Durkee09} but the complete formulation of the geodesic part has been proven in \cite{DurRea09}. The main result is the following proposition. 
\begin{prop}[Geodesic part of the HD GS theorem \cite{DurRea09}]
\label{prop_geodGS}
 An $n>4$ Einstein spacetime admits a multiple WAND if, and only if, it admits a geodesic multiple WAND. 
\end{prop}
Thanks to this result, there is no loss of generality in restricting attention to geodesic multiple WANDs. 

To arrive at this result, the authors of~\cite{DurRea09} proved an intermediate result which is of interest in its own right, since it characterizes spacetimes that admits a {\em non-geodesic} mWAND, namely 

\begin{prop}[Spacetimes with a non-geodesic mWAND \cite{DurRea09}]
 \label{prop_umbilic}
An $n>4$ Einstein spacetime that admits a non-geodesic multiple WAND is
foliated by totally umbilic, constant curvature, Lorentzian submanifolds of dimension
3 or greater and any null vector field tangent to the leaves of the foliation is a
multiple WAND.
\end{prop}

By combining propositions~\ref{prop_geodGS} and \ref{prop_umbilic}, it immediately follows that {\em a spacetime admitting a non-geodesic mWAND is necessarily of type D,\footnote{More specifically, the type is D and purely electric (cf. the definition in section \ref{subsec_PEPM}); see \cite{Wylleman12} and Remark~3.15 of \cite{HerOrtWyl12}.}  and possesses a continuous infinity of mWANDs} {(this infinity being $\infty^1$ or greater -- cf. also footnote~\ref{foot_Wyll})}.

The result of \cite{DurRea09} is in fact stronger in the special case $n=5$:

\begin{prop}[Geodesic part of the HD GS theorem for $n=5$ \cite{DurRea09}]
\label{prop_geodGS_5D}
 A five-dimensional non-conformally flat Einstein spacetime admits a non-geodesic multiple WAND if, and only if, it is locally isometric to one of the following: 
 \begin{enumerate}[(i)]
  \item a direct product dS$_3\times$S$^2$ or AdS$_3\times$H$^2$,
  \item a spacetime with metric
  	\beqn
 			& & \d s^2=f(r)dz^2+f(r)^{-1}dr^2+r^2\Omega^2(-\d t^2+\d x^2+\d y^2) , \\
 			& & \mbox{with } \ f(r)=k-\mu r^{-2}-\lambda r^2, \quad \Omega^{-1}=1+\frac{k}{4}(-t^2+x^2+y^2) ,
		\eeqn  
		where $\mu\neq0$, $k \in \{1,0,-1 \}$, $\lambda$ is (proportional to) the cosmological constant, and the coordinate 		$r$ takes values so that $f(r)>0$. 
 \end{enumerate}
\end{prop}

\subsection{Shearfree part}

As seen above, the standard shearfree condition $\sigma=0$ ($\Leftrightarrow \rho_{(ij)}\propto\delta_{ij}$, {cf.~\eqref{opt_matrices}}) is by far too restrictive when $n>4$ \cite{MyePer86,FroSto03,Pravdaetal04,OrtPraPra07,PodOrt06}. On the other hand, various geometric conditions can be considered which are different from $\sigma=0$ for a generic $n$, but which all reduce to $\sigma=0$ if one makes the special choice $n=4$ \cite{RobTra83,HugMas88,NurTra02,Trautman02a,Trautman02b,MasTag08,OrtPraPra09,Taghavi-Chabert11,Taghavi-Chabert11b,OrtPraPra12}.  Dealing with the shearfree part of the GS theorem is thus more delicate since one should first find what a proper formulation of a ``generalized shearfree condition'' is.

There appear to be conditions on $\rhob$ that are sufficient for $\lb$ to be a multiple WAND, namely $\rhob=0$ or $\rho_{ij}\propto\delta_{ij}$ \cite{PodOrt06,OrtPraPra07}, which are however clearly non-necessary. On the other hand, certain necessary conditions on $\rhob$ follow from the multiple WAND condition. Here we will summarize the latter results. However, explicit examples {\cite{Ortaggioetal12,OrtPraPra12}}  imply that (at least some of) these necessary conditions on $\rhob$ are {\em not sufficient} for the spacetime to be algebraically special.

\subsubsection{Results in five dimensions}

\label{subsubsec_GS_5d}

Let us first discuss the $n=5$ case, for which we have a clear-cut result. In~\cite{Ortaggioetal12}, the following proposition is proved.

\begin{prop}[Shearfree part of the HD GS theorem for $n=5$ \cite{Ortaggioetal12}]
\label{prop_shearGS_5D}
 In a five-dimensional algebraically special Einstein spacetime that is not conformally flat, there exists a geodesic multiple WAND $\lb$ and one can choose the orthonormal basis vectors $\mb{i} $ so that the optical matrix of $\bl$ takes one of the forms 
\bea
\label{5dform1}
 &&i) \ \ \ \  \rhob=b\left( \begin{array}{ccc} 1 & a & 0 \\ -a & 1 & 0 \\ 0 & 0& 1+a^2 \end{array} \right), 
 \\
 \label{5dform2}
 &&ii) \ \ \ \ \rhob=b\left( \begin{array}{ccc} 1 & a & 0 \\ -a & 1 & 0 \\ 0 & 0& 0 \end{array} \right),\qquad 
 \\
\label{5dform3}
 &&iii) \ \ \ \ \rhob=b\left( \begin{array}{ccc} 1 & a & 0 \\ -a & -a^2 & 0 \\ 0 & 0& 0 \end{array} \right) 
 ,
\eea
 {where $a$, $b$ are functions that may vary in spacetime.} If the spacetime is type III or type N, then the form must be $ii)$ {\rm \cite{Pravdaetal04}}.
\end{prop}

Kundt spacetimes ($\rhob=0$) are the only spacetimes belonging to more than one of the above classes {(in the trivial case $b=0$)}. For $b\neq0$, $\rhob$ is of rank 3, 2, 1 in case  $i)$, $ii)$, $iii)$, respectively. Only in case $iii)$ with $a\neq0\neq b$ does $\bl$ violate the optical constraint \eqref{OC}. Recently, it has been proven that  case $iii)$ with $b\neq0$  cannot occur for genuine type II spacetimes \cite{Wylleman_priv} {(see also  \cite{Harvey-in-prep})}, so that {\em in five dimensions the mWAND of a genuine type II  Einstein spacetime always obeys the optical constraint}. 
Furthermore, all type D {(and thus all)} solutions are known in case $iii)$ with $a\neq0\neq b$, since one has the following further result:

\begin{prop}[$n=5$ type D violating the OC \cite{Ortaggioetal12}]
A five-dimensional type D Einstein spacetime admits a geodesic multiple WAND violating the optical constraint $($i.e. case $iii)$ of proposition~\ref{prop_shearGS_5D} with $a\neq0\neq b$$)$ if, and only if, it admits a non-geodesic multiple WAND $($and is thus comprised of the spacetimes of proposition~\ref{prop_geodGS_5D}$)$.
\label{prop_nongeod}
\end{prop}

And thus combining these results one has the following proposition.

\begin{prop}[OC in $n=5$ {\cite{Ortaggioetal12,Wylleman_priv}}]
	{Any} five-dimensional algebraically special Einstein spacetime admits a geodetic multiple WAND obeying the optical constraint \eqref{OC}.
\end{prop}

In the twisting case, examples for each of the above canonical forms $i)$, $ii)$ and $iii)$ are, respectively, Myers-Perry black holes  (with $b=\sqrt{x} / (x+\chi^2)$, $a={\chi} / \sqrt{x}$, where $\chi=\sqrt{\alpha^2\cos^2\theta+\beta^2\sin^2\theta}$, 
$\alpha$ and $\beta$ being rotational parameters \cite{PraPraOrt07}); Kerr black strings;  and dS$_3\times$S$^2$ (in the latter case see \cite{Ortaggioetal12} for the form the corresponding mWAND $\bl$). 
When the twist vanishes ($a=0$) corresponding examples are, respectively, Robinson-Trautman (RT) spacetimes \cite{PodOrt06} (which in fact exhaust the case $i)$ with $a=0$); Schwarzschild black strings dS$_3\times$S$^2$ (with a suitable choice of a non-twisting $\bl$ \cite{Ortaggioetal12}). 

In \cite{Ortaggioetal12}, an example was given of an Einstein spacetime that is not algebraically special and yet admits a geodesic null vector field with an optical matrix of the form iii) above {(with $a=0$)}, showing that the above necessary conditions on $\bl$ are {\em not sufficient}.

Consequences of proposition~\ref{prop_shearGS_5D} have a geometrical meaning in terms of integrability properties of certain totally null distributions \cite{Ortaggioetal12,OrtPraPra12}. Combining the results of \cite{Ortaggioetal12} with the refinements of \cite{Wylleman_priv} (about type II) and of \cite{OrtPraPra12} (about Kundt spacetimes), we arrive at sufficient conditions for the existence of an ``optical structure'' \cite{Taghavi-Chabert11} in five-dimensional spacetimes (strengthening proposition~4.6 of  \cite{Ortaggioetal12}), namely

\begin{prop}[Sufficient conditions for an OS when $n=5$ {\cite{Ortaggioetal12,OrtPraPra12,Wylleman_priv} }]
\label{prop_integrab}
All five-dimensional algebraically special Einstein spacetimes admit an {\em optical structure}. In the case of type D spacetimes, there exist in fact $($at least$)$ two optical structures. 
\end{prop}
Observe that the converse of proposition~\ref{prop_integrab} does not hold. This can be seen by taking a black ring as a counterexample, since this admits a region where the Weyl type is I$_i$ \cite{PraPra05} while still possessing an optical structure \cite{Taghavi-Chabert11}.

A more detailed comparison between the results of \cite{Ortaggioetal12} and those of \cite{Taghavi-Chabert11} (where a definition of ``algebraically special''  is used which is different from the one of the present paper, see also sections~\ref{subsubsec_TC} and \ref{subsec_OSGS}) can be found in \cite{Ortaggioetal12}.

\subsubsection{Results in $n\ge 6$ dimensions for a non-twisting multiple WAND $\bl$}

There is a qualitative difference between $n=4,5$ and $n>5$, i.e. the Weyl tensor possesses new ``degrees of freedom'' in the components $\Phi_{ijkl}$ in the latter case. This makes, in particular, the study of the Bianchi identities more difficult. At present, there is thus no analogue of  proposition~\ref{prop_shearGS_5D} for $n>5$. Nevertheless, in the case of a non-twisting mWAND $\bl$ (i.e. $A_{ij}=0$), several interesting conclusions have been obtained that constrain the possible form of $\rhob$ \cite{OrtPraPra12}:
 
\begin{prop}[{Eigenvalue structure of $\rhob$ for $n\ge 6$ and $A_{ij}=0$}]\label{prop_GSHD}
 In an algebraically special Einstein spacetime of dimension $n\ge 6$ that is not conformally flat, the $($symmetric$)$ optical matrix of a non-twisting multiple WAND has at least {\em one double eigenvalue}. In {the following} special cases, stronger conditions hold and the most general {permitted} forms of $\rhob$  are, respectively:
\begin{enumerate}[(i)]
\item
 if $\Phi^A_{ij} \ne 0$: $\{a,a,0,\ldots , 0\}$ {\rm \cite{Ortaggioetal12}}, 
\item
 if $\det \rhob\not= 0$, $\Phi_{ij}\not=0$: $\{a,a,\ldots , a\}$ (Robinson-Trautman, $\Phi_{ij}\propto \delta_{ij}$, type D(bd)) {\rm \cite{OrtPraPra12}},  \label{RT}
 \item
 if  $\Phi_{ij}=0$ (type II(abd)): $\{a,a,b,b,c_1\ldots , c_{n-6}\}$  \label{Phi0} {\rm \cite{OrtPraPra12,Wylleman_priv}}, 
\item
 for types N, III: $\{a,a,0,\ldots , 0\}$ {\rm \cite{Pravdaetal04}}. 
 \label{NIII}
 \end{enumerate}
 \end{prop}

Note that in the above cases (i)--(iv) the matrix $\rhob$  possesses at least {\em two} double eigenvalues. This is, however, not true in general, as shown by further special cases studied in \cite{OrtPraPra12}. In particular, it is worth observing that explicit solutions for which all the non-zero eigenvalues of $\rhob$ are distinct and rank$(\rhob)>1$ have been constructed  in more than five dimensions \cite{OrtPraPra12}. These clearly violate the OC \eqref{OC} and are forbidden for $n=4,5$. 
Examples constructed in \cite{OrtPraPra12} demonstrate that {the condition that $\rhob$ has a double eigenvalue}, given in proposition~\ref{prop_GSHD}, is again not sufficient for $\bl$ to be a mWAND 
(whereas {the stronger} condition on $\rhob$ in (\ref{RT}) is \cite{PodOrt06}).
Integrability properties of certain totally null distributions that follow from the results presented above have also been analyzed in \cite{OrtPraPra12}. In particular, the case $n=6$ has been discussed in some detail there.

\subsubsection{Results for type III/N in any dimensions}

\label{subsubsec_GS_III_N}

In the case of type  N Einstein spacetimes, the five-dimensional result of proposition~\ref{prop_shearGS_5D} (case ii) therein) extends in fact to any dimensions:
\begin{prop}[Shearfree part of the HD GS theorem for type N \cite{Pravdaetal04}]
\label{prop_shearGS_III_N}
 In a type N Einstein spacetime of dimension $n\ge4$, one can choose the orthonormal basis vectors $\mb{i} $ so that the optical matrix of the unique multiple WAND $\lb$ $($which is geodesic$)$ takes the form
\bea
 \ \ \ \ \rhob=b\left( \begin{array}{cccc} 1 & a & 0 \ldots & 0 \\ -a & 1 & 0  \ldots & 0  \\ 0 & 0& 0  \ldots & 0  \\ & & \vdots &  \\ 0 & 0& 0  \ldots & 0 \end{array} \right) .
\eea
where $a$ and $b$ are functions that may vary in spacetime. 
\end{prop}

Note that rank$(\rhob)=2$ or 0 in any dimensions and that $\rhob$ obeys the OC. In the non-twisting case ($a=0$), proposition~\ref{prop_shearGS_III_N} is contained in proposition~\ref{prop_GSHD}  
(case~(\ref{NIII})).  Proposition~\ref{prop_shearGS_III_N} holds also for type III Einstein spacetimes provided {\em any} of the following conditions is satisfied: a) $n=5$; b) $\bl$ is non-twisting; c) $n>5$, $\bl$ is twisting but certain ``genericity'' assumptions on the Weyl tensor  hold (see \cite{Pravdaetal04} for the Ricci-flat case, while a generalization to proper Einstein spacetimes is straightforward).
 More general results including those types II and D that possess a twisting mWAND in more than five dimensions  are presently not known (except in special cases, see sections \ref{subsec_asym_special} {and \ref{subsec_KS}}).

\subsection{Optical structure approach}

\label{subsec_OSGS}

In~\cite{Taghavi-Chabert11,Taghavi-Chabert11b}, a higher dimensional extension of the Goldberg-Sachs theorem different from the one discussed above is studied. Instead of looking for ``canonical forms'' of $\rhob$, a generalized shearfree condition was phrased there in terms of the existence of an optical structure. We have already briefly discussed this in section~\ref{subsubsec_TC}, and we refer the reader to \cite{Taghavi-Chabert11,Taghavi-Chabert11b} for more details (see also \cite{Ortaggioetal12} for additional comments in five dimensions).

\section{Curvature invariants and VSI/CSI spacetimes}

\label{Sec_invars}

Curvature invariants provide  a coordinate-independent characterization of {spacetime} curvature and they are  {thus} often used to study {its invariant properties}. A textbook example is the use of the Kretschmann scalar $R_{abcd} R^{abcd}$ to distinguish between coordinate and curvature  singularities in the Schwarzschild spacetime. {Furthermore,} in diffeomorphism-invariant theories (such as general relativity), the metric tensor (or, more generally, any tensor) takes different forms in different coordinate systems: 
given two metrics in different coordinates, one is faced with the {\em equivalence problem} (see, e.g., \cite{Stephanibook}), i.e. how to decide whether they represent the same geometric object or not. Since it is clearly desirable to have simple invariant (and possibly algorithmic) criteria to answer this question, one is naturally led to study spacetime invariants, and in particular {\em scalar polynomial curvature invariants}, defined as follows.

\begin{defn}[Curvature invariants of order $p$]
A scalar polynomial curvature invariant of order $p$ is a  scalar polynomial invariant obtained by contracting polynomials in the metric, the Riemann tensor and its covariant derivatives up to order $p$.

\label{def_inv}
\end{defn}
For brevity ``scalar polynomial'' will always be understood in what follows and we will simply talk of ``curvature invariants''. Note that curvature invariants also comprise invariants constructed from the Weyl and Ricci tensors (and their derivatives).

The Riemann tensor and (a finite number of) its covariant derivatives wrt a fixed frame uniquely characterize the metric (up to isometries) \cite{Stephanibook} (see also a recent discussion in \cite{HerCol10}). Therefore, the issue of characterizing a metric by its curvature invariants eventually amounts to the possibility of reconstructing the Riemann tensor and its covariant derivatives (hereafter just ``the curvature tensors'') from the curvature invariants. Obviously, if two metrics possess different (curvature) invariants, they are not equivalent. But the converse is not true, since, in general, curvature invariants contain less information than the curvature tensors. A striking illustration of this fact is provided by the so-called VSI (vanishing scalar invariants) spacetimes (section~\ref{sec_VSI}), for which all curvature invariants of all orders are zero -- in this specific sense, VSI cannot even be distinguished from flat space. 
 In addition to VSI, a {subset} of the CSI (constant scalar invariants) class, for which all scalar curvature invariants of all orders are constant, also contains spacetimes which can{\em not} be characterized by their invariants, even if these are non-zero (section~\ref{sec_CSI}).
Even before considering curvature invariants to address the equivalence problem, it is therefore important to clarify in what cases this can in principle be done, i.e. to determine {\em spacetimes characterized by their curvature invariants}. Recent relevant results are summarized in section~\ref{sec_characterized}. 
Let us note, in passing, that apart from the equivalence problem, VSI and CSI {spacetimes} contain exact solutions in supergravity (see, e.g., for discussions and references, \cite{Coleyetal07,ColFusHer09}) and are also important in the context of so-called universal metrics \cite{Coleyetal08}, which are of special interest since they have vanishing quantum corrections \cite{Gibbons75,Deser75,AmaKli89,HorSte90}.

\subsection{Algebraic VSI theorem for general tensors}

{Before focusing on curvature invariants,} let us first look at the {\em polynomial} invariants of an arbitrary type III tensor $\bT$. By definition~\ref{def_algT}, $\bT$ possesses only 
negative b.w.  components in an appropriate frame. In other words,
$T_{ab\dots c} = e_a^{(\hat{a})} e_b^{(\hat{b})} \dots e_c^{(\hat{c})} T_{\hat{a} \hat{b}\dots \hat{c}}$ is a linear combination of terms $e_a^{(\hat{a})} e_b^{(\hat{b})} \dots e_c^{(\hat{c})}$ containing always
more vectors $e_a^{(1)} = \ell_{a}$ than $e_{a}^{(0)} = n_{a}$. Consequently, all polynomial invariants of a type III tensor $\bT$ vanish: {indeed}, any arbitrary {\em full} contraction of powers of $\bT$, such as $T_{ab\dots c} T^{ab\dots c}$, vanishes since at least one vector $\ell_{a}$ {has to be} contracted with {either} $\ell^{a}$ or $m^{(i)a}$. In \cite{Coleyetal04vsi}, it was conjectured that the type III condition is not only a sufficient but also a necessary condition for  all  polynomial
invariants of $\bT$ to vanish {(``algebraic VSI conjecture'')}. This conjecture has been recently proved by Hervik: 

\begin{prop}[Algebraic VSI theorem \cite{Hervik11}]
\label{algVSI}
All polynomial invariants $($of order zero$)$ of a tensor $\bT$ of arbitrary rank on a Lorentzian manifold of arbitrary dimension vanish if, and only if, $\bT$ is of type III or more special. 
\end{prop}

\subsection{VSI spacetimes}
\label{sec_VSI}

Proposition~\ref{algVSI} applies, in particular, to the curvature tensor, telling us that 
{a spacetime is of Riemann type III/N iff all curvature invariants of order zero vanish.}
One can still try to construct invariants from covariant derivatives of the curvature tensor. However, there exist metrics for which {\em all} such invariants are zero, defined by

\begin{defn}[VSI spacetimes]
We say that a manifold $M$ with a metric of arbitrary signature is VSI $($vanishing scalar invariants$)$ if all curvature invariants of all orders vanish at all points of $M$.
\end{defn}

The VSI condition is obviously very restrictive and the only such space in the case of a positive-definite metric is flat space {(this immediately follows already from the vanishing of the Kretschmann scalar, which is a sum of non-negative quantities in Riemannian signature)}. However, in the Lorentzian case, the set of VSI metrics is non-trivial. Obviously, using proposition \ref{algVSI}, a spacetime is VSI if, and only if, {\em the Riemann tensor and its}  covariant derivatives are of type III or more special {at all orders}. The precise conditions under which this occurs are given by

\begin{prop}[VSI theorem \cite{Coleyetal04vsi}] A Lorentzian manifold of arbitrary dimension is VSI if, and only if, the following two conditions are {both} satisfied: 
\begin{enumerate}[(A)]
\item The spacetime possesses a non-expanding, twistfree, shearfree, geodesic null vector field $\bl$, {i.e. it} belongs to the Kundt class. 
\item Relative to $\bl$, {the Riemann tensor is of type III} or more special.
\end{enumerate}
\end{prop}

(Note that requiring the curvature invariants up to order 2 to vanish is in fact equivalent to VSI, as follows from the results of \cite{Coleyetal04vsi}.) The four-dimensional version of this theorem was proven in \cite{Pravdaetal02}, and subsequently extended to any~$n$ in \cite{Coleyetal04vsi}. To be precise, in  \cite{Coleyetal04vsi}, it has been proven that conditions (A) and (B) imply VSI in arbitrary dimension. The part of the proof showing that the VSI property implies (A) and  (B) is incomplete in \cite{Coleyetal04vsi} since it holds only under the assumption that the algebraic VSI conjecture is valid -- however, the proof is now {complete} thanks to proposition~\ref{algVSI} (proven in \cite{Hervik11}) and footnote \footnote{\label{foot_vsi}An additional  problem already discussed in \cite{Coleyetal04vsi} is that, {in order to} show that type III Ricci-flat non-Kundt spacetimes always possess a non-vanishing curvature invariant, the form of $\rhob$ for a ``generic'' type III (as discussed in section~\ref{subsubsec_GS_III_N}) was assumed so that some special type III cases could have evaded the proof of \cite{Coleyetal04vsi}. {However,} this can be now circumvented by showing that, for $\rhob\not=0$, the first covariant derivative of {any} type III Weyl tensor, 
{for which the (unique) mWAND is geodesic (which holds in particular in the Einstein-space case, see \cite{Pravdaetal04}),} necessarily contains non-vanishing b.w. 0 terms and therefore, by proposition \ref{algVSI}, possesses a non-vanishing curvature invariant. More specifically, some b.w. 0 components of the first derivative of the Weyl tensor are as follows: 
{(a) $C_{abcd;e} n^a \ell^b n^c \ell^d m^{(i)e} \propto \psi'_j\rho_{ji}$,  
(b) $C_{abcd;e}  \ell^a m^{(i)b} n^c  m^{(k)d} m^{(j)e}\propto \rho_{ik}\psi'_j-\rho_{lk}\psi'_{jil}$,  
(c)  $C_{abcd;e}  m^{(i)a} m^{(j)b}m^{(k)c}m^{(l)d} m^{(h)e}\propto\psi'_{jkl}\rho_{ih}+\psi'_{ilk}\rho_{jh}+\psi'_{lij}\rho_{kh}+\psi'_{kji}\rho_{lh}$. }
For type III, the vanishing of {all} these b.w. 0 components implies $\rhob=0$. (This can be seen by multiplying 
{(b) by $\rho_{ik}$ and (c) by $\rho_{ih}$.}) Thus, type III  Ricci-flat spacetimes with $\rhob\not=0$ admit a non-vanishing Weyl invariant of order 1 and consequently are not VSI. }.

VSI spacetimes admit a special form \cite{Coleyetal06} of the {general} Kundt metric \eqref{Kundt_gen}, namely,
\be
 \d s^2 =2\d u\left[\d r+H(u,r,x)\d u+W_{\alpha}(u,r,x)\d x^\alpha\right]+ \delta_{\alpha\beta} \d x^\alpha\d x^\beta , \label{VSI}
\ee
with
\bea
W_{\alpha}(u,r,x)&=&-\delta_{\alpha, 2} \frac{2 \epsilon}{x_2} r + W^{(0)}_\alpha (u,x), \\
H(u,r,x)& = &\frac{\epsilon r^2}{2 (x^2)^2}+ r H^{(1)} (u,x) + H^{(0)} (u,x) \qquad (\epsilon=0,\ 1). \label{VSIH}
\eea
{Note that the transverse space is necessarily flat.} {The value of $\epsilon=0,1$ specifies to which of the two main subclasses $\taub=0$ and $\taub\neq0$  of the Kundt family spacetime \eqref{VSI} belongs (see section~\ref{sec_Kundt}).}
The Riemann {(and thus also the Ricci and Weyl)} tensors of  metric \eqref{VSI}--\eqref{VSIH} are {indeed} of type III or more special.
Various subcases of type III, N and O Ricci and Weyl tensors are studied in \cite{Coleyetal06,FusterThesis} {(where the vacuum equations are also given)}. {In particular,} the case $\epsilon=0=H^{(1)}$ represents VSI \pp-waves ({not all \pp waves are VSI}\footnote{Not even in 4D, unless some conditions on the matter content are imposed.}
-- see also section~\ref{subsubsec_pp}).

\subsection{CSI spacetimes}

\label{sec_CSI}

As an extension of the VSI family, it is also of interest to consider CSI spacetimes, defined by

\begin{defn}[CSI spacetimes]
We say that a manifold $M$ with a metric of arbitrary signature is CSI (constant scalar invariants) if all curvature invariants of all orders are constant at all points of $M$.
\end{defn}

Obviously, VSI is a proper subset of CSI. In 4D the CSI class is fully determined by

\begin{prop}[$n=4$ CSI theorem \cite{ColHerPel09b}] In four dimensions with Lorentzian signature, the CSI class consists of all (locally) homogeneous spacetimes and of a (proper) subset of the degenerate Kundt metrics, i.e. the ``degenerate CSI$_K$'' metrics.\footnote{Note that these two subsets of CSI spacetimes are not disjoint: the Bertotti-Robinson and Nariai-like spacetimes \cite{Stephanibook} are homogeneous (and symmetric) and belong to degenerate CSI$_K$, cf.~\cite{ColFusHer09,ColHerPel09b,HerCol10}; similarly, the Kaigorodov spacetime \cite{Stephanibook} is also homogeneous and degenerate CSI$_K$ \cite{ColFusHer09} (homogeneous \pp waves \cite{Stephanibook} are another such example, however falling into the VSI class -- clearly VSI$\subset$CSI$_K$). More trivial examples are given by constant curvature spacetimes, which are both degenerate CSI$_K$ (in the dS/AdS case, and VSI in the Minkowski case) and homogeneous.}
\label{prop_4DCSI}
\end{prop}

(``Degenerate Kundt metrics'' are defined in arbitrary dimension in section~\ref{subsubsec_deg_Kundt}, while  ``degenerate CSI$_K$'' are given below in \eqref{CSI_Kundt} and \eqref{CSI_Kundt_2}.) Homogeneous 4D spacetimes can be of any Weyl type \cite{Stephanibook}, and thus, the Petrov type of CSI is unrestricted. However, degenerate CSI$_K$ cannot be of Petrov type I (see the discussion below in $n\ge 4$ dimensions), but all 
{algebraically} special types are permitted {\cite{ColHerPel09b}.}

Now, let us  discuss what is known in arbitrary dimension. Obviously, all locally homogeneous spacetimes are CSI. 
There exist also higher dimensional degenerate CSI$_K$ spacetimes (some of which may be locally homogeneous). But, contrary to the case $n=4$, it has   not {yet} been proven that these two families exhaust the CSI class when $n>4$ (this is due to the fact that proposition~\ref{4Dnondeg} has not been proven in higher dimensions, see \cite{ColHerPel09a,Hervik11}). That is, there might in principle still exist some ({non} \ind, as defined in~section~\ref{sec_characterized}) CSI which are neither homogeneous nor degenerate CSI$_K$ (necessarily of Riemann type II or more special at all orders, and with all b.w. zero components being constant \cite{Hervik11}). Nevertheless, let us  briefly describe  the degenerate CSI$_K$ class here.

In arbitrary dimensions $n\ge4$, the degenerate CSI$_K$ metrics are of the form \cite{ColHerPel06,ColHerPel09b} 
\be
 \d s^2 =2\d u\left[\d r+H(u,r,x)\d u+W_{\alpha}(u,r,x)\d x^\alpha\right]+ g_{\alpha\beta}(x)\d x^\alpha\d x^\beta , \label{CSI_Kundt}
\ee
where $g_{\alpha\beta}(x)$ (note $g_{\alpha\beta,u}=0$) is a (locally) homogeneous space, and 
\bea
W_{\alpha}(u,r,x)&=&rW^{(1)}_\alpha (u,x) + W^{(0)}_\alpha (u,x), \nonumber \\
H(u,r,x)& = &\frac{r^2}{8}\left(a+W^{(1)}_\alpha W^{(1)\alpha}\right)+ r H^{(1)} (u,x) + H^{(0)} (u,x) . \label{CSI_Kundt_2}
\eea
The index of $W^{(1)\alpha}$ is raised with the inverse of $g_{\alpha\beta}(x)$ and $a$ is a constant. Eq.~\eqref{CSI_Kundt_2} provides only {\em necessary} conditions for degenerate CSI$_K$: further conditions on $W^{(1)}_\alpha$ 
 must also be fulfilled {(in 4D see  eqs.~(5)--(8) of \cite{ColHerPel09b})}. 
Degenerate CSI$_K$ are of {\em Riemann type II $($or more special$)$ at all orders}, since they are a subset of the degenerate Kundt metrics (i.e. precisely those Kundt spacetimes for which the Riemann tensor and all its covariant derivatives are of type II, or more special, and are all aligned with $\bl$ \cite{ColHerPel09a,ColHerPel10}, see section~\ref{subsubsec_deg_Kundt}). Moreover, {\em all b.w. zero components of the curvature tensors are constant}. 
Conversely, however, not all degenerate Kundt metrics are CSI$_K$, since \eqref{CSI_Kundt_2} is more restrictive than \eqref{deg_Kundt} (take, e.g., the Melvin spacetime for an explicit counterexample).
If $W^{(1)}_\alpha=a=H^{(1)}=0$ in \eqref{CSI_Kundt_2}, one is left with CSI \pp waves (a proper subset of degenerate CSI$_K$). In \cite{ColHerPel06} it is discussed how to construct examples of {degenerate} CSI$_K$ by warping certain VSI.

Similarly as for the VSI class, also CSI spaces are more restricted for positive-definite metrics, and coincide with the class of locally homogeneous spaces \cite{PruTriVan96}.

\subsection{Spacetime characterized by their  curvature invariants}

\label{sec_characterized}

VSI are thus clear examples of spacetimes that cannot be uniquely characterized in terms of their curvature invariants (in the sense of the equivalence problem). The same is true for degenerate CSI$_K$ (as discussed below). A natural question is thus to determine all spacetimes which have the same property. Certain important results are now summarized.

By definition \cite{ColHerPel09a}, a metric $\bg$ is said to be {\em \ind} if there does not exist a (one-parameter) metric deformation $\bg'$ of $\bg$ that possesses the same set of curvature invariants as $\bg$ (see \cite{ColHerPel09a} for more details -- in particular, $\bg'$ must not be diffeomorphic to $\bg$).\footnote{{
See \cite{ColHerPel09a,HerCol10} 
for more details on these definitions and related discussions.}}
Obviously, if a metric is not {\ind}, then it cannot be uniquely determined by its curvature invariants (since other metrics will have the same curvature invariants). It is thus of interest to determine what metrics are (not) \ind. In four dimensions, the issue is settled by

\begin{prop}[$n=4$ \ind\ spacetimes \cite{ColHerPel09a}] In four dimensions, the degenerate Kundt metrics are not \ind, and they are the only metrics with this property. 
\label{4Dnondeg}
\end{prop}

This includes, in particular, all degenerate CSI$_K$ (discussed in section~\ref{sec_CSI}). {\em Degenerate Kundt metrics are not \ind\ also in higher dimensions} \cite{ColHerPel10}, and inequivalent degenerate Kundt metrics can thus have identical {curvature} invariants. However, it has not yet been proven that these are the {\em only} {non-\ind\ metrics when $n>4$. If proposition~\ref{4Dnondeg} turns out to hold also in higher dimensions, then proposition~\ref{prop_4DCSI} can also be extended to any $n\ge4$ \cite{Hervik11}. 

{Spacetime characterized by their curvature invariants have been recently determined in terms of the alignment type of curvature tensors \cite{Hervik11} (see also \cite{HerOrtWyl12} for related discussions).} Let us finally note that when the signature is Riemannian all spaces are characterized by their invariants \cite{HerCol10}.

\section{Asymptotic properties of the gravitational field}

\label{sec_asympt}

In this section, we mainly focus on  asymptotic properties of asymptotically flat spacetimes at null infinity (see \cite{TanTanShi09} for a study  at spacelike infinity). This is of particular interest in the study of gravitational radiation, and} in four dimensions, it is related to the algebraic classification of the Weyl tensor via {the well-known} peeling theorem. There are essentially
two approaches towards studying asymptotic flatness at null infinity. The method based on conformal  compactification can be used only in even dimensions (in odd dimensions smoothness at null infinity fails because of a half-integer power in {the conformal factor} \cite{HolIsh05,HolWal04}). On the other hand, a generalization \cite{Tanabe:2011es} of the Bondi approach can be applied in all dimensions. Combined with the GHP formalism \cite{Durkeeetal10}, it allowed  Ref. \cite{GodRea12} to determine the peeling behaviour of the Weyl tensor {for any asymptotically flat $n>4$ spacetime}, as described in section \ref{subsec_peeling}. In section \ref{subsec_asym_special}, we focus {instead} on asymptotically flat spacetimes {that are, additionally, vacuum and algebraically special}, arriving at considerable differences between  $n=4$ and $n>4$. Finally, in section~\ref{subsec_III_N}, we briefly discuss {the asymptotic behaviour of} (not asymptotically flat) type N and III spacetimes.

\subsection{Peeling of the Weyl tensor {and gravitational radiation}}

\label{subsec_peeling}

{In four-dimensional asymptotically flat spacetimes,} the well-known ``peeling'' theorem 
\be
	C_{\hat{a}\hat{b}\hat{c}\hat{d}}=r^{-1}C^{(N)}_{\hat{a}\hat{b}\hat{c}\hat{d}}+r^{-2}C^{(III)}_{\hat{a}\hat{b}\hat{c}\hat{d}}
					+r^{-3}C^{(II)}_{\hat{a}\hat{b}\hat{c}\hat{d}}+r^{-4}C^{(I)}_{\hat{a}\hat{b}\hat{c}\hat{d}}+{\cal O}(r^{-5}) \qquad {(n=4)} ,
\label{peel-4D}
\ee
closely connects the  behaviour of the Weyl tensor near null infinity with the Petrov classification -- in eq.~\eqref{peel-4D} the {terms $C^{(N)}_{\hat{a}\hat{b}\hat{c}\hat{d}}$, $C^{(III)}_{\hat{a}\hat{b}\hat{c}\hat{d}}$, etc., are of Petrov type N, III, II and I, respectively, and their components are} {expressed} in a frame parallelly propagated (p.p.) along an outgoing {affinely parametrized} geodesic null vector $\bl=\pa_r$.
In particular, the leading $1/r$-term represents gravitational radiation. 

This 4D result can be obtained using a definition of asymptotic flatness {given} either via {expansions in} suitable coordinates \cite{NP,BBM,sachsasympt} or via a conformal compactification \cite{Penrose63,Penrose65prs}. {However,} {as mentioned above}, in higher dimensions {the conformal method can be employed} only in even dimensions \cite{HolIsh05,HolWal04}.  {Hence,} a higher dimensional generalization of Bondi coordinates \cite{Tanabe:2011es} has been used in \cite{GodRea12} for deriving the peeling fall-off of the Weyl tensor near null infinity of asymptotically flat spacetimes in arbitrary $n>4$ dimension, {as we now review}.

\subsubsection{Bondi coordinates}

{Let us first recall the setup of \cite{Tanabe:2009va,Tanabe:2011es} for asymptotically flat spacetimes in higher dimensions}. A spacetime is said to be {\em asymptotically flat at future null infinity} if Bondi coordinates $(u,\tilde{r},x^I)$  can be introduced outside some cylindrical world tube 
{such that the line element reads}
\be
	\d s^2=-A \mathrm{e}^B \d u^2-2\mathrm{e}^B \d u \d \tilde{r}+\tilde{r}^2 h_{IJ} (\d x^I +C^I \d u)(\d x^J + C^J \d u) . 
	\label{Bondi_metric}
\ee
{The} metric functions are assumed to allow  for an asymptotic expansion {(for $\tilde{r}\to\infty$)} in inverse powers of $\tilde{r}$ [$\sqrt{\tilde{r}}$] for even [odd] $n$, {i.e.,} 
\bea
	A&=&  
		1+\sum_{k\geq 0} \frac{A^{(k+1)}(u,x)}{\tilde{r}^{n/2+k-1}}, \qquad
  B= 
 		\sum_{k\geq 0}\frac{B^{(k+1)}(u,x)}{\tilde{r}^{n/2+k}},\nonumber\\
\ C^I&=&
				\sum_{k\geq 0}\frac{C^{(k+1)I}(u,x)}{\tilde{r}^{n/2+k}}, \qquad  
				h_{IJ}=
						\omega_{IJ}{(x)}+\sum_{k\geq 0} \frac{h_{IJ}^{(k+1)}(u,x)}{\tilde{r}^{n/2+k-1}} . \label{peel_asymp_metric}
\eea
{Here} $k\in Z$ for even $n$ and $2k\in Z$ for odd $n$, $\tilde{r} $ is a (non-affine) parameter along null geodetic generators ($u,x^I=$const) $\bl=-\pa_{\tilde{r}} $  of the null surfaces  $u=$const and $\det h_{IJ}=\det\omega_{IJ}$, {where} $\omega_{IJ}$ is the metric of a unit sphere $S^{n-2}$.
For $n=4$, this reduces to the standard definition proposed in \cite{BBM,sachsasympt}.

Some of these coefficients have a {direct} physical meaning, e.g. $A^{(n/2-1)}$ enters the definition of the Bondi mass, and  $h_{IJ}^{(1)}$ {determines (in vacuum) the} Bondi mass decrease \cite{Tanabe:2011es}:

\bea
	M(u)&=&-\frac{n-2}{16\pi}\int_{S^{n-2}}A^{(n/2-1)}\d \omega, \label{M} \\
	\dot{M}(u)&=&-\frac{1}{32\pi}\int_{S^{n-2}}\dot{h}_{IJ}^{(1)}\dot{h}^{(1)IJ}\d \omega, \label{Md} 
\eea
where $S^{n-2}$ is a sphere, $\d\omega$ its volume element and (from now on) a dot denotes differentiation wrt $u$.

So far no field equations have been imposed. However, the functions $h_{IJ}^{(1)}$ remain freely specifiable even if one further demands that the asymptotic  {vacuum} Einstein's equations  {are satisfied} \cite{Tanabe:2011es}. {Their $u$-}derivatives $\dot{h}_{IJ}^{(1)}$ {represent} a {generalization of} Bondi's ``news'' function. {If the spacetime contains (outgoing) {\em  gravitational radiation},  there is mass decrease and $\dot{h}_{IJ}^{(1)}$  is non-zero.}\footnote{{Let us observe that the Schwarzschild-Tangherlini metric in Robinson-Trautman coordinates \eqref{geo_metric fin} with \eqref{Hvacuum}, $K=1$, $\Lambda=0$ is an example (albeit ``trivial'') of a vacuum spacetime admitting the Bondi form \eqref{Bondi_metric}, \eqref{peel_asymp_metric}: in this case, one immediately finds $M=(n-2)\mu(Vol)_{S^{n-2}}/(16\pi)$ (which here coincides with the ADM mass) and $\dot{M}=0$ (recall that indeed the spacetime is static).}}

\subsubsection{Weyl tensor and Bondi flux}

\label{subsubsec_peel_weyl}

Now, let us  summarize the main results of \cite{GodRea12} for $n>5$ {(the case $n=5$ is ``exceptional'' and will be discussed below)}, i.e. the behaviour of the Weyl tensor in spacetimes \eqref{Bondi_metric} with \eqref{peel_asymp_metric} ``far away'' along the direction of the outgoing  twistfree null (and consequently geodesic) {vector field}  $\bl$.

In an {adapted} frame \cite{GodRea12} (not p.p. along $\bl$),  {one finds} 
\bea
 & & \Omega'_{ij}=-\frac{\hat{e}_i^I \hat{e}_j^J\ddot{h}_{IJ}^{(1)}}{2\tilde{r}^{n/2-1}}+{\cal O}(\tilde{r}^{-n/2}),
\nonumber\\
 & & \Phi^S_{ij}=-\frac{(n-4)\hat{e}_i^I\hat{e}_j^J\dot{h}_{IJ}^{(1)}}{4\tilde{r}^{n/2}}+{\cal O}(\tilde{r}^{-(n/2+1)}), \qquad 
				\Phi_{ijkl}=(\hat{e}_i^I\hat{e}_{[k}^J\delta_{l]j}-\hat{e}_j^I\hat{e}_{[k}^J\delta_{l]i})\frac{\dot{h}_{IJ}^{(1)}}{\tilde{r}^{n/2}}+{\cal O}(\tilde{r}^{-(n/2+1)}),\ \nonumber\\
  & & \Psi'_{ijk}={\cal O} (\tilde{r}^{-n/2}), \qquad \Psi'_i={\cal O} (\tilde{r}^{-n/2}), \qquad\qquad\qquad {(n=4,\ n>5)} \label{Weyl_exp} \\
 & & \Psi_{ijk}={\cal O} (\tilde{r}^{-(n/2+1)}), \quad  \Psi_{i}={\cal O} (\tilde{r}^{-(n/2+1)}),  \quad  \Phi^A_{ij}={\cal O} (\tilde{r}^{-(n/2+1)}),  \quad 
\Phi={\cal O} (\tilde{r}^{-(n/2+1)}),\ \nonumber\\
 & & \Omega_{ij}=-\frac{(n-2)(n-4)\hat{e}_i^I \hat{e}_j^J{h}_{IJ}^{(1)}}{8\tilde{r}^{n/2+1}}+{\cal O}(\tilde{r}^{-(n/2+2)}),
\nonumber
\eea
where  $\hat{e}^I_{i}$ is {an} orthonormal basis for the metric $\omega_{IJ}$, i.e. $\omega_{IJ}=\hat{e}_{iI}\hat{e}_{jJ}\delta_{ij}$. 
The leading-order behaviour of the Ricci tensor is also given in \cite{GodRea12}.

A parallelly transported frame is {then} introduced using a simple rescaling {(a boost)} $\hat\bl=e^{-B}\bl$, {a suitable null rotation and spin, while} an affine parameter $r$ along {$\hat\bl=-\pa_r$ is given by}
\be
 r = \int \mathrm{e}^B\d \tilde{r} =\tilde{r}+c +{\cal O}(\tilde{r}^{-(n/2-1)}), \qquad c,_{\tilde{r}}=0 . 
\ee
One can {finally} conclude  that the {\em higher dimensional ``peeling theorem''} reads \cite{GodRea12} 
\be
	C_{\hat{a}\hat{b}\hat{c}\hat{d}}=r^{-({n/2-1})}C^{(N)}_{\hat{a}\hat{b}\hat{c}\hat{d}}
	+r^{-({n/2})}C^{(II)}_{\hat{a}\hat{b}\hat{c}\hat{d}}+r^{-({n/2+1})}C^{(G)}_{\hat{a}\hat{b}\hat{c}\hat{d}}+\dots \qquad {(n>5)} ,
	\label{peel_HD}
\ee
where the ellipsis corresponds to {terms of} order $r^{-({n/2+2})}$ [$r^{-({n/2+3/2})}$] for even [odd] $n$. The algebraic type of each term is indicated as a superscript (note that $C^{(II)}_{\hat{a}\hat{b}\hat{c}\hat{d}}$ is not of the most general type II but it is in fact of type II(ad) \cite{GodRea12}). This result should be contrasted with the different 4D behaviour \eqref{peel-4D}. In particular, the type III term is absent in higher dimensions.

In even dimensions, this behaviour follows immediately from the asymptotic expansions \eqref{peel_asymp_metric} (with no use of Einstein's equations). By contrast, in odd dimensions, additional assumptions are necessary, namely it is required that $h_{IJ}^{3/2}=0$  (to avoid e.g. a term of order $\tilde{r}^{-(n/2-1/2)}$ in the expansion of $\Omega'_{ij}$)
 and
that the Ricci tensor decays ``sufficiently'' fast near null infinity. Einstein's equations then eliminate {additional} terms of order 
 ${\tilde r}^{-(n/2+1/2)}$  in the Weyl tensor expansion.

Note that in a particular case when  $\bl=-\partial_{\tilde{r}}$ is a {(twistfree)} WAND, {i.e. $\Omega_{ij}=0$, eq.~\eqref{Weyl_exp} implies for $n>4$ that ${h}_{IJ}^{(1)}=0$}  and thus also $\dot{M}(u)=0$, and the spacetime is non-radiative \cite{GodRea12} (see also \cite{OrtPraPra09b} and section~\ref{subsec_asym_special}).

The case of five dimensions is exceptional {in that} the order $\tilde{r}^{-(n/2+1/2)}=\tilde{r}^{-3}$  term cannot be eliminated \cite{GodRea12}. {The resulting peeling is thus described by}
\be
	C_{\hat{a}\hat{b}\hat{c}\hat{d}}=r^{-3/2}C^{(N)}_{\hat{a}\hat{b}\hat{c}\hat{d}}+r^{-5/2}C^{(II)}_{\hat{a}\hat{b}\hat{c}\hat{d}}
	+r^{-3}C^{(N)'}_{\hat{a}\hat{b}\hat{c}\hat{d}}+r^{-7/2}C^{(G)}_{\hat{a}\hat{b}\hat{c}\hat{d}}+
	{\cal O}(r^{-4}) \qquad {(n=5)}, \label{peel_5D}
\ee 
where {the new type N term} $C^{(N)'}_{\hat{a}\hat{b}\hat{c}\hat{d}}$ 
is in general  non-vanishing in radiating spacetimes. {The type II term is in fact type II(acd).}

{The above results for the Weyl tensor also enable one to rewrite} the decrease of the Bondi mass at future null infinity {in terms of the ``radiative'' Weyl components. Namely, assuming that} no Bondi flux is present in the {far} past (i.e. $\dot h_{IJ}^{(1)}\rightarrow 0$  for $u\rightarrow -\infty$), {from \eqref{Md}  and the first of \eqref{Weyl_exp} one arrives (for any $n\ge4$) at} \cite{GodRea12} 
\be
{\dot{M}}(u)=-\lim_{\tilde{r}\rightarrow \infty}\frac{\tilde{r}^{n-2}}{8\pi}
\int_{S^{n-2}}\left( \int^u_{-\infty}\Omega'_{ij}(\hat{u},\tilde{r},x)\d \hat{u}\right)^2\d\omega ,
\ee
where {on the r.h.s. a compact notation $(Y_{ij})^2=Y_{ij}Y_{ij}$ is used}.

To conclude, let us observe that Ref.~\cite{GodRea12} has also shown that the Bondi approach is equivalent to the conformal definition for even $n$, which is stable under linearized metric perturbations with initial data on a compact support \cite{HolIsh05}.

\subsection{Asymptotically flat algebraically special vacuum spacetimes}

\label{subsec_asym_special}

It was noticed by Sachs \cite{Sachs61} that four-dimensional algebraically special spacetimes, while leading to considerable mathematical simplification, still asymptotically retain the essential features of more realistic (outgoing) radiation fields. As mentioned {in section~\ref{subsubsec_peel_weyl}}, one now expects significant differences when $n>4$.

Prior to the work \cite{GodRea12}, in \cite{OrtPraPra09b},  the asymptotic behaviour of the Weyl tensor along a geodetic {\em multiple} WAND $\bl$ in an algebraically special vacuum spacetime was studied, and {a distinct behaviour for $n>4$} was indeed found.  
Moreover, the leading Weyl components at null infinity were also   characterized in \cite{OrtPraPra09b}, as we now summarize. Although not proven rigorously, it was argued in \cite{OrtPraPra09b} that, due to asymptotic flatness, $\bl$ must be non-degenerate, i.e. $\det\rhob\neq 0$. On the other hand, $\bl$ was not restricted to be twistfree (so the results of \cite{OrtPraPra09b} are, in this narrow sense, complementary to those of \cite{GodRea12}). Thanks to the conclusions of \cite{Pravdaetal04} {(cf.~section~\ref{subsubsec_GS_III_N})}, this condition rules out spacetimes of type III and N, leaving us in the following with the genuine type II (D) only. An affine parameter $r$ is defined along $\bl=\pa_r$, {and a frame parallelly transported along $\bl$ is employed}. The mWAND conditions read $\Psi_{ijk}=0=\Omega_{ij}$.

Now, with the above assumptions from the Sachs equations (section~\ref{subsec_ricci}), one immediately fixes the $r$-dependence of $\rhob=(rI-b)^{-1}$, where $I$ and $b$ represent $(n-2)\times(n-2)$ matrices, $I$ being the identity and $b_{,r}=0$. Note, in particular, that the (possible) vanishing of twist is determined by $A_{ij}=0 \Leftrightarrow b_{[ij]}=0$. Then, the behaviour of $\rhob$ for large $r$ follows (after restoring matrix indices)
\be
 \rho_{ij}=\sum_{m=0}^p\frac{(b^m)_{ij}}{r^{m+1}}+{\cal O}(r^{-p-2})=\frac{1}{r}\delta_{ij}+\frac{1}{r^2}b_{ij}+{\cal O}(r^{-3}) , 
 \label{L_infty}
\ee
which, to the leading order, simply becomes $\rho_{ij}\approx r^{-1}\delta_{ij}$ {(and the expansion $\theta$ is clearly non-zero)}.

One can then fix the $r$-dependence of the Weyl tensor by integrating the Bianchi identities containing $D$-derivatives. We assume that the Weyl components can be expanded in non-positive integer powers of $r$ (positive powers would lead to a p.p.~curvature singularity and can thus be excluded, {while half-integer powers for b.w. 0 components can be excluded by inspection of Bianchi equations (6) and (7) of \cite{OrtPraPra09b} {-- see also   \cite{OrtPraPra12}}).

It is convenient to first study b.w. zero components, for which we further assume the following condition for {\em asymptotic flatness}\footnote{By this we do not mean that terms of order $r^{-3}$ in b.w. zero components have to be non-zero (in fact for $n>4$ they will vanish, as explained below), but rather that terms of order $1/r^2$ vanish. This is natural to demand since this is the case already in four dimensions and the fall-off will be faster in higher dimensions. This is confirmed more rigorously (at least in even dimensions) by a study of the asymptotic behaviour of gravitational perturbations \cite{HolIsh05,HolWal04,Ishibashi08} {and is consistent with the results of \cite{GodRea12} summarized in section~\ref{subsec_peeling}}. Recall instead that in the case of $n=4$ vacuum algebraically special spacetimes, this is not an assumption but follows from the Ricci/Bianchi identities \cite{Sachs61}. For general vacuum $n=4$ spacetimes, it is related to the vanishing of the unphysical Weyl tensor on scri, which holds if asymptotic flatness is assumed \cite{Penrose63}.}
\be
 \Phi_{ijkl}\sim {\cal O}(r^{-3}).
\ee 

With these assumptions one arrives at the leading terms of the b.w. zero Weyl components \cite{OrtPraPra09b}
\bea
& & \Phi_{ij}^S=\frac{\phi^{(n-1)}\delta_{ij}}{n-2}\frac{1}{r^{n-1}}+ {\cal O}(r^{-n}), \qquad 						
										\Phi_{ij}^A=\frac{(n-1)\phi^{(n-1)}b_{[ij]}}{(n-2)(n-3)}\frac{1}{r^{n}}+ {\cal O}(r^{-n-1}), \nonumber \\ 
& & \Phi_{ijkm}=\frac{2\phi^{(n-1)}(\delta_{jk}\delta_{im}-\delta_{jm}\delta_{ik})}{(n-2)(n-3)}\frac{1}{r^{n-1}} + {\cal O}(r^{-n}), \label{zero_b.w.}
\eea
where $\phi^{(n-1)}_{,r}=0$. Subleading terms can similarly be determined to any desired order \cite{OrtPraPra_prep}, and $\phi^{(n-1)}$ and $b$ are the only integration ``constants'' characterizing the full expansions.

For negative b.w. components, one can show that for $n>4$ \cite{OrtPraPra09b}
\be
 \Psi'_{ijk}\sim {\cal O}\left(\frac{1}{r^{n-1}}\right) , \qquad \Omega'_{ij}\sim {\cal O}\left(\frac{1}{r^{n-1}}\right) \qquad (n>4) . 
 \label{negative_b.w.}
\ee
Again higher order terms can be determined to any desired order \cite{OrtPraPra_prep} (in particular, one finds that if $\phi^{(n-1)}=0$, the spacetime is flat) -- they all vanish in the non-twisting case, so that the spacetime is of type D (and the second mWAND is parallelly transported along $\bl$), see also proposition~\ref{prop_asym} below.

It is worth observing that as a byproduct of the expansion of $\Phi_{ijkm}$, one also finds (if $\phi^{(n-1)}\neq 0$) the important condition  $(n-2)b_{(ij)}=b_{kk}\delta_{ij}$. This is equivalent to $\rhob$ obeying the optical constraint. In particular, if $\bl$ is non-twisting (i.e. $b_{ij}=b_{(ij)}$), then $\rhob$ is automatically shearfree, so that the considered spacetime is Robinson-Trautman. Combing these observations with the results of \cite{PodOrt06} one can formulate the following
\begin{prop}[mWANDs in asymptotically flat spacetimes \cite{OrtPraPra09b}]
\label{prop_asym}
 In an asymptotically flat algebraically special vacuum spacetime, a non-degenerate geodesic mWAND $\bl$ obeys the optical constraint. If $\bl$ is twistfree and $n>4$, then the spacetime is Schwarzschild-Tangherlini. 
\end{prop}

The second part of this proposition should be contrasted with the 4D case \cite{Stephanibook}; see \cite{OrtPraPra09b} for more comments {(note that, in fact, one arrives at the same conclusion also with weaker assumptions, see proposition~{\ref{prop_GSHD}} and \cite{OrtPraPra12})}. Moreover, also the fact that all b.w. components display the same fall-off rate (eqs.~\eqref{zero_b.w.} and \eqref{negative_b.w.}) is peculiar of algebraically special spacetimes {in} higher dimensions. In particular, there is no peeling and radiative terms are absent, as opposed to the $n=4$ case \cite{Sachs61}. This is in agreement with the results of \cite{GodRea12} (section~\ref{subsec_peeling}), at least in the non-twisting case. Contrast again \eqref{zero_b.w.} and \eqref{negative_b.w.} with the general behaviour  
{  \eqref{peel_HD} and \eqref{peel_5D}}, 
which occurs instead when $\bl$ is not a WAND.

\subsection{Type III/N Einstein spacetimes}

\label{subsec_III_N}

Algebraically special vacuum spacetimes that are asymptotically flat can be only of type II or D, as discussed in section~\ref{subsec_asym_special}. Here, we consider instead Einstein spacetimes of type III and N \cite{Pravdaetal04}. For these, the (unique) mWAND is necessarily geodesic  and $\rhob$ is degenerate.  This may be of interest for spacetimes with, e.g., Kaluza-Klein asymptotics. With no need of assumptions on the asymptotics of the Weyl tensor, its full $r$-dependence can now be fixed in the closed form by using suitable Ricci and Bianchi identities \cite{OrtPraPra10}. Again, we take an affinely parametrized mWAND $\bl=\pa_r$ {and a parallelly transported frame}.

Concerning the optical matrix $\rhob$ of $\bl=\pa_r$, {type N  and (``generic'') type III} Einstein spacetimes allow for just two possibilities. Either $\rhob=0$ (Kundt spacetimes), or $s\equiv\rho_{22}=\rho_{33}$ and $A\equiv\rho_{23}=-\rho_{32}$ can be taken as the only non-zero components of $\rhob$ \cite{Pravdaetal04} (see {section \ref{subsubsec_GS_III_N}} and recall that $A\neq0$ requires $s\neq0$ \cite{OrtPraPra07}). In the latter case from the Sachs equations (section~\ref{subsec_ricci}), one readily gets
\be
 \rho\equiv s+iA=\frac{1}{r-ia_0} ,
 \label{rho2block}
\ee
where $a_{0,r}=0$. These two subcases will be now  discussed separately.

\subsubsection{Kundt spacetimes}

\label{subsubsec_III_N_Kundt}

When $\rhob=0$, one arrives at the general behaviour for type III Einstein spacetimes \cite{PraPra08,OrtPraPra10}
\be
 {\Psi'}_{ijk}={\Psi'}_{ijk}^0, \qquad {\Psi'}_{i}={\Psi'}_{i}^0={\Psi'}_{jij}^0 , \qquad {\Omega'}_{ij}=\tilde{\Omega'}^0_{ij}r+{\Omega'}^0_{ij} ,
\ee
where quantities with superscript $^0$ do not depend on $r$. These ``integration constants'' are related to certain Ricci rotation coefficients and must satisfy some relations due to ${\Omega'}_{ii}=0={\Omega'}_{[ij]}$ \cite{PraPra08,OrtPraPra10} {(in particular, for \pp waves, {discussed in section \ref{subsubsec_pp},} it follows from \cite{OrtPraPra10} that $\tilde{\Omega'}^0_{ij}=0={\Psi'}_{i}^0$)}. This is the typical peeling-off of type III Kundt spacetimes, cf.~\cite{Sachs61} in four dimensions.  
By setting ${\Psi'}_{ijk}^0=0=\tilde{\Omega'}^0_{ij}$, one finds the (``plane wave-like'') behaviour of type N spacetimes.

Recall that Einstein spacetimes of type III/N belonging to the Kundt class are VSI (if $\Lambda=0$) or CSI (if $\Lambda\neq 0$) \cite{Coleyetal04vsi,Pravdaetal02,OrtPraPra10} (section \ref{Sec_invars}), and no physically useful information can thus be extracted from their invariants.

\subsubsection{Expanding spacetimes}

\label{subsubsec_III_N_expand}

When $\rhob\neq 0$, in the type N case the Weyl $r$-dependence is given by \cite{OrtPraPra10}
\be
 \Omega'_{22}+i\Omega'_{23}=\frac{{\Omega'}^0_{22}+i{\Omega'}^0_{23}}{{r-ia_0}} \qquad \mbox{(type N)}.
 \label{Psi_ij_N}
\ee
This is similar to well-known results in 4D -- see \cite{OrtPraPra10} for comments, e.g., on the ``rotation'' of the polarization modes. Knowing the $r$-dependence of the Weyl tensor is also useful for studying possible spacetime singularities. Namely, for the simplest non-trivial
curvature invariant {admitted by} type N spacetimes in four \cite{BicPra98} and higher \cite{Coleyetal04vsi} dimensions (i.e. $I_{N} \equiv C^{a_1 b_1 a_2 b_2  ; c_1 c_2} C_{a_1 d_1 a_2  d_2 ; c_1 c_2} C^{e_1 d_1 e_2 d_2 ;f_1 f_2} C_{e_1 b_1 e_2 b_2  ; f_1 f_2}$)
one finds
\be
I_N=\frac{A^0}{(r^2+a_0^2)^6} .
\ee
If $a_0$ vanishes at some spacetime points, then there will be a curvature singularity at $r=0=a_0$ (this always occurs in the non-twisting case \cite{PraPra08}), see \cite{OrtPraPra10} for more details. Note also that $I_N\to0$ for $r\to\infty$, i.e. far away along the mWAND.

For type III the Weyl $r$-dependence is more complicated and we refer to \cite{OrtPraPra10} for full details. Let us just present here the $r$-dependence of the simplest non-trivial curvature invariant (namely \newline $I_{{III}} = C^{a_1 b_1 a_2 b_2;e_1} C_{a_1 c_1 a_2 c_2;e_1} C^{d_1 c_1 d_2 c_2;e_2} C_{d_1 b_1 d_2 b_2;e_2}$, see \cite{Pravda1999,Coleyetal04vsi}). One finds \cite{OrtPraPra10}
\be
 	I_{III}=\frac{A^0}{(r^2+a_0^2)^6}+\frac{B^0}{(r^2+a_0^2)^5}+\frac{C^0}{(r^2+a_0^2)^4} .
	\label{Inv3}
\ee
As in the type N case, there may be curvature singularities localized at points where $r^2+a_0^2=0$, which may or may not exist, in general (but they always do in the non-twisting case \cite{PraPra08}). Again, $I_{III}\to0$ for $r\to\infty$.

Explicit examples of expanding type N/III Einstein spacetimes were also given \cite{OrtPraPra10}.

\section{Shearfree solutions}

\label{sec_shearfree}

{As we have seen in section~\ref{sec_GS}, for higher dimensional Einstein spaces, the condition $\sigma=0$ does not in general follow from assuming the existence of an mWAND. Nevertheless, there still exist classes of shearfree spacetimes that are {necessarily} algebraically special and satisfy the vacuum Einstein equations. Below we review the known examples with these properties. By converse, recall also the general result that in {\em odd} spacetime dimensions a geodesic twisting WAND is necessarily shearing \cite{OrtPraPra07}.}

\subsection{Kundt spacetimes}

\label{sec_Kundt}

{Here, we give  the line element of general Kundt spacetimes and discuss its algebraic type and its main subclasses, also in relation with the VSI and CSI spacetimes described in sections~\ref{sec_VSI} and \ref{sec_CSI}.}

\subsubsection{General line element and alignment type}

Kundt spacetimes are {defined as} spacetimes admitting a null geodesic vector field $\bl$ with vanishing shear, expansion and twist, {i.e. $\kappa_i=0=\rhob$}.  These spacetimes admit a metric in the form \cite{Coleyetal03,ColHerPel06,PodZof09} 
\be
 \d s^2 =2\d u\left[\d r+H(u,r,x)\d u+W_\alpha(u,r,x)\d x^\alpha\right]+ g_{\alpha\beta}(u,x) \d x^\alpha\d x^\beta , \label{Kundt_gen}
\ee
where $\alpha,\beta=2 \dots n-1$ and $\bl=\partial_r$. 
In an adapted frame with  $n_a\d x^a=\d r+H\d u+W_\alpha\d x^\alpha$ (so that the vectors $\bmd{i} $ live in the ``transverse space'' spanned by the $x^\alpha$), the covariant derivative of $\bl$ reduces to $\ell_{a;b}=L_{11}\ell_a\ell_b+\tau_i(\ell_a m^{(i)}_b+m^{(i)}_a\ell_b)$ \cite{Coleyetal09}, where $L_{11}=H_{,r}$ and $\taub=0\Leftrightarrow W_{\alpha,r}=0$ {(recall the definition of the Ricci rotation coefficients in section~\ref{subsec_ricci} and table~\ref{tab:weights}).}

{From the Sachs equations {\cite{Pravdaetal04,OrtPraPra07}} (see eqs.~\eqref{sachs_she} and \eqref{sachs_exp}) with $\rhob=0$, it follows immediately  that} b.w.~+2 components of the Ricci and Weyl tensors identically vanish  (see also \cite{FusterThesis,PodZof09}) {so that, in particular,} $\bl$ is always a WAND. 
Furthermore, it follows directly from the Ricci identity (11k) of \cite{OrtPraPra07} that $\bl$ is an mWAND iff $R_{ab}\ell^b \propto \ell_a$, i.e., for Kundt spacetimes, the Weyl type II condition and the Ricci type II condition are {\em equivalent} (this has been noticed in \cite{PodSvarc_Kundt}).
We can thus state the following proposition:

\begin{prop}[Weyl type of Kundt spacetimes \cite{OrtPraPra07,FusterThesis,PodZof09}]\label{prop_Kundt}
An arbitrary Kundt spacetime is of Riemann type I or more special, and the non-expanding, non-twisting and non-shearing geodesic null congruence $\bl$ automatically defines a WAND (as well as an AND of the Ricci tensor). 
$\bl$  is a multiple WAND $($i.e. the Weyl tensor is also of  type II or more special$)$ if, and only if, the Ricci tensor is of type II or more special wrt $\bl$ $($i.e. $R_{ab}\ell^b\propto\ell_a$, equivalent to $W_{\alpha,rr}=0$ in \eqref{Kundt_gen}$)$.
 \end{prop}

The above condition for the Ricci tensor  
$R_{ab}\ell^b\propto\ell_a$ is satisfied, e.g., by Einstein spacetimes (see section~\ref{subsubsec_vectors_etc}).

The Kundt class is very rich and contains several interesting subclasses, {some of which are discussed in what follows} {and in figure~\ref{fig_Kundt}} (see \cite{FusterThesis,Coleyetal09,PodZof09,Coleyetal06} for studies of various aspects).  
First of all, it contains two invariantly defined subfamilies with either $\taub\neq0$ or $\taub=0$ (see \cite{Coleyetal03,Coleyetal06,PodZof09} for more details in special cases). {In the subclass $\taub\neq0$, one can always set $L_{11}=0$ by a null rotation about $\bl$. In the subclass $\taub=0$, which can also be} equivalently defined by the presence of a recurrent null vector field (see \cite{Walker50,GibPop08}  and section  \ref{subsubsec_othersymm}), {$L_{11}$ is invariant under null rotations about $\bl$ -- it vanishes iff $\bl$ is covariantly constant (so that the corresponding spacetime is a \pp wave, see section~\ref{subsubsec_pp}).
Both families $\taub\neq0$ and $\taub=0$ contain spacetimes} of all algebraically special types {(see, e.g., \cite{Coleyetal07,JakTaf09,GodRea09} and the examples mentioned in sections~\ref{sec_VSI}, \ref{sec_CSI} and \ref{subsubsec_pp}).} 
Recall that the Kundt spacetimes of {Riemann type III/N} give precisely the VSI class, already discussed in section~\ref{sec_VSI}: these spacetimes fall into both subclasses $\taub\neq0$ or $\taub=0$. {On the other hand, the overlap of Kundt spacetimes with the CSI class is more complicated, see section~\ref{sec_CSI}.}

The general form of the Riemann and Ricci tensors corresponding to the Kundt metric \eqref{Kundt_gen} has been given in \cite{PodZof09}, where the vacuum Einstein equations (with $\Lambda$) as well as the Einstein-Maxwell equations with an aligned electromagnetic field have also been explicitly presented. In particular, {\em all Einstein spacetimes of the Kundt class must satisfy $W_{\alpha,rr}=0=H_{,rrr}$} (these are necessary, but not sufficient, conditions). For these, in the case  of Weyl type III and N, the $r$-dependence of the Weyl tensor has been given in section~\ref{subsubsec_III_N_Kundt}. Ricci-flat Kundt spacetimes of type N coincide with non-expanding Ricci-flat Kerr-Schild metrics of type N (see proposition~\ref{KS_N}).

{Note also that Kundt Einstein spacetimes of type III/N are VSI (if $\Lambda=0$) or CSI (if $\Lambda\neq 0$) \cite{Coleyetal04vsi,Pravdaetal02,OrtPraPra10} (see sections \ref{sec_VSI}, \ref{sec_CSI}).}

\subsubsection{Degenerate Kundt metrics}

\label{subsubsec_deg_Kundt}

A {\em degenerate} Kundt metric is defined as a Kundt spacetime in which the Riemann tensor and all its covariant derivatives are of type II or more special (and all aligned with $\bl$). This happens iff $W_{\alpha,rr}=0=H_{,rrr}$ \cite{ColHerPel09a,Coleyetal09}, i.e. substituting into~\eqref{Kundt_gen} the functions
\be
 W_{\alpha}(u,r,x)=rW_{\alpha}^{(1)}(u,x)+W_{\alpha}^{(0)}(u,x) , \qquad  	
 H(u,r,x)=r^2H^{(2)}(u,x)+rH^{(1)}(u,x)+H^{(0)}(u,x) .
 \label{deg_Kundt}
\ee
In particular, {\em all VSI spacetimes, all degenerate CSI$_K$ spacetimes and all Kundt Einstein spacetimes are thus degenerate Kundt} (but not vice versa).

Degenerate Kundt metrics are spread over both subclasses $\taub\neq0$ and $\taub=0$ and can have any algebraically special Weyl type {(already in 4D vacua  \cite{Stephanibook})}. They are important in the context of the equivalence problem (see section~\ref{Sec_invars}) since they are {\em not \ind} \cite{ColHerPel10} {and, in general, } one cannot use (only) {curvature} invariants to distinguish among {them}, since
\begin{prop}[{Degeneracy of degenerate Kundt spacetimes}\cite{ColHerPel10}]
 In a degenerate Kundt spacetime, the b.w. 0 components of all curvature tensors $($and thus also their curvature invariants$)$ are independent of the functions $W_{\alpha}^{(0)}(u,x)$, $H^{(1)}(u,x)$ and $H^{(0)}(u,x)$ in \eqref{deg_Kundt}. 
	\label{prop_deg_Kundt}
\end{prop}
Proposition~\ref{prop_deg_Kundt} is relevant, in particular, to Kundt spacetimes representing gravitational waves (typically type II or N) propagating on a given background (typically type D or O);\footnote{Provided the metric functions satisfy the corresponding Einstein (typically vacuum) equations.} see, e.g., \cite{Obukhov04,JakTaf08,JakTaf09,GodRea09} for some special examples.

\subsubsection{\pp waves}

\label{subsubsec_pp}

\pp waves are defined as spacetimes admitting a covariantly constant null vector (CCNV) field $\bl$, i.e. $\ell_{a;b}=0$, {which is thus, in particular, a Killing vector field (see also {sections~\ref{subsubsec_null} and \ref{subsubsec_othersymm}}). They obviously belong to the Kundt class, with $L_{11}=0=\taub$.
The corresponding metrics were introduced already in 1925 in arbitrary dimension by Brinkmann \cite{Brinkmann25} and are given by the Kundt line element \eqref{Kundt_gen} {(in which $\ell_{a;b}=\pul g_{ab},_r$ \cite{PodOrt06,PodZof09})} but with metric functions independent of the affine parameter $r$ along the CCNV $\bl$
\be
 \d s^2 =2\d u\left[\d r+H(u,x)\d u+W_{i}(u,x)\d x^i\right]+ g_{ij}(u,x) \d x^i\d x^j. 
 \label{pp_gen}
\ee
They are thus in fact a (proper) {\em subset of degenerate Kundt metrics}. Moreover, they can (but need not) be VSI (section~\ref{sec_VSI}) or CSI (section~\ref{sec_CSI}). The functions $W_{i}$ and $H$ can be (locally) set to zero by a coordinate transformation \cite{Brinkmann25} (although this choice is not always convenient since it will affect the form of $g_{ij}$).

From $\ell_{a;b}=0$ and the definition of the Riemann tensor, one finds 
\be
R_{abcd} \ell^d = 0, \label{typeNnecc}
\ee
which {immediately} implies that b.w. +2 and +1 components of the Riemann tensor, {and thus also of the Ricci and Weyl} tensors, vanish (this can be also seen by inspecting explicit expressions for {the Riemann tensor} of the metric \eqref{pp_gen} given in \cite{PodZof09}), {along with some b.w. 0 and -1 components \cite{PraPra05,Ortaggio09}. In particular, $\bl=\partial_r$ is thus a multiple  WAND. {It is also easy to see that proper Einstein spacetimes cannot occur \cite{Brinkmann25}. Due to the comments relative to certain subtypes in section~\ref{subsubsec_spin}},  eq.~\eqref{typeNnecc} is more restrictive in four and five dimensions than in higher dimensions. Let us summarize {and extend results \cite{Coleyetal06,Ortaggio09,OrtPraPra09} on possible algebraic types of \pp waves:}

\begin{prop}[Alignment types of \pp waves]
A generic \pp wave is of Weyl type II(d) and Ricci type II with $R_{ab}\ell^b=0$ {(or more special)}. 
\pp waves cannot be properly Einstein. Ricci-flat {$($non-conformally flat$)$} \pp waves are {of Weyl type}:
\begin{enumerate}[(i)]
\item
N for dimension $n=4$,
\item
III(a) {or N} for $n=5$,
\item
II'(abd), {D(abd), III(a), or N} for $n \geq 6$.
\end{enumerate}

{In particular, Ricci-flat \pp waves of type III(a)/N are VSI}.
	\label{prop_pp}
\end{prop}

Note that these bounds cannot be improved since, apart from the well-known type N case, {also} type III(a) and II'/D(abd) Ricci flat \pp waves indeed do exist. A simple example of a type III(a)  Ricci-flat \pp wave {in five dimensions} \cite{OrtPraPra09} can be obtained by specializing results of \cite{Coleyetal06} to the vacuum case:
\bea
 & & W_2=0, \quad W_3=h(u) x^2 x^4,  \quad W_4 = h(u) x^2 x^3,\quad  g_{ij} = \delta_{ij} , \nonumber \\
 & & H=H_0 = h(u)^2 \left[ \frac{1}{24} \left( (x^3)^4+ (x^4)^4 \right)+ h^0( x^2,x^3,x^4) \right], \label{ppIII}
\eea
where $h^0(x^2,x^3,x^4)$ is linear in $x^2$, $x^3$, $x^4$. See proposition~\ref{prop_pp_products} for examples of vacuum \pp waves of type II'(abd) or D(abd) \cite{OrtPraPra09,Ortaggio09}.

 \pp waves are also briefly mentioned in section~\ref{sec_VSI}  in the context of VSI spacetimes {(which they intersect)} and in section~\ref{sec_QG}, where it is pointed out that while type N {Ricci flat} \pp waves are {also} exact vacuum solutions of  quadratic gravity, type III {Ricci flat} \pp waves are not.  {The general field equations for Ricci-flat \pp waves were given in \cite{Brinkmann25} (in particular, it is necessary that also $g_{ij}$ is Ricci-flat).} Isometries of \pp waves and the subset of \pp waves which are also CSI have been studied in \cite{McNColePel09}.

\subsubsection{Gyratons}

{Let us finally} briefly mention {\em gyratons}, which  describe the gravitational field of a localized spinning source propagating at the speed of light. These solutions are constructed by matching two different Kundt solutions {(which may be of various types, e.g., either $\taub=0$ or not, etc.)} -- the ``external'' solution is typically taken to be vacuum,  while the ``internal'' one possesses  non-vanishing $T_{uu}$ and $T_{u\alpha}$ components of the stress-energy tensor {(in the coordinates of~\eqref{Kundt_gen})}, which represent spinning ``gyraton matter''. {Gyratons were first introduced in 4D by Bonnor \cite{Bonnor70} and extended to higher dimensions in \cite{FroFur05,FroIsrZel05} -- cf., e.g., \cite{Coleyetal06,PodZof09,Krtousetal12}  for comments and more references}.

\begin{figure}[t]
\centering
\includegraphics[width=.6\textwidth]{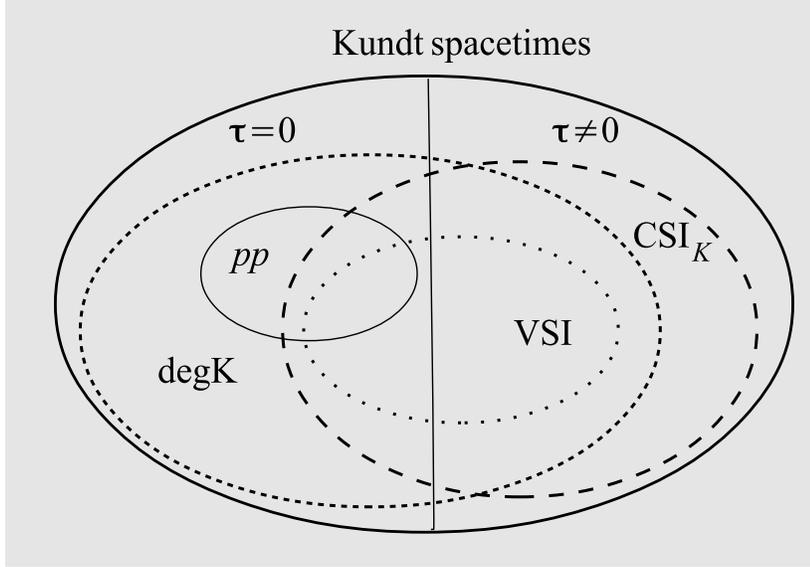}
\caption{Various subsets of the Kundt class of spacetimes in $n\ge 4$ dimensions. There is no meaning associated to the use of different line formats --  the only purpose of this is to make the picture better readable. ``{\em pp}'' stands for ``\pp waves'' while ``degK'' for ``degenerate Kundt''. The set 
CSI$_K\setminus$(CSI$_K\cap$degK) 
 might be empty, i.e. no \ind\ CSI$_K$ spacetime seems to be presently known (but if it exists it is necessarily homogeneous). All depicted subsets are proper subsets and all depicted intersections are non-empty. Recall that Kundt Einstein/Ricci-flat spacetimes are a proper subset of degK intersecting {\em pp}, VSI and CSI$_K$, as well as both subclasses $\taub\neq0$, $\taub=0$. {For $n=4,5$, Ricci-flat \pp waves are a proper subset of VSI.} See the text for more details.\label{fig_Kundt}} 
\end{figure}

\subsection{Robinson-Trautman spacetimes}

\label{sec_RT}

Robinson-Trautman spacetimes are defined by the existence of an hypersurface-orthogonal, null (and thus automatically geodesic) vector field $\bl$ with zero shear and non-zero expansion, i.e. $\kappa_i=0$, $\rho_{ij}=\frac{\theta}{n-2}\delta_{ij}$. Natural coordinates can be defined such that $\ell_a\d x^a=-\d u$ and $\ell^a\pa_a=\pa_r$ (i.e. $r$ is an affine parameter along $\bl$), along with $(n-2)$ ``transverse'' spatial coordinates {$x^\alpha$} which are constant along the null geodesics generated by $\bl$ (cf.~\cite{Brinkmann25,RobTra62}).  The resulting metric is \cite{PodOrt06} 
\be
 \d s^2=p^{-2}\gamma_{\alpha\beta}\left(\d x^\alpha+ K^{\alpha}\d u\right)\left(\d x^\beta+ K^{\beta}\d u\right)+2\d u\d r-2H\d u^2 ,
 \label{geo_metric}
\ee
where $\alpha,\beta=2 \dots n-1$, the matrix $\gamma_{\alpha\beta}$ is unimodular and independent of $r$, while $p$, $K^{\alpha}$ and $H$ are arbitrary functions of $(x,u,r)$, and the expansion of $\bl$ {is} given by $\theta=-(\ln p)_{,r}$. Such a metric is left invariant by a coordinate transformation $x^\alpha=x^\alpha(\tilde x,\tilde u)$, $u=u(\tilde u)$, $r=r_0(\tilde x,\tilde u)+\tilde r/\dot u(\tilde u)$.

\subsubsection{Vacuum solutions}

By imposing that the spacetime is Einstein and using the coordinate freedom, one can simplify the above line element and arrive at \cite{PodOrt06}
\be
 \d s^2=r^2h_{\alpha\beta}(u,x)\d x^\alpha\d x^\beta+2\d u\d r-2H(u,r,x)\d u^2 .
 \label{geo_metric fin}
\ee
The $(n-2)$-dimensional Riemannian metric $h_{\alpha\beta}$ is of the factorized form $h_{\alpha\beta}=P^{-2}(u,x)\gamma_{\alpha\beta}(x)$ and must be Einstein, with Ricci scalar ${\cal R}=(n-2)(n-3)K$ and $K=0,\pm 1$. Here, $\theta=1/r$. In order to specify $H$, one needs to consider the following two possibilities:
\begin{enumerate}[(i)]
\item\label{RTvacuum} $h_{\alpha\beta}=h_{\alpha\beta}(x)$ (i.e. $P=P(x)$): 
	\be 
		2H=K-\frac{2\Lambda}{(n-1)(n-2)}\,r^2-\frac{\mu}{r^{n-3}}  ,
		\label{Hvacuum} 
	\ee 
	where $\mu$ is a constant (parametrizing the mass).
 
\item\label{RTvacuum2} $h_{\alpha\beta}=P^{-2}(u,x)\gamma_{\alpha\beta}(x)$: 
	\be
 		2H=K-2r(\ln P)_{,u}-\frac{2\Lambda}{(n-1)(n-2)}\,r^2 .
		\label{Hvacuum2}
	\ee
\end{enumerate}

{\em Spacetimes (\ref{RTvacuum})} are warped products with a two-dimensional Lorentzian factor {and an $(n-2)$-dimensional Einstein factor} and are thus of type D(bd) (see proposition~\ref{prop: warp 2D factor}), as one can also check by explicitly computing the Weyl tensor \cite{PodOrt06}. The vector field $\pa_u$ is clearly Killing. When the transverse space is compact, these solutions describe various well-known static black holes in Eddington-Finkelstein coordinates, with ``Einstein'' horizon geometries, non-standard asymptotics and, possibly, non-spherical horizon topology \cite{Birmingham99,GibIdaShi02prl,GibHar02}. In particular, if the horizon has constant curvature, one obtains Schwarzschild-Kottler-Tangherlini black holes \cite{Tangherlini63}, of type D(bcd) (cf.~propositions~\ref{prop: warp 2D factor} and \ref{prop_spherical}) -- these are constant-curvature spacetimes iff $\mu=0$. 
For {\em spacetimes (\ref{RTvacuum2})} the non-vanishing components of the Weyl tensor are $C_{\alpha\beta\gamma\delta}=r^2{\cal C}_{\alpha\beta\gamma\delta}$ \cite{PodOrt06}, {where ${\cal C}_{\alpha\beta\gamma\delta}$ is the Weyl tensor of $h_{\alpha\beta}$,} so that now the type is D(abd) (in five dimensions ${\cal C}_{\alpha\beta\gamma\delta}=0$ identically and these spacetimes are of constant curvature). A technique to obtain explicit examples of metrics~(\ref{RTvacuum2}) was illustrated in \cite{Ortaggio07} and is based on \cite{Brinkmann25}. These spacetimes seem not to have a clear physical interpretation, yet they have been useful in the context of the Goldberg-Sachs theorem as first known examples of Einstein spacetimes with a non-geodesic mWAND (of course different from $\bl$) \cite{PraPraOrt07}.

Let us remark that, in contrast to the four-dimensional case, vacuum Robinson-Trautman spacetimes of {genuine} types II, III and N are not permitted in higher dimensions, nor does an analog of the $C$-metric exist within the Robinson-Trautman class. This seems to be related to the fact that the condition $\sigma=0$ is too restrictive when $n>4$ (see also section~\ref{sec_GS}). More general Robinson-Trautman spacetimes admitting aligned pure radiation are also of type D and were considered in \cite{PodOrt06}.

\subsubsection{Electrovac solutions}

The authors of \cite{OrtPodZof08} extended the above solutions by allowing for an aligned Maxwell field, i.e. $F_{ab}\ell^b\propto\ell_a$. 
After integration of the Einstein-Maxwell system, one arrives again at metric \eqref{geo_metric fin}, with $h_{\alpha\beta}=h_{\alpha\beta}(x)$ being an Einstein space with Ricci scalar ${\cal R}=(n-2)(n-3)K$ and $K=0,\pm 1$. Now, $H=H(r)$ and the associated Maxwell field are given by
\beqn
	& & \hspace{-1cm} 2H = K-\frac{2\Lambda}{(n-1)(n-2)}\,r^2-\frac{\mu}{r^{n-3}} + \frac{2 Q^2}{(n-2)(n-3)} \frac{1}{r^{2(n-3)}} - \frac{F^2}{(n-2)(n-5)}  \frac{1}{r^{2}} , \label{H_charged} \\
  & & \hspace{-1cm} \mbox{\boldmath$F$} = \frac{Q}{r^{D-2}} \d r \wedge \d u + {\frac{1}{2}}F_{\alpha\beta}(x) \, \d x^\alpha \wedge \d x^\beta , \label{Max_even}
\eeqn
where $\mu$, $Q$ and $F^2\equiv F_{\alpha\gamma} F_{\beta\delta} h^{\alpha\beta} h^{\gamma\delta}$ are constants. In {\em odd dimensions} necessarily $F_{\alpha\beta}(x)=0$, and $\mbox{\boldmath$F$}$ obeys both the Maxwell and the Maxwell-Chern-Simons equations. {\em In even dimensions}, if $F^2\neq 0$, then $h_{\alpha\beta}$ must be not only Einstein but (almost-)K\" ahler Einstein, with $F_{\alpha\beta}(x)$ related to the (almost-)complex structure by $J^\alpha_{\, \beta}=|F|^{-1}(n-2)^{1/2}F^\alpha_{\ \beta}$  (which guarantees that indeed $F^2$ is a constant). As argued above in the vacuum case, all these spacetimes are of type D(bd). The Maxwell field is also of type D (wrt the same null directions), whereas in 4D it can also be of type N (cf. also \cite{Ortaggio07,Durkeeetal10} for related results). When $F^2=0$, these solutions represent generalizations of the familiar Reissner-Nordstr\" om-(A)dS spacetimes \cite{Tangherlini63,GibWil87}, where $\mu$ parametrizes the mass and $Q$ is the electric charge (see \cite{OrtPodZof08} for more details and related references). A non-zero $F^2$ gives rise to a magnetic ``monopole-like'' term, however, in this case, $h_{\alpha\beta}$ cannot be a sphere of constant curvature. The case $n=6$ is special in that there can also be some solutions (not included above) with $F^2=F^2(u,x)$ and $h_{\alpha\beta}=h_{\alpha\beta}(u,x)$, in which case $\pa_u$ need not be a Killing vector field; see \cite{OrtPodZof08} for details {(the algebraic type of these special solutions, which are not warped products, has not been studied)}. The inclusion of aligned pure radiation is also discussed in \cite{OrtPodZof08}.

\subsection{Shearfree twisting solutions (even dimensions)}

\label{subsec_NUT}

All shearfree geodetic WANDs also satisfy, in particular, the optical constraint~\eqref{OC} {(see \eqref{sachs_she})}. In sections~\ref{sec_Kundt} and \ref{sec_RT}, we described Einstein spacetimes with a shearfree and twistfree (thus geodesic) mWAND. Here, we summarize the known results for the case of shearfree but twisting, geodesic mWANDs. First, an obvious consequence of the Sachs equation (section~\ref{subsec_ricci}) is that the twist of 
a geodesic WAND (multiple or not) is constrained by $(n-2)A_{ik}A_{kj}=-\omega^2\delta_{ij}$ so that {\em a twisting geodesic WAND can be shearfree only in even spacetime dimensions} \cite{OrtPraPra07}. If, additionally, $R_{00}\equiv R_{ab}\ell^a\ell^b=0$ (such as for Einstein spaces), then $\bl$ must necessarily be expanding, so that $\det\rhob\neq0$ \cite{OrtPraPra07}. To our knowledge, so far only one solution with a shearfree, twisting, geodesic mWAND has been identified \cite{OrtPraPra12}. This is the six-dimensional Ricci flat static Taub-NUT metric \cite{AwaCha02} 
\beqn
 \d s^2= & & -F(r)(\d t-2k\cos\theta_1\d\phi_1-2k\cos\theta_2\d\phi_2)^2+\frac{\d r^2}{F(r)} \nonumber \\
 & & {}+(r^2+k^2)(\d\theta_1^2+\sin\theta_1^2\d\phi_1^2+\d\theta_2^2+\sin\theta_2^2\d\phi_2^2) ,
 \label{MasSte6D}
\eeqn
where $k$ is a constant and 
\be
 F(r)=\frac{r^4/3+2k^2r^2-2mr-k^4}{(r^2+k^2)^2} .
\ee
We observe that this is a spacetime of type D (for $k=0$, it reduces to a generalized Schwarzschild-Tangherlini solution with S$^2\times$S$^2$ transverse space). A geodesic mWAND is given by
\be
 \ell_a\d x^a=\d t+F(r)^{-1}\d r-2k(\cos\theta_1\d\phi_1+\cos\theta_2\d\phi_2) , 
\ee 
while a second one can simply be obtained by reflecting $\bl$ as $t\to-t$, $\phi_1\to-\phi_1$, $\phi_2\to-\phi_2$ \cite{PraPraOrt07}. These mWANDs are both shearfree, twisting (for $k\neq0$) and expanding \cite{OrtPraPra12}.  One can also show that in the spacetime~\eqref{MasSte6D} there exists an optical structure, as well as certain other integrable totally null distributions with totally geodesic integral surfaces (see \cite{OrtPraPra12} for details).

\section{Other invariantly defined classes of solutions}

In this section, we describe certain classes of spacetimes that admit an invariant definition and that are interesting from various perspectives. In particular, we discuss their permitted Weyl types.

\subsection{Kerr-Schild spacetimes and generalizations}

\label{subsec_KS}

Generalized Kerr-Schild (GKS) metrics have the form 
\be
g_{ab} = \bgmetric a b - 2 {\cal{H}} k_a k_b, \label{GKS}
\ee
where the background metric $\bgmetric a b$ can be any spacetime of constant curvature,  ${\cal{H}}$ is a scalar function and  $\bk$ is a null vector wrt $\bgmetric a b$ -- {and thus automatically also wrt the} full metric $g_{ab}$.\footnote{In fact, one can consider an even more general GKS ansatz in which the background metric $\bgmetric a b$ can be an {\em arbitrary} spacetime, but we will not discuss this possibility here -- see \cite{Stephanibook} for results and references in 4D, and \cite{DerGur86} for some results in higher dimensions. A different extension of the KS ansatz has been studied in \cite{Aliev2009,EttKas10}.\label{foot_GKS}} {The background line element is of course conformally flat, i.e. $\bgmetric a b = \Omega \eta_{ab}$ for a suitable $\Omega$, and represents the (A)dS spacetime -- trivially, GKS includes Kerr-Schild (KS) for $\Omega=1$, and this will be understood in the following, unless stated otherwise.} 

The KS class of spacetimes has played an important role in 4D containing, e.g., Kerr and Kerr-Newman black holes, the Vaidya metric, vacuum \pp waves and type N Kundt metrics \cite{Stephanibook}. In arbitrary higher dimensions, the KS ansatz led to the discovery of rotating vacuum black holes -- the Myers-Perry black holes \cite{MyePer86}. In addition, the GKS class of metrics  contains, e.g., rotating black holes with de Sitter and anti-de Sitter backgrounds in four and higher dimensions \cite{Carter68cmp,HawHunTay99,Gibbonsetal04_jgp}.

In general, the null vector $\bk$ can be arbitrary. However, it turns out that

\begin{prop}[Geodeticity of the GKS vector $\bk$ \cite{OrtPraPra09,MalPra11}]
\label{prop_GKS_geod}
 The null vector $\bk$ in the GKS  metric (\ref{GKS}) is geodetic if, and only if, it is an AND of the Ricci tensor, i.e. \mbox{$R_{ab} k^a k^b=0$}. 
\end{prop}

Note that if Einstein's field equations are employed, this is equivalent to the condition ${T_{ab} k^a k^b}=0$ on the matter content. This includes, in particular, Einstein spacetimes and aligned matter fields such as an aligned Maxwell field ($F_{ab} k^a \propto k_b$) or aligned pure radiation ($T_{ab} \propto k_a k_b$). From now on, we will thus focus on GKS spacetimes in which $\bk$ is geodetic. Note also that $\bk$ is geodesic wrt to $g_{ab}$ iff it is geodesic wrt $\bgmetric a b$. {Moreover, $\bk$ has the same optical properties both in the background and in the full geometry, i.e. the matrix $\rhob$ is unchanged \cite{OrtPraPra09,MalPra11}.}

{Assuming that $\bk$ is geodesic,} in \cite{OrtPraPra09}, it was shown that KS spacetimes are algebraically special. These results were subsequently generalized to the GKS class in \cite{MalPra11}:
\begin{prop}[Multiple WAND in GKS spacetimes  \cite{OrtPraPra09,MalPra11}]
\label{GKS-Weyltypes}
GKS spacetime (\ref{GKS}) with a geodetic KS vector $\bk$ is algebraically special with $\bk$ being {a} multiple WAND. Moreover, $\bk$ is also a multiple AND of the Ricci tensor (i.e. $R_{ab} k^a \propto k_b$), {so that the Riemann type is II}.
\end{prop}

In what follows, we will {restrict to GKS spacetimes that solve the vacuum Einstein equations with a possible cosmological constant. For these, 
the algebraic type can be further constrained according to whether the KS vector $\bk$ is expanding ($\theta\neq0$) or not ($\theta=0$). Let us discuss these subfamilies separately.}

\subsubsection{Non-expanding (Kundt) subclass}

\begin{prop}[Algebraic type of Einstein GKS spacetimes with $\theta=0$  \cite{OrtPraPra09,MalPra11}]
\label{Prop_nonexp}
{$($Non conformally flat$)$} Einstein GKS spacetimes $($\ref{GKS}$)$ with a non-expanding KS congruence $\bk$ are of type N, with $\bk$ being the multiple WAND. Twist and shear of the KS congruence $\bk$ necessarily vanish and these solutions thus belong to the Kundt class.
\end{prop}

In fact, in the Ricci-flat case {also the converse result {has been proven}, so that:}

\begin{prop}[Ricci-flat KS spacetimes with $\theta=0$  \cite{OrtPraPra09}]
{$($Non conformally flat$)$} Ricci-flat KS spacetimes with a non-expanding KS vector $\bk$ coincide with the class of {Ricci-flat} Kundt spacetimes of type N.
\label{KS_N}
\end{prop}
{This case includes type N vacuum \pp waves.} {The} explicit metric for Ricci-flat type N Kundt spacetimes can be found in \cite{Coleyetal06}.

\subsubsection{Expanding subclass}

{As it turns out,} the algebraic types III and N are not compatible with an expanding $\bk$ in the GKS ansatz: 

\begin{prop}[Einstein GKS spacetimes with expanding $\bk$ \cite{OrtPraPra09,MalPra11}]
\label{prop_exp}
{$($Non conformally flat$)$} Einstein GKS spacetimes $($\ref{GKS}$)$ with an expanding KS congruence $\bk$ are of Weyl types II or D, and the multiple WAND $\bk$ obeys the optical constraint~\eqref{OC}.
\end{prop}
{This subclass contains all the black hole solutions mentioned above (which are of type D) -- in particular, static black holes lie in the intersection of the KS and Robinson-Trautman classes \cite{OrtPraPra09}. The fact that $\bk$ is generically shearing when $n>4$, and yet obeys the optical constraint, suggested a way towards a possible extension of the Goldberg-Sachs theorem \cite{OrtPraPra09}, later developed elsewhere, as described in section~\ref{sec_GS}. It was also shown in \cite{OrtPraPra09,MalPra11} that $\bk$ has in general caustics corresponding to curvature singularities. If there is another mWAND (type D subcase), this must also be geodesic \cite{OrtPraPra09,MalPra11}.}

Note that it follows from propositions~\ref{Prop_nonexp} and \ref{prop_exp} that type III Einstein spacetimes are incompatible with the GKS ansatz. However, it can be shown \cite{MalekThesis} that Ricci-flat type III spacetimes of the Kundt class are compatible with the {\em extended} KS ansatz of \cite{Aliev2009,EttKas10}.

\subsection{Purely electric and purely magnetic spacetimes}

\label{subsec_PEPM}

\subsubsection{Definitions and general properties}

The standard 4D decomposition of the Maxwell tensor $F_{ab}$ into its electric and magnetic parts $\vec E$ and $\vec B$ 
wrt an observer (identified here with its normalized timelike $n$-velocity $\bu$) can be extended to any tensor in an $n$-dimensional Lorentzian space \cite{Senovilla00,Senovilla01,HerOrtWyl12} (here we will adhere to the terminology of \cite{HerOrtWyl12}, where a comparison with the earlier references \cite{Senovilla00,Senovilla01} can also be found). In the following, we will discuss this in the case of the Weyl tensor (see \cite{Senovilla00,Senovilla01,HerOrtWyl12} for results for general tensors; in particular, in \cite{HerOrtWyl12}, several useful results for the Ricci and Riemann tensors have been also worked out). {In 4D, such an ``electro-magnetic'' splitting of the curvature tensor was first considered in \cite{Matte53}, and there is now an extensive literature on the subject \cite{Stephanibook} (some more recent references can be found in \cite{HerOrtWyl12}). Some of the 4D results have been extended to higher dimensions, but there are also some differences, as shown in the following.}

The ``electric'' and ``magnetic'' parts of $C_{abcd}$ can be defined, respectively, as the tensors \cite{HerOrtWyl12}
\begin{eqnarray}
&&(C_+)^{ab}{}_{cd}=h^{ae}h^{bf}h_c{}^gh_d{}^h C_{efgh}+4u^{[a}u_{[c}C^{b]e}{}_{d]f}u_eu^f ,\label{C+}\\
&&(C_-)^{ab}{}_{cd}=2h^{ae}h^{bf}C_{efk[c}u_{d]}u^k+2u_ku^{[a}C^{b]kef}h_{ce}h_{df} , \label{C-}
\end{eqnarray}
where 
\be
	h_{ab}=g_{ab}+u_au_b , 
	\label{space_h}
\ee
is the projector orthogonal to $\bu$. These extend the well-known 4D definitions \cite{Matte53,Stephanibook}. 
In any orthonormal frame adapted to $\bu$, the electric [magnetic] part
accounts for the components of the Weyl tensor with an even [odd] number of indices $u$ (i.e. those which are invariant [change sign] under a time reflection $\bu\to-\bu$ \footnote{Note that the opposite rule applies if one defines the electric/magnetic components of the Maxwell tensor $F_{ab}$ \cite{Senovilla00,Senovilla01,HerOrtWyl12}.}). There are $(n^2-1)(n-3)/3$ independent magnetic components and, due to the tracefree property,  $(n^2-2n+4)(n+1)(n-3)/12$ independent electric components, which together add up to the $(n+2)(n+1)n(n-3)/12$ independent components of the Weyl tensor. We can give the following
\begin{defn}[PE/PM spacetimes]
\label{def PE/PM} 
At a spacetime point $($or in a region$)$, the Weyl tensor is called purely electric $[$magnetic$]$ $($from now on, PE $[$PM$])$ wrt $\bu$ if $C_-=0$ $[C_+=0]$. The corresponding spacetime is also called PE $[$PM$]$.
\end{defn}
According to the above definition, a zero Weyl tensor is (trivially) both PE and PM. In no other cases can a Weyl tensor be both PE and PM (even wrt different timelike directions) \cite{HerOrtWyl12}.

There exist frame-independent criteria, similar to the Bel-Debever conditions considered in section~\ref{Sec_BelDeb}, to assess whether a Weyl tensor is PE or PM:
\begin{prop}[Weyl PE/PM Bel-Debever criteria \cite{HerOrtWyl12}] 
A Weyl tensor $C_{abcd}$ is
\begin{itemize}
\item PE wrt $\bu$ iff $u_ag^{ab}C_{bc[de}u_{f]}=0$, 
\item PM wrt $\bu$ iff $u_{[a}C_{bc][de}u_{f]}=0$.
\end{itemize}
\label{prop_Bel-Debever}
\end{prop}

In connection to the bivector approach of section~\ref{subsec_bivector}, the following necessary conditions are also of interest:
\begin{prop}[Eigenvalues of PE/PM Weyl operators \cite{HerOrtWyl12}] 
\label{th_Weyl_PE} 
A PE or PM
Weyl operator is diagonalizable, i.e. a basis of eigenvectors for
$\wedge^2T_pM$ exists. A PE $[$PM$]$ Weyl operator has only real $[$purely
imaginary$]$ eigenvalues. Moreover, a PM Weyl
operator has at least $\frac{(n-1)(n-4)}{2}$ zero eigenvalues.
\end{prop}
Note that in 4D (only) the above conditions are also sufficient (see \cite{Stephanibook,WylVan06} and references therein).

\subsubsection{Relation with the null alignment classification}

The electric/magnetic splitting is defined wrt a timelike vector field $\bu$, and there is no obvious reason, a priori, that this has anything to do with the null alignment classification. However, it turns out that only certain Weyl types are permitted for a PE/PM Weyl tensor. More specifically one has the following proposition:
\begin{prop}[Algebraic types of PE/PM Weyl tensors \cite{HerOrtWyl12}] 
\label{PE_PM_types}
A PE/PM Weyl tensor wrt a certain $\bu$ can be only of type G, I$_i$, D or O. In the type I$_i$ and D cases, the vector $\bu$
``pairs up'' the space of WANDs, in the sense that the second null
direction of the timelike plane spanned by $\bu$ and any WAND is
also a WAND with the same multiplicity.
Furthermore, a type D Weyl tensor is PE iff it is type D$($d$)$, and PM
iff it is type D$($abc$)$.
\end{prop}
This can thus be also viewed as another possible refinement, for some subcases closely related to the spin type one \cite{Coleyetal12}. The (non-)uniqueness of $\bu$ is described for different Weyl types by the following proposition:
\begin{prop}[Uniqueness of $\bu$ \cite{HerOrtWyl12}] 
\label{prop uniqueness}
A PE $[$PM$]$ Weyl tensor is PE $[$PM$]$ wrt 
\begin{itemize}
 \item a unique $\bu$ $($up to sign$)$ in the type I$_i$ and $G$ cases, 
 \item any $\bu$ belonging to the space spanned by all mWANDs $($and only wrt such $\bu$s$)$ in the type D case $($noting also that if there are more than two mWANDs the Weyl tensor is necessarily PE $($type D$($d$))$ {\rm \cite{Wylleman12}}$)$.
 \end{itemize}
\end{prop}

\subsubsection{PE spacetimes}

\label{subsubsec_PE}

Here, we {present} a class  of PE spacetimes. {More examples will be given in section~\ref{subsec_products}.} 

\begin{prop}[Spacetimes with a shearfree and twistfree observer \cite{HerOrtWyl12}] 
 \label{prop_shear_twist_free}
All spacetimes admitting a shearfree, twistfree, unit timelike vector field $\bu$ are PE wrt $\bu$. These admit a line element of the form 
\begin{equation}
    \d s^2=-V(t,x)^2\d t^2 + P(t,x)^2\xi_{\alpha\beta}(x)\d x^\alpha \d x^\beta.
    \label{sheartwistfree}
\end{equation}
In these coordinates one has $\bu=V^{-1}\pa_t$, and the remaining kinematic quantities are $\Theta={V}^{-1}(\ln P)_{,t}$ and $\dot{u}_\alpha=
(\ln V)_{,\alpha}$.\footnote{Recall that, in general, the kinematic quantities of a unit timelike vector field are defined as the irreducible parts of $u_{a;b}$, i.e. $u_{a;b}=-\dot u_au_b+\omega_{ab}+\sigma_{ab}+\frac{\Theta}{n-1}h_{ab}$, where $\omega_{ab}u^a=\Theta_{ab}u^a=\sigma_{ab}u^a=\dot{u}_au^a=0$ (cf.~\eqref{space_h}). See, e.g., appendix~C of \cite{HerOrtWyl12} for more details.}
\end{prop}
The above class of spacetimes includes, in particular, direct, warped and doubly warped products with a one-dimensional timelike factor, and thus all {static} spacetimes (see also \cite{PraPraOrt07} and propositions~\ref{prop_warped1} and \ref{prop_static}).

Also the presence of certain (Weyl) isotropies (e.g., $SO(n-2)$ for $n>4$) implies that the spacetime is PE, see \cite{ColHer09,HerOrtWyl12} for details and examples. Certain spacetimes possessing also a purely electric Riemann tensor are discussed in \cite{HerOrtWyl12}.

\subsubsection{PM spacetimes}

\label{sec_PM}

PM spacetimes are most elusive, also in four dimensions (see, e.g., \cite{WylVan06} and references therein). In particular, no properly PM Einstein spacetimes are presently known (even for $n=4$). Although there is no general non-existence proof of PM Einstein spacetimes, one has the following restriction (extending the 4D results of \cite{Hall73,McIntoshetal94})

\begin{prop}[{No type D PM Einstein spacetimes} \cite{HerOrtWyl12}]
Einstein spacetimes with a type D, PM Weyl tensor do not exist.
\end{prop}

However, examples of PM spacetimes can be constructed by taking appropriate products, as described {in \cite{HerOrtWyl12}, where explicit (non-Einstein) metrics have also been constructed} (to our knowledge, the only PM examples known in higher dimensions so far).

To conclude this part, let us just mention that PE/PM tensors (in particular, PE/PM  Weyl or Riemann tensors) provide examples of {\em minimal tensors} \cite{RicSlo90}. Thanks to the recently proved {\em alignment theorem} \cite{Hervik11}, the latter are of special interest since they are precisely the {\em tensors characterized by their invariants} \cite{Hervik11} (cf. also \cite{HerOrtWyl12}).  This in turn sheds new light on the algebraic classification of the Weyl tensor, providing a further invariant characterization that distinguishes the (minimal) types G/I/D from the (non-minimal) types II/III/N.

\subsection{Direct product and warped spaces}

\label{subsec_products}

Several known solutions (e.g., Schwarzschild-Tangherlini) are described by a warped line element. Moreover, direct/warped product spaces represent a useful tool to generate examples with relative ease, which can  be employed to test or falsify certain conjectures about properties of higher dimensional gravity (see, e.g., \cite{GodRea09,OrtPraPra12} and section~\ref{sec_GS} for instances in the context of the Goldberg-Sachs theorem, or the construction \cite{HerOrtWyl12} of PM spacetimes mentioned in section~\ref{sec_PM}). It is thus the purpose of this section to summarize the main properties of such spacetimes, especially in connection with the Weyl tensor classification. 

\subsubsection{Definition and general properties}

Let us start with the following definition:

\begin{defn}[Product spacetimes]
The $n$-dimensional spacetime $(M,\bg)$ is called a ``doubly warped product'' if 
\begin{itemize}
\item $M$ is a direct product manifold $M=M^{(n_1)}\times M^{(n_2)}$ of factor spaces $M^{(n_1)}$ and $M^{(n_2)}$, where $n=n_1+n_2\ge 4$, and $M^{(n_1)}$ represents the Lorentzian (timelike) factor, 
\item  $\bg$ is conformal to a direct sum metric $\bg=e^{2(f_1+f_2)}\left(\bg^{(n_1)}\oplus \bg^{(n_2)}\right)$,
where $\bg^{(n_i)}$ is a metric on $M^{(n_i)}$ ($i=1,2$) and $f_1$, $f_2$ are smooth scalar functions on $M^{(n_1)}$ and $M^{(n_2)}$, respectively.
\end{itemize}
If {either} $f_1$ or $f_2$ are constant, the spacetime is dubbed as a ``warped product'', and if both are constant  then it is dubbed as a ``direct product''. 
\end{defn}

(Doubly) warped spacetimes are clearly conformal to direct products, so that the Weyl tensors of both will have the same properties. {From now on, quantities denoted by an index $n_i$ will refer to the factor geometry $(M^{(n_i)},\bg^{(n_i)})$.} 
A basic property known for a long time is the following:
\begin{prop}[Decomposability of {Riemann, Ricci and Weyl} tensors \cite{Ficken39}]
\label{prop_decomp}
In a direct product spacetime, the Riemann and Ricci tensors and the Ricci scalar are decomposable; the Weyl tensor is decomposable iff both product spaces are Einstein spaces and $n_2(n_2-1)R_{(n_1)}+n_1(n_1-1)R_{(n_2)}=0$.  
\end{prop}
Note that the latter condition is identically satisfied whenever $n_1=1$ or $n_2=1$, while for $n_1=2$ [$n_2=2$] it implies that $(M^{(n_1)},\bg^{(n_1)})$ [$(M^{(n_2)},\bg^{(n_2)}$)] must be of constant curvature. Proposition~\ref{prop_decomp} has the following important consequences:
\begin{prop}[Conformally flat products \cite{Ficken39}]
A direct product is conformally flat iff both product spaces are of constant curvature and $n_2(n_2-1)R_{(n_1)}+n_1(n_1-1)R_{(n_2)}=0$.
\end{prop}
\begin{prop}[Einstein direct products \cite{Ficken39}]
A direct product is an Einstein space iff each factor is an Einstein space and their Ricci scalars satisfy $R_{(n_1)}/n_1=R_{(n_2)}/n_2$.
\end{prop}
The latter result and proposition~\ref{prop_decomp} implies that a direct product with a decomposable Weyl tensor cannot be properly Einstein (and it is Ricci-flat iff both factor spaces are Ricci-flat).

\subsubsection{Relation with the null alignment classification and PE/PM character}

Here, we consider various cases of product spacetimes, which describe several important classes of metrics and in which the Weyl type can be constrained. First, (doubly) warped and direct {products} with a one-dimensional timelike factor are special cases of metric~\eqref{sheartwistfree} (see \cite{HerOrtWyl12}), so that from propositions~\ref{PE_PM_types} and \ref{prop_shear_twist_free} we immediately have the following proposition:

\begin{prop}[Warps with $n_1=1$ \cite{PraPraOrt07,HerOrtWyl12}]
\label{prop_warped1}
	A (doubly) warped spacetime with a one-dimensional timelike factor can be only of type G, I$_i$, D(d) or O, and it is PE.
\end{prop}

In particular, static spacetimes belong to the above family (see also propositions~\ref{prop_static} and \ref{prop_timelike}). If the ``timelike'' (Lorentzian) factor is two dimensional, we instead have

\begin{prop}[Warps with $n_1=2$ \cite{PraPraOrt07,HerOrtWyl12}] 
\label{prop: warp 2D factor} 
	A (doubly) warped spacetime with a two-dimensional Lorentzian factor is either type O, or type D(d) and PE wrt {any} unit timelike vector living in $M^{(n_1)}$, the uplifts of the null directions of the tangent space to $(M^{(n_1)},\bg^{(n_1)})$ being double WANDs of the complete spacetime $(M,\bg)$. If $(M^{(n_2)},\bg^{(n_2)})$ is Einstein, the type specializes to D(bd) and if it is of constant curvature to D(bcd).
\end{prop}
{In particular, all spherically, hyperbolically or plane symmetric spacetimes belong to the latter special case (see also proposition~\ref{prop_spherical}).}

For warped products in which the Lorentzian factor is at least three dimensional the above proposition does not hold, in general. However, {if $(M^{(n_1)},\bg^{(n_1)})$ is three dimensional and Einstein} one has the following proposition:

\begin{prop}[Warps with $n_1=3$ \cite{PraPraOrt07,HerOrtWyl12}] 
\label{prop_warped3} 
	A (doubly) warped spacetime  in which $(M^{(n_1)},\bg^{(n_1)})$  is a three-dimensional Einstein spacetime can be only of type D(d) or O. The uplift of {\em any} null direction of the tangent space to $(M^{(n_1)},\bg^{(n_1)})$ is a double WAND of the complete spacetime $(M,\bg)$, which is PE wrt {any} unit timelike vector living in $M^{(n_1)}$.
\end{prop}
The above are thus examples of spacetimes admitting a continuous infinity ($\infty^1$) of (including non-geodesic) mWANDs, such as dS$_3\times$S$^{(n-3)}$, see also \cite{GodRea09,Durkee09,DurRea09}.

If $n_1>3$, one needs stronger assumptions to arrive at a similar result:
\begin{prop}[Warps with $n_1>3$ \cite{PraPraOrt07,HerOrtWyl12}] 
\label{prop_warped>3} 
In a (doubly) warped spacetime with $n_1>3$
	\begin{enumerate}[(a)]
	\item if $(M^{(n_1)},\bg^{(n_1)})$ is an Einstein spacetime of type D, $(M,\bg)$ can be only of type D (or O) and the uplift of a double WAND of $(M^{(n_1)},\bg^{(n_1)})$ is a double WAND of $(M,\bg)$, 
	\item if $(M^{(n_1)},\bg^{(n_1)})$ is of constant curvature, $(M,\bg)$ can only be of type D(d) (or O) and the uplifts of {\em any} null direction of the tangent space to $(M^{(n_1)},\bg^{(n_1)})$ is a double WAND of $(M,\bg)$; $(M,\bg)$ is PE wrt {any} unit timelike vector living in $M^{(n_1)}$.
	\end{enumerate}
\end{prop}

Examples of the class {(a)} are Kerr black strings, which are indeed of type D but not D(d). Spacetimes of the class {(b)} (clearly a subfamily of {(a)}) contain a continuous infinity {($\infty^{n_1-2}$)}  of mWANDs. {A vacuum example of the latter is the KK bubble \cite{Witten82} (see also \cite{GodRea09}), while an Einstein-Maxwell example is the Melvin fluxbrane \cite{Gibbons86,GibWil87}, {which belongs to the Kundt class} -- thus, both these spacetimes are of type D(d).}

More generally, necessary and sufficient conditions for a direct product spacetime to be PE or PM  (wrt a unit timelike vector either living in $M^{(n_1)}$ or not) have been given in \cite{HerOrtWyl12}, along with some examples.

Certain vacuum \pp waves of type more general than N can be easily constructed as direct products {(recall also proposition~\ref{prop_pp} for a more general result)}:
\begin{prop}[Direct product vacuum \pp waves \cite{OrtPraPra09}] 
\label{prop_pp_products} In a  direct product spacetime $M^{(n)}=M^{(n_1)}\times M^{(n_2)}$ 
\begin{enumerate}[(a)]
	\item if $(M^{(n_1)},\bg^{(n_1)})$ is a vacuum \pp wave of type N or III(a) and $(M^{(n_2)},\bg^{(n_2)})$ is Ricci-flat (but non-flat), then $(M^{(n)},\bg^{(n)})$ is a vacuum \pp wave of type II'(abd) and not more special 	  (here $n_1,n_2\ge4$, and $n_1\ge5$ if the $n_1$-dimensional \pp wave is of type III(a)),
	\item if $(M^{(n_1)},\bg^{(n_1)})$ is flat and $(M^{(n_2)},\bg^{(n_2)})$ is Ricci-flat (but non-flat), then $(M^{(n)},\bg^{(n)})$ is a vacuum \pp wave of type D(abd), and not more special (with $n_1\ge2$, $n_2\ge4$).
\end{enumerate}
\end{prop}
The above specific subtype D(abd) was mentioned in \cite{Ortaggio09}.

Einstein spacetimes given by a warp with $n_1=1$ and $f_2=$const, or $n_1=n-1$ and $f_1=$const are described by Brinkmann's metrics \cite{Brinkmann25}. In the former case, they are comprised by proposition~\ref{prop_warped1}; in the latter case, they can be of any Weyl type and have been analyzed in detail in \cite{OrtPraPra11}, as reviewed in section~\ref{subsubsec_Brink}.

\subsubsection{Brinkmann's warps}

\label{subsubsec_Brink}

As mentioned above, a special class of warped metrics with a single spacelike ``extra dimension'' (i.e. $n_1=n-1$ and $f_1=$const) is particularly interesting if one restricts to Einstein spacetimes. These were first considered by Brinkmann in his study of Einstein spaces that can be mapped conformally on other Einstein spaces \cite{Brinkmann25}. From our perspective, the interest in such metrics is twofold. On the one hand, Brinkmann's line element can be used to generate Einstein spacetimes with given algebraic properties and optics (see {\cite{Ortaggio07,PraPraOrt07,OrtPraPra10,OrtPraPra11}} for some applications). On the other hand, 
it essentially consists of a slicing of an Einstein spacetime by hypersurfaces which are, in turn, also Einstein. It is thus of interest also in brane-world scenarios, where it provides a consistent embedding of $(n-1)$-dimensional Einstein gravity in $n$-dimensional Einstein gravity (with various possible values for the bulk and lower dimensional cosmological constants), see, e.g., \cite{RanSun99a,RanSun99b,LuPop01,CveLuPop01,ParPopSad02} and references therein, where various supergravity extensions (relying on the same metric ansatz) have also been studied. 

Brinkmann's line element reads 
\be
 \d s^2=\frac{1}{f(z)}\d z^2+f(z)\d\tilde s^2 ,  
 \label{ansatz} 
\ee
where $\d\tilde s^2$ is an $(n-1)$-dimensional metric. Assuming that d$s^2$ is an {\em Einstein spacetime}, it follows that 
\be
 f(z)=-\lambda z^2+2dz+b , \qquad \lambda=\frac{2\Lambda}{(n-1)(n-2)} ,
 \label{fz}
\ee
with $b$ and $d$ being constant parameters. The  {``seed''} metric $\d\tilde s^2$ must also be Einstein, with the Ricci scalar (hereafter tildes will denote quantities referring to the geometry of $\d\tilde s^2$)
\be
 \tilde R=(n-1)(n-2)(\lambda b+d^2).
\label{ricci-n-1}
\ee 
These are precisely the Einstein spacetimes that can be mapped conformally on other Einstein spacetimes by a {\em proper} map \cite{Brinkmann25}. 
In order to preserve the signature, we require $f(z)>0$, which may restrict possible parameter values and (possibly) the range of $z$ \cite{OrtPraPra11}. In particular, not all possible combinations of the signs of $R$ and $\tilde R$ are permitted. Note that (\ref{ansatz}) is form invariant under a redefinition $z=\alpha z'+\beta$, under which $\lambda'=\lambda$, $d'=\alpha^{-1}(d-\lambda\beta)$,  $b'=\alpha^{-2}[b+\beta(-\lambda\beta+2d)]$ and $\d\tilde s'^2=\alpha^{2}\d\tilde s^2$ (so that $\tilde R'=\alpha^{-2}\tilde R$). Alternative coordinates for the metric~(\ref{ansatz}) have also been presented in \cite{OrtPraPra11}. These may sometimes be more convenient for certain applications, e.g., for Kaluza-Klein reduction.

Using coordinates $x^a=(x^\mu,z)$ {(with Greek indices ranging from 0 to $n-2$)}, for the coordinate components of the Weyl tensor, it is straightforward to show that \cite{Brinkmann25} 
\be
 C_{\mu\nu\rho\sigma}=f\tilde C_{\mu\nu\rho\sigma} , \qquad C_{z\mu\nu\rho}=0=C_{z\mu z\nu} .
 \label{weyl}
\ee
First, it is obvious that $\d s^2$ is conformally flat (and thus of constant curvature since Einstein) iff $\d\tilde s^2$ is such \cite{Brinkmann25}. Therefore, in four dimensions, metric~(\ref{ansatz}) describes only spaces of constant curvature (a three-dimensional Einstein space $\d\tilde s^2$ is necessarily of constant curvature). For $n>4$, this is generically not the case, and one can show \cite{OrtPraPra11} that $\d s^2$ inherits WANDs from $\d\tilde s^2$ with the same multiplicity. In particular cases, $\d s^2$ can however also possess additional WANDs unrelated to those (if any) of $\d\tilde s^2$. Consequently, $\d s^2$ is in general of the same Weyl type of $\d\tilde s^2$, but in {special} cases (see \cite{OrtPraPra11} for details and some examples) it can be more special. In particular, if $\d\tilde s^2$ is algebraically special, then $\d s^2$ is of the same algebraic type {(but not vice versa)}. This is summarized in table~\ref{tab_types}.

\begin{table}[t]

\begin{minipage}[b]{0.5\linewidth}

\centering

\begin{center}
	\begin{tabular}{c|c}
			 Given type & Possible type \\
			 of $\d\tilde s^2$ & of $\d s^2$ \\
			 &\\[-3mm]
			 \hline &\\[-10pt]
   G	& G, I$_i$, D$_{abd}$    \\ [1pt]
		\hline &\\[-10pt]
		I & I, I$_i$  \\[1pt]
		\hline &\\[-10pt]
	  I$_i$ & I$_i$, D$_{abd}$  \\[1pt]
		\hline &\\[-10pt]
		II & II  \\[1pt]
		\hline &\\[-10pt]
		D & D  \\[1pt]
		\hline &\\[-10pt]		
		III & III \\[1pt]
		\hline &\\[-10pt]
		N & N \\[1pt]
		\hline &\\[-10pt]
		O & O \\[1pt]
		\end{tabular}
		\end{center}

\end{minipage}
\begin{minipage}[b]{0.1\linewidth}

\centering

\begin{center}
	\begin{tabular}{c|c}
			 Given type  & Possible type \\
			 of $\d s^2$ & of $\d\tilde s^2$ \\
			 &\\[-3mm]
			 \hline &\\[-10pt]
   G	& G   \\[1pt]
		\hline &\\[-10pt]
		I & I  \\[1pt]
		\hline &\\[-10pt]
		I$_i$ & G, I, I$_i$  \\[1pt]
		\hline &\\[-10pt]
		II & II  \\[1pt]
		\hline &\\[-10pt]
		D & G, I$_i$, D \\[1pt]
				\hline &\\[-10pt]
		III & III \\[1pt]
		\hline &\\[-10pt]
		N & N \\[1pt]
		\hline &\\[-10pt]
		O & O \\[1pt]		
		\end{tabular}
		\end{center}
	
\end{minipage}

\caption{Possible relation between the Weyl type of the seed spacetime $\d\tilde s^2$ and the full spacetime $\d s^2$ for metric (\ref{ansatz}) with (\ref{fz}) (from \cite{OrtPraPra11}). {In the case $n=5$ (only), if d$s^2$ is type D, then d$\tilde{s}^2$ must be type D as well \cite{Ortaggio09,OrtPraPra11}.}}

\label{tab_types}

\end{table}

Using scalar curvature invariants and components of the Weyl tensor in a p.p. frame, it has also been shown \cite{OrtPraPra11} that all spacetimes $\d s^2$ (\ref{ansatz}) with (\ref{fz}) contain (p.p.) curvature singularities arising in the full spacetime $\d s^2$ due to the warp factor (at points where $f(z)=0$), except when both the cosmological constant of $\d\tilde s^2$ and that of $\d s^2$ are negative (and in the trivial case of a direct product spacetime $f=$const).  Two explicit {five-dimensional} examples of warped metrics without naked singularities have also been discussed in \cite{OrtPraPra11}: an AdS black string sliced by an AdS spinning black hole, and an accelerated AdS black string generated from the four-dimensional AdS C-metric. Many more general solutions can be easily obtained.

\subsection{Spacetimes with symmetries}

\label{sec_symmetries}

{In this section,} we summarize  classes of spacetimes with isometries (sections~\ref{subsubsec_timelike}--\ref{subsubsec_axisymm}) or other ``symmetries'' (section~\ref{subsubsec_othersymm}), for which the intersection with the Weyl classification has been studied (see, e.g., \cite{Stephanibook,Hallbook} and references therein for corresponding results in 4D). 
Properties of spacetimes admitting certain isotropies are discussed in \cite{ColHer09,HerOrtWyl12}.

{First of all, let us} recall that an $n$-dimensional spacetime is of constant curvature iff it admits a group of motions of dimension $n(n+1)/2$. Furthermore, an $n$-dimensional spacetime (with $n\ge 3$) is of constant curvature (and thus of type O) iff it admits an isotropy group of dimension $n(n-1)/2$. These are local results and hold, in fact, for any signature.  The corresponding maximally symmetric spacetimes are of course only Minkowski or (A)dS (see, e.g., \cite{Stephanibook}). In the following, we first discuss consequences of the presence of a single Killing vector field\footnote{Namely, a timelike or null one. A single spacelike Killing vector does not impose any constraints on the algebraic type of the Weyl tensor, in general, and all types are in fact possible \cite{PraPraOrt07}.} and then review results on spacetimes admitting larger isometry groups, such as {spherical/hyperbolic/plane or ``axial'' symmetry} (note that the considered classes of spacetimes are not necessarily disjoint). Most of the results are purely geometric, i.e. no field equations are initially assumed, except for section~\ref{subsubsec_axisymm} and, in part, \ref{subsubsec_othersymm}.

\subsubsection{Spacetimes with a timelike Killing vector field} 

\label{subsubsec_timelike}

If the timelike Killing vector field is hypersurface orthogonal, we have {\em static} spacetimes, for which 

\begin{prop}[Static spacetimes \cite{PraPraOrt07,HerOrtWyl12}]
\label{prop_static}
	An $n\ge4$ static spacetime can be only of type G, I$_i$, D(d) or O, and it is PE.
\end{prop}

This can also be seen as a special case of propositions~\ref{PE_PM_types} and \ref{prop_shear_twist_free}. Indeed, the general line element of static spacetimes is \eqref{sheartwistfree} with $V=V(x)$ and $P=P(x)$. See table~\ref{tab_summary} for corresponding examples of all permitted Weyl types. Obviously, the conclusions of propositions~\ref{prop_static} apply also to spacetimes that are conformal to a static spacetime, such as (doubly) warped products with a one-dimensional timelike factor {(proposition~\ref{prop_warped1})}. 

When the timelike Killing vector field is not hypersurface orthogonal, the corresponding spacetime is {\em stationary}. By adding some extra assumptions, one arrives again at the types G, I$_i$, D(d) or O (but counterexamples, not obeying the extra assumptions, are also known) \cite{PraPraOrt07}. However, stationary spacetimes in general are neither PE nor PM \cite{HerOrtWyl12}. Typical examples are rotating black holes/rings (table~\ref{tab_summary}).

\subsubsection{Spacetimes with a null Killing vector field} 

\label{subsubsec_null}

In this case, one can extend well-known 4D results \cite{Stephanibook} by the following proposition:

\begin{prop}[Null Killing vector field \cite{PraPraOrt07,OrtPraPra07}]
\label{prop_null_kill}
	A null Killing vector field $\bk$ is necessarily geodesic, shearfree and non-expanding, with twist given by $\omega^2=R_{ab}k^a k^b$. If $\bk$ is twistfree (i.e. $R_{ab}k^a k^b=0$), then it must be a WAND (and vice versa if $n$ is odd), so that the spacetime is of {Weyl (and Riemann)} type I (or more special) and belongs to the Kundt class. If, additionally, $R_{ab}k^a\propto k_b$, then the Weyl {(and Riemann)} type is II (or more special). 
\end{prop}

Note that both conditions on the Ricci tensor (the latter implying the former) hold, in particular, in the case of Einstein spacetimes, while the existence of a twisting null Killing vector field requires the presence of ``null'' matter (see \cite{Stephanibook} for examples in 4D). On the other hand, a special subfamily admitting a twistfree null Killing vector field is represented by the full class of \pp waves (for which $k_{a;b}=0$), see also section~\ref{subsubsec_pp}. See instead \cite{JakTaf09} for an example of a (Ricci-flat) spacetime of type II$_i$ that belongs to the Kundt class with $\taub\neq0$  (and thus is not a \pp wave) {and admits a null Killing vector field}.

\subsubsection{Spherically, hyperbolically or plane symmetric spacetimes} 

A line element with spherical, hyperbolic or plane symmetry can be written as a warped spacetime with a two-dimensional Lorentzian factor (see, e.g., \cite{KraPle80,Liang90}), which implies (cf. the more general proposition~\ref{prop: warp 2D factor}) 

\begin{prop}[Spherical, hyperbolic and plane symmetry \cite{PraPraOrt07,HerOrtWyl12}]
\label{prop_spherical}
 An $n\ge 4$ spherically, hyperbolically or plane symmetric spacetime is of type {D(bcd)} (in particular, PE) or O. 
\end{prop}

If one adds the assumption that the spacetime is Einstein, a {\em generalized Birkhoff theorem} \cite{KraPle80,Liang90} implies that the only non-conformally flat solutions are Schwarzschild-Tangherlini metrics~\eqref{geo_metric fin} {with \eqref{Hvacuum} and $h_{\alpha\beta}$ representing a space of constant curvature} (with $K=\pm1, 0$ depending on the isometry group), or Nariai-like solutions, i.e. dS$_2\times$S$^{n-2}$ (spherical symmetry) or AdS$_2\times$H$^{n-2}$ (hyperbolic symmetry).

\subsubsection{Axially symmetric Einstein spacetimes} 

\label{subsubsec_axisymm}

In~\cite{GodRea09},  higher dimensional Einstein spacetimes with axial symmetry (i.e. invariance under an $SO(n-2)$ isometry group whose orbits are $(n-3)$--spheres)were studied. The authors of \cite{GodRea09}  first considered the {\em static} subclass and gave conditions on the metric functions (in adapted coordinates) for these spacetimes to be of type G, I$_i$ or D (other types are not permitted \cite{PraPraOrt07}, see proposition~\ref{prop_static}). Examples here include the type D Schwarzschild-Tangherlini black hole and Schwarzschild black string, the type G static Kaluza-Klein (KK) bubble (i.e. a direct product of a flat time direction with Euclidean Schwarzschild), and other spacetimes of type G or I$_i$. Some of these solutions have the interesting property of being of type G in an open subset of spacetime and of type I$_i$ in another open subset. See also \cite{ChaGre04} for other properties of this class of metrics.

Next, in~\cite{GodRea09},  general (i.e. not necessarily static) axisymmetric Einstein spacetimes were studied and  all such solutions for which the type is II or more special
were determined. These comprise (cf. also table~\ref{tab_summary}, {where some of the following Weyl types are further constrained}): (generalized) Schwarzschild black holes (type D), (warped) black strings (type D), axisymmetric Kundt solutions (II, D, or N),\footnote{Note that axisymmetric Kundt solutions of type D also include, e.g., the products dS$_3\times$S$^{n-3}$, dS$_2\times$S$^{n-2}$, AdS$_2\times$H$^{n-2}$. However, within the axisymmetric Kundt class, only the type N subclass solutions can be fully determined in {an} analytic form and it represents non-expanding axisymmetric gravitational waves propagating on a constant-curvature background. See also \cite{JakTaf08,JakTaf09} for examples of type II and N.} dS$_{n-2}\times$S$^2$ (type D, $\Lambda>0$), and an analytic continuation of Schwarzschild-Tangerlini, {i.e.,} KK bubbles (type D, arbitrary $\Lambda$). The latter two solutions are unique in that they possess (a continuous infinity $\infty^{n-4}$ of) {\em non-axisymmetric mWANDs}, some of which are non-geodesic (see also \cite{DurRea09}). On the other hand, the only axisymmetric Einstein spacetime admitting a {\em non-geodesic, axisymmetric mWAND} is dS$_3\times$S$^{n-3}$ (type D, with $\Lambda>0$). All remaining solutions possess geodesic, axisymmetric mWANDs. 
Note in particular that type III spacetimes are not permitted, and the only possible type II/N spacetimes belong to Kundt. All type D solutions turn out to admit an isometry group larger than the assumed $SO(n-2)$. The analysis of \cite{GodRea09} also shows that, within the considered class, a geodesic, axisymmetric WAND must be an mWAND, and must be twistfree.

\subsubsection{Other symmetries} 

\label{subsubsec_othersymm}

So far we have considered metrics with certain isometries. We now consider spacetimes that admit a vector or tensor field {with other} special properties and discuss consequences in terms of the permitted Weyl types. 

First, let us mention spacetimes whose holonomy is {(a subgroup of)} Sim($n-2$). These are spacetimes which admit a {\em recurrent null vector field} {(included in the class of metrics studied in \cite{Walker50})}, for which we have the following proposition:
\begin{prop}[Recurrent null vector field \cite{GibPop08,Ortaggio09}]
\label{prop_recurr}
 Spacetimes that admit a recurrent null vector field, i.e. $k_ak^a=0$, $k_{a;b}=k_ap_b$, {coincide with the subclass $\taub=0$ of the Kundt family}; they are of type II(bd) if Einstein, and II(abd) if Ricci-flat. 
\end{prop}
More explicitly, upon defining a suitable rescaling $\bl=\alpha\bk$, one can always arrive at $\ell_{a;b}=L_{11}\ell_a\ell_b$, which indeed defines the particular subset of Kundt metrics having $\taub=0$, and the corresponding line element is thus \eqref{Kundt_gen} with {$W_{\alpha,r}=0$} \cite{Walker50,GibPop08} {(so that these spacetimes intersect the degenerate Kundt family, cf. section~\ref{subsubsec_deg_Kundt}), see also \cite{Coleyetal06} in the VSI subcase.} In particular, constraints on the metric functions arising in the case of Einstein spacetimes have been studied in \cite{GibPop08} (see also \cite{PodZof09}) {--} in 4D these metrics are ``universal'' \cite{Coleyetal08}.
A special subfamily of the spacetimes of proposition~\ref{prop_recurr} {is obtained if one restricts the holonomy to (a subgroup of) $E(n-2)$ \cite{GibPop08}. Then, the null vector field must be covariantly constant, i.e. also $L_{11}=0$, and one is left with the already discussed} \pp waves \eqref{pp_gen} (i.e. \eqref{Kundt_gen} with {$W_{\alpha,r}=0=H_{,r}$}).

{Next, spacetimes with a {\em covariantly constant timelike vector field} ($u_{a;b}=0$) are those} for which the holonomy is (a subgroup of) $SO(n-1)$. {They are, in particular, static,} and one has the following proposition:
\begin{prop}[Covariantly constant timelike vector field \cite{Ortaggio09,HerOrtWyl12}]
\label{prop_timelike}
Spacetimes that admit a covariantly constant timelike vector field can only be of Weyl type G, I$_i$, D(d) or O, {are PE}, and cannot be properly Einstein. For Ricci-flat spacetimes, the possible types reduce to  G, I$_i(a)$, D(abd) (and O, this being the only possibility if $n=4,5$).
\end{prop}
{The corresponding line element is \eqref{sheartwistfree} with $V=V(t)$, $P=P(x)$, i.e. a direct product spacetimes with a one-dimensional timelike factor (cf.~also propositions~\ref{PE_PM_types}, \ref{prop_shear_twist_free} and \ref{prop_static}). In fact, to arrive at the conclusions of proposition~\ref{prop_timelike}, it suffices to assume $R_{abcd}u^d=0$ (and not necessarily $u_{a;b}=0$) \cite{Ortaggio09,HerOrtWyl12}.}

{Apart from considering vector fields with certain properties, it is sometimes also useful to characterize spacetimes by the existence of special tensor fields. Having in mind the Weyl tensor classification, here we briefly summarize some results for spacetimes admitting a rank-2 {\em conformal Killing-Yano (CKY) tensor}, i.e. a 2-form $h_{ab}=-h_{ba}$ that satisfies}
\be
 h_{ab;c}=\tau_{cab}+\frac{2}{n-2}g_{c[a}\xi_{b]} ,
\ee
{where $\tau_{cab}$ is a 3-form. If $h_{ab}$ is closed, i.e. $\tau_{cab}=0$, {and ``non-degenerate'' (i.e. of maximal matrix rank), it} is dubbed {\em principal CKY tensor}. First, the presence of a CKY tensor restricts the algebraic type of the Weyl tensor:}
\begin{prop}[Conformal Killing-Yano tensor \cite{MasTag08}]
\label{prop_Killing-Yano}
 A spacetime of dimension $n\ge4$  that admits a non-degenerate rank-2 conformal Killing-Yano tensor with distinct eigenvalues is of Weyl type D.
\end{prop}

For spacetimes admitting a principal CKY tensor, the authors of~\cite{HouOotYas07,Krtousetal08} went a step further (see also  \cite{Krtousetal07}). First, they proved the existence of preferred coordinates in which the line element takes a canonical form. Next, by enforcing the vacuum Einstein equations (with a possible $\Lambda$) they proved that the only solution is represented by the Kerr-NUT-(A)dS spacetimes (including special subcases such as Myers-Perry black holes). {Note however that this uniqueness result has been proven only for the Euclidean signature ``continuation'' of those spacetimes. Further, certain technical assumptions such as functional independence of the eigenvalues of $h_{ab}$ were made.}

{Let us just mention that the interest in CKY tensors arises in connection with the integrability of geodesic motion and separability of certain PDEs (such as Hamilton-Jacobi, Klein-Gordon and Dirac) in the corresponding spacetimes. See, e.g., the recent reviews {\cite{YasHor11,CarKrtKub12,Frolov12}} for a number of related references.}

\section{Miscellanea of examples and summarizing table}

In table~\ref{tab_summary}, we present examples of higher dimensional spacetimes of all Weyl types.
{We mainly focus} on Einstein spacetimes but a few solutions with matter/electromagnetic field are also mentioned. BR and BH stand for ``black ring'' and ``black hole'', respectively. Certain spacetimes may have different Weyl types in different regions; in those cases, we have generally indicated only one of those types (for example, black objects are always of type II or more special at Killing horizons, provided the null generator is an eigenvector of the Ricci tensor \cite{LewPaw05,PraPraOrt07}). Similarly, ``static'' and ``PE'' spacetimes are often such only in a certain region -- this is understood. In the last column, we have indicated references where the Weyl type was discussed, whereas references where the corresponding solution was first presented (when different) are given in the second column. Note that some of these classes of solutions overlap, and some are subsets or others, yet it seems convenient to display them separately due to their own importance: for instance, Robinson-Trautman includes Schwarzschild-Tangherlini BH, and the same does Myers-Perry BH if one sets rotation to zero; all \pp waves belong to the Kundt class, etc.  When more details of some examples are discusses in the main text, the corresponding sections are given in the last but one column. The symbol ``$\subset$'' means that the mentioned family is a subset of a larger class of solutions (e.g., Kundt), whereas ``$\in$'' means that the considered spacetime belongs to a certain class.

\begin{footnotesize}
\begin{sidewaystable}[!htbp]
\footnotesize
 \begin{center}
   \begin{tabular}{|c l l r l r r|}
    \hline Type & Spacetime & Matter & $n$ & Comments & Cf. section & Ref. \\ \hline
     G 		& static KK bubble   & vac & $\ge 5$  & PE & \ref{subsubsec_axisymm} & \cite{GodRea09} \\ [1mm]
	     		& ``homog. wrapped object'' \cite{DeSmet02}\footnote{{This metric is a special case of a family of solutions given in \cite{ChoDet82}.} Note that the type is more special according to \cite{ColPel06}, however the WANDs considered in this reference are complex (cf. their eqs.~(30)--(32)), which is of no relevance to the present discussion.} &  vac & 5 & static, PE & {\ref{subsubsec_spinors}} & \cite{GodRea09} \\ [1mm]     
      		& static charged BR \cite{IdaUch03,KunLuc05,Yazadjiev05}  & electrovac & 5 & PE & & \cite{OrtPra06} \\ [2mm] \hline 
    I$_i$ & (non-)rotating BR \cite{EmpRea02prd,EmpRea02prl}\footnote{To be precise, in the rotating case it has been proven only in certain spacetime regions that the type is I$_i$ \cite{PraPra05}.} & vac & 5 &  & {\ref{subsubsec_GS_5d}} & \cite{PraPra05} \\[1mm]      
     			& Gross-Perry \cite{ChoDet82,GroPer83}\footnote{The Weyl type is G in certain spacetime regions \cite{GodRea09}. These metrics contain, for a special choice of parameters, the Gross-Perry-Sorkin soliton \cite{GroPer83,Sorkin83}, while they degenerate to type D for other parameter choices (including static black strings) \cite{GodRea09}.} & vac & 5 & static, PE & & \cite{ColPel06,JakTaf09,GodRea09} \\[1mm]     
		      & lifted Lozanovski \cite{Lozanovski07,HerOrtWyl12} & {imperfect fluid} & 5
		       & PM & \ref{subsec_PEPM} & \cite{HerOrtWyl12} \\ [2mm]  \hline                 			 
		D & Myers-Perry BH \cite{MyePer86} & vac & $\ge 5$ &  & \ref{subsec_KS},\ref{subsubsec_othersymm} & \cite{Pravdaetal04,PraPraOrt07,Hamamotoetal06} \\[1mm]
		  & Kerr-NUT-(A)dS \cite{HawHunTay99,Gibbonsetal04_jgp,CheLuPop06}  & $\Lambda$ & $\ge 5$ &  & \ref{subsubsec_othersymm} & \cite{ColPel06,Hamamotoetal06,PraPraOrt07} \\[1mm]            
		  & Tangherlini BH \cite{Tangherlini63} & (electro)vac, $\Lambda$ & $\ge 5$ & D(bcd) (PE); $\sigma=0$ & \ref{sec_RT},\ref{subsec_products},\ref{sec_symmetries} & \cite{Pravdaetal04,ColPel06,PraPraOrt07,OrtPodZof08} \\[1mm]      
      & Robinson-Trautman\footnote{{Note that for a ``special'' {electrovac} 6D case {(cf. section~4.3 of \cite{OrtPodZof08})} the {Weyl} type has not been determined.}}  & (electro)vac, 												$\Lambda$, pure rad. & $\ge 5$ & D(bd) (PE); $\sigma=0$ & \ref{sec_RT} & \cite{PodOrt06,OrtPodZof08} \\ [1mm]  
 			& black strings/branes  &  vac & $\ge 5$ & {static, D(d) (PE)} & \ref{subsec_products},{\ref{subsubsec_axisymm}} & \cite{ColPel06,PraPraOrt07} \\[1mm]
 			& warped black strings \cite{ChaHawRea00,HirKan01,GodRea09} & $\Lambda$ & $\ge 5$ & static, D(d) (PE) & \ref{subsec_products},{\ref{subsubsec_axisymm}} & \cite{GodRea09,OrtPraPra11} \\[1mm]
 			& (A)dS$_{n_1}\times$(Einst)$^{n_2}$ \cite{Ficken39} &  $\Lambda$ (if $\frac{R_{(n_1)}}{n_1}=\frac{R_{(n_2)}}{n_2}$) & $\ge 5$ & D(d) (PE);  {$\infty^{n_1-2}$ mWANDs}; {$\in$Kundt} & \ref{subsec_products} & \cite{PraPraOrt07,GodRea09,Durkee09,OrtPraPra12}  \\[1mm]                   
 			& Schwarzschild KK bubble \cite{Witten82}  & vac, $\Lambda$ & $\ge 5$ & {D(d) (PE); $\infty^{n-4}$ mWANDs} & \ref{subsec_products},\ref{subsubsec_axisymm} & \cite{GodRea09,PraPraOrt07} \\[1mm]
 			&  Myers-Perry KK bubble \cite{Dowkeretal95,Aharony:2002cx}  & vac& $ 5$ &  &  & \cite{Ortaggioetal12} \\[1mm]
      & Taub-NUT \cite{AwaCha02,ManSte04}  & vac & 6  & $\sigma=0$, $\rhob\neq0$ & \ref{subsec_NUT} & \cite{OrtPraPra12} \\[1mm]
	 	  & $\subset$ \pp waves & vac & $\ge 6$ & D(abd) & \ref{sec_Kundt},\ref{subsec_products} & \cite{OrtPraPra09,Ortaggio09} \\[1mm]  
	 	  & {Melvin \cite{Gibbons86,GibWil87}} & electrovac & $\ge 5$ & static, D(d) (PE) & \ref{subsec_products} & \cite{PraPraOrt07} \\[2mm]  \hline

	II & $\subset$ \pp waves & vac & $\ge 8$ & II'(abd) & \ref{sec_Kundt},\ref{subsec_products} & \cite{OrtPraPra09,Ortaggio09} \\[1mm]            
		 & $\subset$ Kundt & vac & $\ge 5$ & II$_i$ & \ref{sec_symmetries} & \cite{JakTaf09} \\[1mm]
		 & $\subset$ warps of 4D \cite{Brinkmann25} & vac, $\Lambda$ & $\ge 5$ & $\rhob\neq 0$ & \ref{subsec_products} & \cite{OrtPraPra11} \\[2mm]  \hline

 III & $\subset$ Kundt/\pp waves & vac, pure rad.  & $\ge 5$ & VSI; {also $\Lambda$ for Kundt} & \ref{sec_VSI},\ref{sec_Kundt} & \cite{Coleyetal06,OrtPraPra09} \\[1mm]
     & $\subset$ warps of 4D \cite{Brinkmann25}  & vac, $\Lambda$ & $\ge 5$ & $\rhob\neq 0$ & \ref{subsec_products} & \cite{OrtPraPra10,OrtPraPra11} \\[2mm]  \hline

  N  & $\subset$ Kundt/\pp waves & vac, pure rad. & $\ge 5$ & VSI, KS; {also $\Lambda$ for Kundt} & \ref{sec_VSI},\ref{sec_Kundt},\ref{subsec_KS} & \cite{Coleyetal06,OrtPraPra09,MalPra11} \\[1mm]            
     & $\subset$ warps of 4D \cite{Brinkmann25}  & vac, $\Lambda$ & $\ge 5$ & $\rhob\neq 0$ & \ref{subsec_products} & \cite{OrtPraPra10,OrtPraPra11} \\[2mm]  \hline

		O & Minkowski/(A)dS  & vac/$\Lambda$ & $\ge 5$ & & \ref{sec_symmetries} &  \\[1mm]      
		  & dS$_{n_1}\times$H$^{n_2}$, AdS$_{n_1}\times$S$^{n_2}$ \cite{FreRub80,CarDiaLem04}  & $n_1$-form (up to duality) & $\ge 5$ & 					
		  																							   	$\frac{R_{1}}{n_1(n_1-1)}=-\frac{R_{2}}{n_2(n_2-1)}$; {CSI$_K$} & \ref{subsec_products} & \cite{Ficken39,PraPraOrt07} \\	           
			& Einstein universe $\mathbb{R}\times S^{n-1}$  & perfect fluid, $\Lambda$ & $\ge 5$ & static, PE & \ref{subsec_products} & \cite{Ficken39,PraPraOrt07} \\[1mm]
			& LFRW-like universes & perfect fluid, $\Lambda$ & $\ge 5$ & PE & \ref{subsec_products} & \cite{Ficken39,PraPraOrt07,ColHer09,HerOrtWyl12} \\[1mm] \hline           
  \end{tabular}
  \caption{Examples of spacetimes of various Weyl types in higher dimensions. See the text and the indicated sections for more details.
  \label{tab_summary}}
 \end{center}
\end{sidewaystable}
\end{footnotesize}

\section{Applications in quadratic gravity}

\label{sec_QG}

As we have seen in section \ref{Sec_invars}, the curvature structure of {Riemann} type III and N spacetimes is so restricted that all {curvature invariants of order 0 (see definition~\ref{def_inv})} vanish.\footnote{{Recall that some curvature invariants of higher order} are non-vanishing even for type III and N spacetimes  as long as $\rhob\not=0$; see also footnote~\ref{foot_vsi}.}  It thus seems natural  to study whether some subclasses of these spacetimes  solve the vacuum field equations of modified gravities with Lagrangians constructed from {curvature invariants of order 0 in $n$ dimensions} \cite{MalekPravdaQG}. 

One important class of such modified gravities is {\em quadratic gravity}, whose action contains, {in addition to the standard Einstein-Hilbert term, also} general quadratic terms in the curvature \cite{Deser2002}
\begin{equation}
  S = \int \d^n x \sqrt{-g} \bigg( \frac{1}{\kappa} \left( R - 2 \Lambda_0 \right)
      + \alpha R^2 + \beta R^2_{ab}
      + \gamma \left(R^2_{abcd} - 4 R^2_{ab} + R^2 \right) \bigg),
  \label{QG:action}
\end{equation} 
in which $\Lambda_0$, $\kappa$, $\alpha$, $\beta$ and $\gamma$ are constants.\footnote{Recall that in four
dimensions one can set $\gamma=0$ since the corresponding (Gauss-Bonnet) term in the Lagrangian does not contribute to the field equations.}
This action leads to the vacuum quadratic gravity field equations \cite{GulluTekin2009}
\begin{eqnarray}
 \!\!\!\!\!&&\! \!\!\!\!\frac{1}{\kappa} \left( R_{ab} - \frac{1}{2} R g_{ab} + \Lambda_0 g_{ab} \right)
  + 2 \alpha R \left( R_{ab} - \frac{1}{4} R g_{ab} \right)
  + \left( 2 \alpha + \beta \right)\left( g_{ab} \nabla^c \nabla_c - \nabla_a \nabla_b \right) R \nonumber \\
 \!&&\! + 2 \gamma \bigg( R R_{ab} - 2 R_{acbd} R^{cd}
  + R_{acde} R_{b}^{\phantom{b}cde} - 2 R_{ac} R_{b}^{\phantom{b}c}
  - \frac{1}{4} g_{ab} \left( R^2_{cdef} - 4 R^2_{cd} + R^2 \right) \bigg) \nonumber \\
 \! && \! + \beta \nabla^c \nabla_c \left( R_{ab} - \frac{1}{2} R g_{ab} \right)
  + 2 \beta \left( R_{acbd} - \frac{1}{4} g_{ab} R_{cd} \right) R^{cd} = 0. \ 
  \label{QG:fieldeqns}
\end{eqnarray}

For Einstein spacetimes,\footnote{There are {at least} three reasons  to study Einstein spacetimes in this context: (i) this assumption leads to a dramatic simplification of the field equations of quadratic gravity; (ii) it is of interest to know which spacetimes are vacuum solutions of both theories, i.e. they are ``immune'' to the corrections of quadratic gravity; (iii) Einstein spacetimes are relatively well studied and one can thus use the know-how of (higher dimensional) Einstein gravity.}
i.e. $R_{ab}=\frac{2\Lambda}{n-2} g_{ab}$, the equations of quadratic gravity~\eqref{QG:fieldeqns} 
reduce to \cite{MalekPravdaQG}\footnote{Note than in general {{the}} cosmological constant $\Lambda_0$ appearing in the Lagrangian \eqref{QG:action} is distinct from the effective cosmological constant of the Einstein spacetime $\Lambda$.} 
\begin{equation}
  \mathcal{B} g_{ab} - \gamma \left( C_a^{\phantom{a}cde} C_{bcde} - \frac{1}{4} g_{ab} C^{cdef} C_{cdef} \right) = 0,
  \label{WNE:fieldeqns}
\end{equation}
where 
\begin{equation}
  \mathcal{B} = \frac{\Lambda - \Lambda_0}{2 \kappa}
  + \Lambda^2 \bigg[ \frac{(n-4)}{(n-2)^2} (n \alpha + \beta)
  + \frac{(n-3) (n-4)}{(n-2) (n-1)} \gamma \bigg]. 
  \label{WNE:fieldeqns:B}
\end{equation}

\subsection{Type N Einstein solutions to quadratic gravity}

For type N spacetimes, {one has} $C_a^{\phantom{a}cde} C_{bcde}=0$ (see eq.~\eqref{eq:rscalars}) and $C^{cdef} C_{cdef}=0$ (section~\ref{Sec_invars}), so that  eqs. (\ref{WNE:fieldeqns}) and (\ref{WNE:fieldeqns:B}) reduce to a simple algebraic constraint $\mathcal{B} = 0$  relating the effective cosmological constant $\Lambda$ to the parameters $\alpha$, $\beta$, $\gamma$, $\kappa$, $\Lambda_0$. {These spacetimes are thus ``immune'' to corrections of quadratic gravity.}
\begin{prop}[Type N solutions of QG \cite{MalekPravdaQG}]
\label{prop_QGN}
In arbitrary dimension, all Weyl type N Einstein spacetimes with cosmological constant $\Lambda$ (chosen to obey  $\mathcal{B} = 0$)
are exact solutions of quadratic gravity (\ref{QG:fieldeqns}).
\end{prop}

Many higher dimensional  type N Einstein spacetimes are known and thus in proposition \ref{prop_QGN} we identified a large class of exact vacuum solutions of quadratic gravity. These include type N Kundt spacetimes (see section~\ref{sec_Kundt}) and expanding, shearing (and possibly twisting) type N solutions obtained by (repeatedly) warping known four-dimensional type N Einstein spacetimes {(see \cite{OrtPraPra10} for examples and section~\ref{subsubsec_Brink} for comments on the warping method)}. 

\subsection{Type III Einstein solutions to quadratic gravity}

Clearly, for type III spacetimes $ C^{cdef} C_{cdef} $ vanishes as in the type N case {(see section~\ref{Sec_invars})}. By contrast, the remaining Weyl term appearing in \eqref{WNE:fieldeqns} is in general non-vanishing. From \eqref{eq:rscalars}, it follows
\be
C_a^{\phantom{a}cde} C_{bcde} = \tilde \Psi  \ell_a \ell_b,
 \label{QG_III}
\ee 
where
\be
\tilde \Psi \equiv  \frac{1}{2} \Psi'_{ijk} \Psi'_{ijk} -  \Psi'_i \Psi'_i. \label{tPsi}
\ee
By tracing  \eqref{WNE:fieldeqns}, we get $\mathcal{B} = 0$, {so that \eqref{WNE:fieldeqns} reduces to $\gamma\tilde \Psi=0$}, and thus, (assuming $\gamma\neq0$)  {\it type III Einstein spacetimes are exact vacuum solutions of QG if, and only if, $\tilde \Psi = 0$ $($assuming of course that the effective cosmological constant $\Lambda$ is chosen to obey \eqref{WNE:fieldeqns:B}$)$}.

Let us first observe that {not all type III Einstein spacetimes are solutions of vacuum QG, since there do exist} type III Einstein spacetimes with $\tilde \Psi \not= 0$.  For instance, $\tilde \Psi$ is clearly non-vanishing for the type III(a) subclass of type III spacetimes, characterized by $\Psi'_i =0$ \cite{Coleyetal04} {(section~\ref{subsubsec_spin})}. Take, for example, type III Ricci-flat \pp waves, which are automatically of type III(a)  (proposition~\ref{prop_pp}), {and thus {\em not} vacuum solutions of QG} (an example of such metric is given by \eqref{ppIII}). 
{Similarly, also type III(b) Einstein spacetimes cannot solve QG (unless $n=4$, cf. the definition given in section~\ref{subsubsec_spin}).}

On the other hand, many {other} type III Einstein spacetimes {do} obey $\tilde \Psi = 0$. {To see that this is indeed the case, note first that} in four dimensions $\tilde \Psi=0$ identically (this follows from symmetries of $\Psi'_{ijk}$). {Next, from the results of section~\ref{subsubsec_Brink}, it follows that for Brinkmann's warped metrics \eqref{ansatz} one has}
\be
C_\mu^{\phantom{\mu}  \nu \rho \sigma} C_{\tau \nu \rho \sigma}  = \frac{1}{f} {\tilde C}_\mu^{\phantom{\mu}  \nu \rho \sigma} {\tilde C}_{\tau \nu \rho \sigma},
\ee 
with all $z$-components being zero. {Therefore} $\tilde \Psi$ also vanishes for all type III Einstein spacetimes {of the form \eqref{ansatz}}, i.e. obtained by warping four-dimensional type III Einstein spacetimes.  Such spacetimes are thus also exact solutions of QG.  Various classes of  such Einstein spacetimes are given in  \cite{OrtPraPra10}.

\subsection{Type III and N spacetimes with aligned null radiation}

The above results for Einstein spacetimes may be partially generalized to the case of a Ricci tensor of the form
\begin{equation}
  R_{ab} = \frac{2\Lambda}{n-2} g_{ab} + \chi \ell_a \ell_b . 
  \label{Ricci}
\end{equation}
Assuming further $\tilde \Psi = 0$,  eq. \eqref{QG:fieldeqns} reduces to  \cite{MalekPravdaQG,MalekThesis},
\be
  (\beta \Box + \mathcal{A} ) (\chi \ell_a \ell_b ) = 0,
  \label{QG:fieldeqns:box}
\ee
with 
\be
  \mathcal{A} = \frac{1}{\kappa}
  + 4 \Lambda \bigg( \frac{n \alpha}{n-2}
  + \frac{\beta}{n-1}
  + \frac{(n-3)(n-4)}{(n-2)(n-1)} \gamma \bigg).
  \label{QG:fieldeqns:A}
\ee
{As it turns out \cite{MalekPravdaQG}}, for $\beta \not=0$ eq.~\eqref{QG:fieldeqns:box} implies that $\rho_{ij}$  vanishes, and these spacetimes  thus belong to the Kundt class (section \ref{sec_Kundt}). Explicit examples of such solutions of QG (obtained by solving  \eqref{QG:fieldeqns:box}) can be found in  \cite{MalekPravdaQG,MalekThesis}. Note that these vacuum solutions of QG are not vacuum solutions of Einstein theory due to the non-vanishing Ricci tensor.

\section*{Acknowledgments}

We are grateful to  Mahdi Godazgar, Sigbj{\o}rn Hervik, David Kubi{z}{\v  n}{\'a}k, Ji\v r\'{\i} Podolsk\'y, Harvey Reall and Lode Wylleman for reading the manuscript and for helpful comments. The authors acknowledge support from research plan {RVO: 67985840} and research grant no P203/10/0749.

\end{document}